# Designing the Haystack:
# Programmable Chemical Space for Generative Molecular Discovery

Yuchen Zhu[1], Donghai Zhao[1], Yangyang Zhang[1], Yitong Li[1], Xiaorui Wang[1], Shuwang Li[2], Yue Kong[2], Beichen Zhang[2], Ricki Chen[3], Chang Liu[2], Xingcai Zhang[3*], Tingjun Hou[1*], Chang-Yu Hsieh[1*]

[1] College of Pharmaceutical Sciences, Zhejiang University, Hangzhou, 310058

[2] LEPU MEDICAL TECHNOLOGY (BEIJING) CO., Ltd, Beijing, 102200

[3] Department of Chemical and Nano Engineering, University of California, San Diego, La Jolla, CA, USA

Corresponding authors

Chang-Yu Hsieh

E-mail: kimhsieh@zju.edu.cn

Tingjun Hou

E-mail: tingjunhou@zju.edu.cn

Xingcai Zhang

E-mail: xiz292@ucsd.edu

## Abstract

Chemical space exploration underlies drug discovery, yet most generative models treat chemical space as a fixed, implicitly learned distribution, focusing on sampling individual molecules rather than deliberately designing the space from which they are drawn. We introduce SpaceGFN, a generative framework that elevates chemical space to a programmable computational object -- a controllable degree of freedom that enables explicit construction and adaptive traversal of structured molecular universes. SpaceGFN decouples space definition from exploration: users specify building blocks and reaction rules to construct chemically and synthetically coherent spaces, while a generative flow network (GFlowNet) performs efficient, property-biased sampling within them. In Discovery mode, we instantiate programmable space design through two representative strategies. A Pseudo-natural product space enables the systematic assembly of natural product-like architectures. An evolution-inspired (Evo) space recombines endogenous metabolite fragments through enzyme-consistent transformations, introducing an evolutionary structural prior into chemical generation. This biologically grounded bias induces statistically favorable shifts in predicted metabolic and toxicological profiles while preserving pharmacological diversity, as evidenced by broad docking enrichment across diverse therapeutic targets. In Editing mode, SpaceGFN enables reaction-consistent lead optimization by applying a curated molecular editing toolkit composed of executable synthetic transformations. This allows local, synthesis-aware modification of existing compounds rather than unrestricted graph mutation. Across 96 drug targets, SpaceGFN achieves robust optimization performance while maintaining structural diversity under synthetic constraints. By integrating programmable chemical universe construction with flow-based exploration and reaction-level editing, SpaceGFN establishes a general paradigm for the deliberate design and navigation of therapeutic chemical space.

# Introduction

The quest for therapeutic molecules has evolved from the empirical discovery of natural products to the era of rational, *de novo* molecular design[1]. The size of drug-like chemical space—often estimated to exceed $10^{60}$ molecules—suggests an immense reservoir for innovation[2]. Yet, despite decades of advancement, drug development remains a high-risk, high-attrition endeavor[3]. The central challenge lies in identifying molecules that simultaneously satisfy potency, bio-compatibility, and synthetic accessibility within this vast landscape[4,5].

In recent years, generative AI has emerged as a transformative force, shifting the paradigm from the passive screening of enumerated libraries to the active exploration of chemical space[6]. From variational autoencoders[7] to diffusion models[8], these tools have significantly improved our ability to sample high-quality molecular candidates. However, a fundamental limitation remains: most current models treat chemical space as a fixed, implicitly learned distribution derived from curated databases[9,10]. They operate within a data-defined manifold, without explicitly controlling the structural organization of the chemical space itself.

Here, we argue that the next frontier of molecular design lies in elevating chemical space to a programmable computational object. We introduce SpaceGFN, a framework that decouples space definition from space exploration, transforming the molecular universe from a static constraint into a controllable variable. By integrating programmable, reaction-defined chemical spaces with Generative Flow Networks (GFlowNets), SpaceGFN enables the construction and adaptive traversal of structured universes tailored to specific therapeutic hypotheses. This shift moves the field beyond finding a "needle in a haystack" toward deliberately shaping the "haystack" so that its intrinsic distribution aligns with desired synthetic and biological profiles.

In Discovery mode, we demonstrate the implications of space-level programming through two distinct strategies. First, we construct a Pseudo-natural product (Pseudo-NP) space that systematically assembles architectures recapitulating the structural complexity of natural products—regions of chemical space that remain sparsely explored in many computational libraries[11]. More importantly, we introduce an evolution-inspired (Evo) space constructed from endogenous metabolites and enzyme-catalyzed transformations. Rather than relying solely on post hoc ADMET filtering, Evo space embeds evolutionary biochemistry as a structural prior within the generative process. By leveraging molecular motifs historically processed by living systems, this prior induces statistically favorable shifts in predicted metabolic and toxicological distributions while preserving pharmacological diversity, as evidenced by broad docking enrichment across diverse therapeutic targets.

Parallel to discovery, lead optimization often encounters a disconnect between AI-driven structural modifications and synthetic feasibility. Many generative models optimize via latent-space perturbations or unconstrained graph edits, which can yield synthetically intractable structures. SpaceGFN's Editing mode addresses this limitation by redefining optimization as reaction-consistent local traversal. Using a curated molecular editing toolkit composed of executable synthetic transformations[12], SpaceGFN enables AI-driven "digital medicinal chemistry," in which each proposed edit corresponds to a chemically valid and synthetically reachable step.

We validated SpaceGFN across 96 diverse drug targets, demonstrating robust optimization performance while substantially expanding topological diversity under explicit synthetic constraints. By unifying programmable chemical universe construction with flow-based exploration and reaction-level editing, SpaceGFN establishes a computational framework for the deliberate design and navigation of therapeutic chemical space, bridging generative AI, synthetic methodology, and biological design principles.

# Results

## Discovery and Editing Framework of SpaceGFN

To address the distinct requirements of hit discovery and lead optimization in drug development, we designed two modes of SpaceGFN, Discovery and Editing (Fig. 1a). In our previous work, we introduced SynGFN[13], a generative model based on "molecular building blocks x chemical reactions," which transforms the molecular generation process into the assembly of building blocks through reactions, essentially constructing explicit chemical spaces via combinatorial chemistry. The primary motivation of SynGFN was to enhance the synthetic feasibility of generated molecules. From another perspective, however, this explicit chemical space construction strategy is also well suited for the deliberate design of chemical spaces. Previous studies have mainly focused on improving exploration efficiency within predefined chemical spaces, while paying comparatively little attention to the proactive design of the spaces themselves.

In the Discovery mode, SpaceGFN provides experimental scientists with a DIY chemical space framework (Fig. 1b), in which users can flexibly select building block libraries and reaction libraries according to their needs, thereby constructing novel chemical spaces and using SpaceGFN to efficiently and purposefully explore them in search of molecules with desired properties (e.g., drug activity). As examples, we present two novel types of chemical spaces—Pseudo-NP and Evo—and validate their effectiveness (see Sections 2, 3 and 4 for details). By enabling the deliberate construction of chemical spaces, we can constrain generative AI more tightly to specific tasks, thereby leveraging its advantage in efficient sampling while avoiding the uncontrollable black-box nature of implicit chemical space construction, which lacks interpretability, and ultimately returning greater design and decision-making power to scientists themselves.

Building on the rationale of SynGFN, we further asked: if de novo molecular design can achieve synthetic feasibility through combinatorial chemistry, could molecular optimization also benefit from this approach? However, directly applying this strategy faces two key limitations: (1) reactions used for de novo design are often two-component, with simple substrates that tend to generate new scaffolds rather than modify existing ones, making them unsuitable for fine modifications of complex molecules; and (2) continued reliance on classical synthetic reactions inevitably leads back to the "limited chemical space paradox," remaining confined within the boundaries of commercial chemical space.

To address this, we introduced the concept of molecular editing into the Editing mode of SpaceGFN. Molecular editing aligns intuitively with the demands of molecular optimization: depending on the modification goal, it can be categorized into scaffold editing and peripheral editing, and is characterized by broad substrate scope, high site selectivity, and mild reaction conditions. More importantly, it can condense multistep syntheses that previously required dozens of transformations into a single step, and even enable structural modifications that were previously inaccessible, thereby greatly enhancing the novelty of optimized molecules. We systematically organized nearly a decade of molecular editing methods into a reaction template dataset, Edit Rule V1 (see Section 5 and Methods for details), and incorporated it as a chemical constraint in the Editing mode (Fig. 1c). In this way, the Editing mode achieves molecular optimization while simultaneously providing concrete synthetic routes, with each reaction step corresponding to a molecular editing transformation. Traditional medicinal chemists are often constrained by the complexity of multistep synthesis and thus limited to a narrow set of optimization options, whereas earlier generative AI–based molecular optimization approaches, although ensuring a certain degree of structural plausibility, lacked synthetic constraints. The Editing mode of SpaceGFN establishes a balance between creativity and feasibility: it retains design options that would otherwise be discarded due to excessive synthetic steps while filtering out virtual modifications that are synthetically implausible.

In terms of algorithmic performance, we further optimized the model architecture of SynGFN to improve exploration efficiency under large-scale building block libraries (Fig. 1d). On a library of approximately 100,000 building blocks, the exploration efficiency was enhanced by about threefold compared with SynGFN (Fig. S1). Detailed strategies for these adjustments and their evaluations are provided in the Methods and Supplementary Information (Fig. S2–S4).

## Construction and Validation of the Pseudo-NP Chemical Space

Natural products (NPs) have long served as an important source of inspiration for drug discovery owing to their unique molecular properties and structural features[14]. However, evolutionary constraints limit their scaffold diversity, making it difficult to cover broader regions of chemical space[15]. To address this, Waldmann and colleagues proposed the concept of pseudo-natural products (PNPs)[11]: by combining fragments derived from natural products, one can generate novel molecules absent in nature, thereby expanding chemical space while retaining the characteristic features of natural products. Previous attempts, however, have largely focused on small-scale fragment assembly, typically involving only a few hundred molecules, which restricts systematic exploration[16,17].

In this study, we employed the Discovery mode of SpaceGFN to explicitly construct a Pseudo-NP chemical space within the DIY chemical space framework and explored it under the guidance of drug activity (Fig. 2a). We selected a structurally diverse fragment set extracted by Waldmann and colleagues from natural product databases[18]. These fragments not only exhibit physicochemical properties closer to those of natural products (e.g., higher oxygen content and lower nitrogen content) but also contain readily connectable chemical "handles," thereby improving synthetic accessibility (Fig. S5, Table S1). As a control, we simultaneously constructed a synthetic chemical space based on commercially available building blocks. The two spaces displayed marked differences in reaction preferences (Fig. 2b), molecular space distribution (Fig. 2c), and physicochemical properties (Fig. 2d, Fig. S6–S7).

To evaluate the potential of the Pseudo-NP space in drug discovery, we selected four important targets—epidermal growth factor receptor (EGFR), fibroblast growth factor receptor 1 (FGFR1), proto-oncogene tyrosine-protein kinase Src (SRC), and vascular endothelial growth factor receptor 2 (VEGFR2)—and compared the sampling results of SpaceGFN in the Pseudo-NP space versus the synthetic space. UMAP analysis revealed a clear separation in the distributions of molecules generated from the two spaces (Fig. 2e), a difference that was also evident at the level of physicochemical properties (Fig. S8–S15, Table S2). Molecules from the Pseudo-NP space exhibited physicochemical properties more consistent with those of natural products. Further comparison with structurally similar active compounds in ChEMBL[19] demonstrated that molecules from the Pseudo-NP space overall showed greater novelty (Fig. 2f).

We subsequently evaluated the activity score distributions of molecules sampled from the two spaces (Fig. 2g) and found that the Pseudo-NP space exhibited an overall tendency toward higher potential activity across multiple targets. Further molecular docking results (Fig. S16) demonstrated that molecules from the Pseudo-NP space were able to adopt reasonable conformations within the binding pockets of multiple targets and displayed favorable binding energies.

Finally, we evaluated the sampled molecules using NP likeness[20] as a metric, and the results showed that molecules generated from the Pseudo-NP space were significantly closer to the characteristics of natural products (Fig. 2h). This indicates that the Pseudo-NP chemical space we constructed not only expands the natural product space but also overcomes the limitations of commercial chemical space. Through the Discovery mode of SpaceGFN, we achieved a large-scale and systematic explicit construction and exploration of the Pseudo-NP chemical space, thereby providing an alternative avenue for discovering molecules with both high novelty and strong natural product–like characteristics.

## Concept and Initial Validation of the Evo Chemical Space

The original motivation for constructing the Pseudo-NP space was to exploit the differences between natural products and commercial chemical space, thereby overcoming the narrow boundaries imposed by conventional building blocks and classical reactions. By broadening the natural product chemical space, we demonstrated that drug discovery can extend into truly uncharted regions. In drug development, structural novelty is often regarded as a key driver of breakthrough innovation. However, even when entirely new molecular scaffolds are identified, they cannot advance to the clinic without favorable ADMET properties[21,22]. Thus, ADMET has long represented a bottleneck in drug discovery. Existing AIDD (artificial intelligence drug discovery) strategies can be broadly divided into two categories[23]: one involves coupling generative models with ADMET prediction tools during the iterative process, using prediction scores as optimization feedback; the other generates large molecular libraries first and subsequently screens candidate molecules with prediction models. Both approaches, however, are fundamentally limited by the scarcity of high-quality ADMET data[21]. For many ADMET-related properties, experimental datasets contain fewer than a thousand cases, and toxicity data in particular are hindered by the prevalence of rare but project-terminating adverse effects. Such issues are difficult to resolve simply by expanding training datasets. Consequently, a sole reliance on prediction-based post hoc evaluation is unlikely to provide a true solution to the ADMET bottleneck.

To overcome this challenge, we propose a reverse-thinking strategy: if, at the very source of chemical space design, one could construct a space naturally predisposed to favorable ADMET properties, the risks of rare adverse effects and unpredictable toxicities might be fundamentally reduced. Based on this rationale, we introduce the concept of the Evo space (Fig. 3a).

The concept of the Evo space is supported by several theoretical considerations. First, metabolites themselves have been shown to be of exceptional value in drug discovery—nearly 50% of approved drugs can be traced back to metabolites or their derivatives[24,25]. These molecules are inherently adapted to the human physiological environment, and their physicochemical properties often align well with ADMET requirements. Second, the human metabolic network is the product of millions of years of evolution and natural selection. Existing metabolic pathways represent the "survivors" under specific selective pressures, but this does not imply that other molecular species never emerged during evolutionary history. Molecules that were eliminated may simply have lacked competitive advantages, rather than being intrinsically incompatible with human physiology. These facts provide a rational basis for the hypothesis underlying the Evo space: if chemical space is constructed from endogenous metabolites and enzymatic reactions as the fundamental units, then such a space may, in aggregate, be inherently biased toward favorable ADMET properties. Of course, whether this hypothesis holds must be validated through systematic experimental studies, and in the following sections we provide supporting evidence through a series of comparative analyses.

In our experimental validation, we constructed two types of Evo spaces based on metabolites from the HMDB database[26] and associated enzymatic reactions[27]: Evo-Narrowspec and Evo-Broadspec, which differ in the scope of metabolite selection (the former includes only endogenous molecules present in samples such as blood, saliva, cerebrospinal fluid, and breast milk). As controls, we designed six additional spaces, including synthetic chemical spaces of comparable block size to Evo (Synthetic-S and Synthetic-M), as well as four types of hybrid spaces formed by cross-combinations (Evo-N/B+SynT and Syn-S/M+EvoT), to test whether the specificity of Evo truly arises from the combination of "endogenous molecules x endogenous reactions." Further experimental details are provided in the Methods section.

We first compared the property distributions of different building block libraries. Among 35 ADMET-related descriptors, the Evo spaces were largely comparable to synthetic building block libraries on most measures, but exhibited superior performance in certain key attributes (Fig. S17–S21). For example, Evo showed more favorable distributions in absorption-related properties such as Pgp-inhibitor and F30%, metabolism-related properties such as CYP1A2-inhibitor and CYP2C19-inhibitor, and toxicity-related properties such as mutagenicity and carcinogenicity. These results provide partial validation of the rationale for using metabolites as inspiration in drug design. In terms of reaction preferences, the building block libraries of the two Evo spaces showed high similarity (Fig. S22). Moreover, in physicochemical property distributions (Fig. S23), Evo and synthetic spaces exhibited marked differences.

We further evaluated the exploration results of SpaceGFN on four targets: EGFR, FGFR1, SRC, and VEGFR2. In terms of ADMET distributions, molecules sampled from the Evo spaces exhibited consistent characteristics (Fig. S24–S43). Among seven absorption-related properties, Evo performed less favorably than the controls in MDCK permeability but was significantly superior to synthetic spaces in Pgp-inhibitor and HIA. For distribution-related properties, Evo spaces and their hybrid variants outperformed in PPB and Fu. Among ten metabolism-related properties, Evo achieved better performance in all except CYP1A2-inhibitor. For thirteen toxicity-related properties, only the Evo spaces consistently maintained values within the desirable range. Taken together, these results indicate that the Evo spaces indeed display distributional profiles distinct from those of synthetic spaces along the ADMET dimension.

In terms of activity distributions, the Evo spaces showed slightly lower activity scores across multiple targets compared with synthetic spaces (Fig. 3c), but molecular docking results (Fig. 3b) demonstrated that Evo was still capable of generating molecules with strong activity. This can be attributed to the efficient exploration ability of SpaceGFN and indicates that the uniqueness of the Evo spaces does not compromise their potential for drug discovery. Comparisons of spatial distribution (Fig. S44), diversity (Fig. S45), and novelty (Fig. 3d) further highlighted the distinct features of the Evo spaces.

## Assessing ADMET Potential of the Evo Chemical Space

To evaluate the overall potential of the Evo space, we enumerated the Evo-Narrow space to construct a subspace containing 1.8 billion molecules and compared it with the Synthetic-S space. The results revealed clear differences in the overall distributions between Evo and synthetic spaces (Fig. 3f). The similarity between Evo molecules and the original metabolite library was found to be low (Fig. S46), indicating that Evo is not merely an extension of metabolites but rather represents an unexplored novel chemical space.

To more intuitively validate the advantages of Evo, we set thresholds for 35 ADMET properties and calculated the proportion of sampled molecules meeting these criteria. The results (Fig. 3e, Fig. S47–S50) showed that the Evo spaces significantly outperformed synthetic spaces on 28 properties, particularly in the key categories of metabolism and toxicity. The main disadvantages were observed in Caco-2 and MDCK permeability. This was further corroborated by physicochemical property analyses: molecules from the Evo spaces exhibited favorable profiles in properties such as logP, logD, and logS, but had markedly higher TPSA (topological polar surface area) compared with synthetic spaces (Fig. 3g–h, Fig. S51), reflecting their overall greater polarity. Importantly, this disadvantage can also be viewed positively. On one hand, it indicates that the Evo space still has room for further refinement and optimization; on the other hand, among current ADMET prediction tasks, absorption-related properties are the most likely to be addressed first. Their critical determinants are largely physicochemical in nature, with clearer rules, stronger modelability, and greater predictive generalizability[28]. Thus, Evo molecules can be effectively filtered with high confidence using accurate absorption prediction tools, while absorption deficiencies can often be compensated by formulation strategies or drug delivery approaches.

The true value of the Evo space lies in its overall potential to reduce ADMET risks. Our results demonstrate that, compared with synthetic chemical spaces, Evo molecules exhibit pronounced advantages in metabolic stability and toxicological safety, while also showing favorable distributions across several other key ADMET properties. Given the inherent uncertainties of current ADMET prediction tools, which stem from the scarcity of high-quality data, greater reliance should be placed on the intrinsic bias of the Evo space—its natural tendency to generate molecules with more favorable ADMET properties. This implies that candidate molecules selected from the Evo space, relative to those from conventional chemical spaces, carry a lower overall ADMET risk and hold a greater likelihood of avoiding development failures caused by unpredictable properties.

## Molecular Editing with SpaceGFN

Through the two representative cases of the Pseudo-NP and Evo spaces, we demonstrated the substantial potential of the Discovery mode of SpaceGFN in DIY chemical space construction. However, from the perspective of the history and practice of drug development, any promising molecule must undergo further structural modifications before it can exhibit desirable performance in wet-lab experiments. As noted above, both medicinal chemists' experience-driven optimization and the generative optimization of existing AI tools are inevitably constrained by synthetic complexity: the former often limits imagination due to synthetic difficulties, while the latter frequently generates molecules lacking practical feasibility. To overcome this bottleneck, we introduced the concept of molecular editing from synthetic chemistry into SpaceGFN and proposed a novel Editing mode.

From the perspective of structural modification, molecular editing can be broadly divided into two categories[29]: scaffold editing and peripheral editing (Fig. 4a). Scaffold editing focuses on modifying the core skeleton of a molecule, including transformations such as ring expansion, ring contraction, and atom replacement within rings. Yu and colleagues further classified this into single-atom editing, multi-atom editing, and macrocyclization[30]. Peripheral editing, in contrast, emphasizes the introduction or exchange of functional groups on the scaffold, with common strategies including C–H activation, functional group exchange, and skeletal remodeling. While these strategies have already demonstrated great potential in synthetic chemistry, this work represents the first systematic introduction of such approaches into generative AI–driven molecular design.

To enable the integration of molecular editing with generative AI, we first constructed the corresponding dataset (Fig. 4b). We systematically collected and reviewed molecular editing literature from the past decade and, based on detailed analysis, retained only those methods that exhibited mild reaction conditions, high selectivity, and broad substrate scope. Considering both the number of reports and practical potential, we ultimately focused on four representative categories of editing reactions: single-/multi-atom editing, C–H activation, and functional group exchange. Through this stringent selection process, we compiled 248 relevant publications. Each method was then manually converted into a reaction template using SMARTS descriptions. Although automated tools are now available to extract SMARTS from reaction text[31], their error rates remain high for complex editing reactions; therefore, we employed manual curation to ensure maximal accuracy. In the end, we constructed, to the best of our knowledge, the first reaction template dataset centered on molecular editing—Edit Rule V1. This dataset currently contains 300 high-quality reaction templates, not only supporting the operation of the Editing mode of SpaceGFN but also providing a transferable resource for downstream AI+chemistry tasks such as retrosynthetic planning.

To demonstrate the effectiveness of the Editing mode, we selected two case studies of drug molecule optimization (Fig. 4c). The results showed that the Editing mode not only ensures synthetic feasibility but also provides clear synthetic routes for molecular optimization, with each reaction step corresponding to a specific molecular editing method. Importantly, unlike previous generative AI–based optimization approaches that typically decompose the task into single objectives such as scaffold hopping or side-chain decoration[32], the Editing mode of SpaceGFN, by leveraging cascade reactions, can naturally realize multiple modification strategies within a single optimization pathway while ensuring that the final molecule remains accessible through practically feasible synthetic routes.

Furthermore, we evaluated the "plug-and-play" capability of the Editing mode (Fig. 4d). We selected two recently reported molecular editing methods[33,34] and, following the same processing workflow, rapidly incorporated them into the Edit Rule V1 dataset. The results showed that these methods could be immediately applied to molecular optimization tasks. Given that molecular editing remains a rapidly evolving research area, it is reasonable to expect that more and more novel editing strategies are bound to emerge. With its modular and open architecture, the Editing mode of SpaceGFN can continuously incorporate the latest editing tools, enabling iterative updates in step with advances in the field. This makes it not merely an algorithmic framework but a sustainable and evolving platform that can expand the possibilities of molecular optimization in parallel with the progress of synthetic methodology.

## Large-Scale Validation of the Editing Mode

Through the case studies presented above, we have highlighted the unique advantages of the Editing mode: it not only provides feasible synthetic routes while optimizing molecules, but also enables multiple structural modifications to be carried out in parallel within a single optimization pathway, and offers plug-and-play extensibility. To further validate the feasibility and generalizability of integrating molecular editing with generative AI in molecular optimization tasks, we conducted a large-scale validation experiment.

Specifically, based on the approved drug target atlas constructed by Santos and colleagues[35], and following rigorous data processing (see Methods), we ultimately selected 96 protein targets covering several major human protein families, including kinases, G protein-coupled receptors (GPCRs), ion channels, and nuclear receptors (Fig. 5a–b). This test set, in terms of target number, distribution of target types, as well as the scale of initial molecules to be optimized (Fig. S52) and their activity score distributions (Fig. S53), effectively simulates the complexity and diversity of real-world drug optimization tasks, thereby offering a realistic yet idealized benchmark for assessing the generalizability and robustness of molecular optimization algorithms.

We first examined the molecular optimization performance of the Editing mode across 96 targets, using Unidock[36] docking scores as the activity evaluation metric. Statistics of optimization success rates (Fig. 5c) showed that SpaceGFN exhibited robust performance under different thresholds: molecular optimization (score improvement > 0.1 kcal/mol) was achieved for 98.80% of targets, with 84.20% of targets showing improvements greater than 1 kcal/mol and 45.18% exceeding 2 kcal/mol. These results indicate that the Editing mode not only possesses broad applicability but also maintains strong stability in multi-target tasks. Further analysis revealed that the optimized molecules exhibited increases of 76% in scaffold count and 98% in Circles number (an important indicator reflecting topological diversity[37]) (Fig. 5d–e, Table S3), while their activity score distributions were significantly improved (Fig. 5f), with an average enhancement of 1.35 kcal/mol (Fig. 5g). Collectively, these findings demonstrate that the Editing mode can concurrently improve molecular activity and expand structural diversity.

After validating the effectiveness of the Editing mode, we further assessed the novelty of the optimized molecules. By comparing the top 10 active molecules for all targets (Fig. 5h), we found that approximately 65% of molecules exhibited structural similarity below 0.4 between the pre- and post-optimization states. This indicates that candidate molecules generated by the Editing mode not only achieve significant activity improvements but also possess structural novelty. Indeed, the primary motivation for introducing molecular editing is to integrate state-of-the-art synthetic methodologies in order to overcome the dependence of traditional molecular optimization on classical synthetic strategies, thereby injecting new chemical creativity into the optimization process.

Finally, we conducted an exploratory statistical analysis to investigate the practical usage preferences of different molecular editing strategies in large-scale, real-world optimization scenarios. Using optimized molecules (score improvement $\geqslant$ 0.5 kcal/mol) as the basis, we calculated the frequencies of editing methods employed (Fig. 5i) and highlighted representative high-frequency reaction templates (Fig. 5j). The results showed that the most frequently applied editing strategies during optimization included: aromatic substitutions (e.g., amination, hydroxylation, halogenation), heterocycle incorporation (particularly nitrogen-containing heterocycles), and cross-coupling of small substituents (Fig. S54). This finding carries important implications: on the one hand, it provides guidance for future updates of the Edit Rule dataset by prioritizing the inclusion of more practically useful editing reactions; on the other hand, it offers synthetic chemists a realistic reference by indicating which molecular editing strategies are most likely to add value in drug optimization, thereby accelerating the transition of molecular editing from methodology to application.

## Discussion

Chemical space has traditionally been treated as an implicit backdrop for molecular discovery—a vast and largely unstructured landscape from which algorithms attempt to retrieve promising candidates. In this work, we propose a fundamental shift in perspective: chemical space itself can be treated as a **programmable computational object**. Rather than sampling from a static, data-defined manifold, SpaceGFN allows researchers to explicitly define the generative universe through reaction rules and building blocks, and then efficiently traverse that structured space using GFlowNets. In this view, construction and exploration become decoupled but coordinated processes.

Our results suggest that this additional degree of freedom—space-level programming—can meaningfully influence the distributional properties of generated molecules and, crucially, escape the "data-anchoring" trap of traditional generative AI. Most data-driven models remain anchored to the implicit distributions of existing human-made libraries, effectively interpolating within known manifolds. By contrast, SpaceGFN enables the explicit construction of uncharted territories. By defining spaces through fundamental biochemical rules (as in Evo space) or structural motifs (as in Pseudo-NP), we can access molecular scaffolds that are structurally distinct from those in public repositories yet remain chemically coherent and familiar to medicinal chemistry logic. This demonstrates that programmable constraints are not merely restrictive; they serve as a scaffold for reaching novel, bio-relevant regions previously overlooked due to the historical biases of synthetic chemistry.

We emphasize that Evo space does not assume that metabolite-derived structures are intrinsically bioactive or pharmacologically superior. Rather, it encodes an evolutionary structural prior that biases exploration toward regions of chemical space historically compatible with biological systems. Our docking and ML-based ADMET analyses indicate statistically favorable shifts in predicted metabolic and toxicological distributions without collapsing pharmacological diversity. However, these conclusions remain computational in nature. Docking scores are approximations of binding affinity, and ML-based ADMET predictions inherit the biases and uncertainties of their training data. Experimental validation will ultimately be required to determine whether the Evo prior translates into improved success rates in practice.

In Editing mode, we address the persistent disconnect between generative modeling and synthetic feasibility. Many contemporary molecular optimization strategies rely on latent-space perturbations or unconstrained graph edits, which frequently yield structures that are difficult or unrealistic to synthesize. By constraining modifications to a curated set of executable transformations, SpaceGFN reframes optimization as reaction-consistent local traversal. Importantly, because every trajectory in SpaceGFN is composed of discrete, valid reaction steps, the framework provides an **implicit 'forward-synthetic' trace**. This significantly simplifies subsequent retrosynthetic validation compared to latent-space models, where the path from the original lead to the optimized candidate is often chemically opaque.

Several limitations warrant consideration. First, all validations in this study are computational; these results reflect *in silico* performance metrics rather than empirical pharmacokinetic outcomes. Second, programmable spaces are only as expressive as the reaction rules and building blocks used to construct them. However, we view this "DIY" nature as a strength: SpaceGFN does not attempt to replace human expertise, but rather provides a medium for medicinal chemists to formalize their domain-specific intuition into executable structural priors. Third, while the current Edit Rule V1 toolkit contains 300 transformations, it is not exhaustive and does not yet capture the full complexity of reaction conditions or selectivity, which remains an important direction for future integration.

Looking forward, programmable chemical space design may benefit from adaptive refinement loops in which experimental feedback reshapes the underlying space definition. One could imagine dynamically adjusting reaction priors or building block pools based on iterative synthesis and biological testing. In this sense, SpaceGFN provides the computational scaffold for a closed-loop paradigm, where chemical universe design and empirical validation co-evolve.

We do not claim that programmable chemical space supersedes existing generative approaches; rather, it complements them by introducing an additional axis of control. By unifying space construction, flow-based traversal, and reaction-level editing, SpaceGFN suggests that future advances in molecular design may depend not only on better sampling algorithms, but also on the deliberate engineering of the spaces those algorithms explore. This perspective shifts the focus of molecular design from searching within fixed universes toward designing chemical universes that are intrinsically aligned with therapeutic or other fucntional objectives.

# Method
# SpaceGFN

Generative Logic

SpaceGFN is built upon the framework of Generative Flow Networks (GFlowNets)[38], modeling the molecular generation process as a trajectory of stepwise reactions. Each molecular generation trajectory is composed of a sequence of states and actions.

(1) State: represents the current partial molecular structure; the initial state is an empty molecule or a predefined substrate, and the terminal state is a fully generated molecule.

(2) Action: at each step, the model first selects a reaction template and then chooses the corresponding reactants.

During training, the objective of SpaceGFN is to learn a sampling distribution such that the probability of generating a molecule is proportional to its reward (e.g., activity). To achieve this, we adopt Trajectory Balance (TB)[39] as the optimization objective, ensuring that sampled molecules are both diverse and biased toward high-reward regions. This framework transforms molecular generation from simple fragment concatenation into a process driven by stepwise chemical reactions, thereby guaranteeing both chemical validity and synthetic feasibility of the generated molecules.

Algorithmic Improvements

SpaceGFN builds upon our previous synthesis-aware molecular generation framework, SynGFN[13], which introduced a reaction-defined GFlowNet formulation for constructing molecules from building blocks under explicit reaction constraints. On this shared foundation, SpaceGFN introduces the following extensions.

(1) Unlike SynGFN, which uses a fixed number of reaction steps, SpaceGFN allows users to flexibly adjust the reaction length according to task requirements.

(2) For the representation of reaction templates, SpaceGFN replaces one-hot encoding with RXNFP[40], a reaction fingerprint derived from BERT (Bidirectional Encoder Representations from Transformers). Compared with one-hot encoding, RXNFP captures similarities between different reactions and improves the chemical semantic relevance of representations.

(3) For the policy network of reaction and reactant selection, SpaceGFN provides two alternative strategies.

(i) Discrete indexing strategy: the policy model outputs probabilities aligned with the size of the reaction/reactant library, directly yielding sampling probabilities. This is the strategy used in SynGFN.

(ii) Fingerprint embedding strategy: the policy model outputs a fixed-dimensional embedding. Specifically, all reaction/reactant fingerprints are precomputed as a matrix, and the output of the policy model is multiplied by the transpose of this matrix to indirectly yield sampling probabilities.

Benchmarking Against SynGFN

We benchmarked SpaceGFN against SynGFN using soluble epoxide hydrolase (sEH) as the target. Under four building block library scales previously constructed in the SynGFN study, the activity of sEH (evaluated with a QSAR model provided by Bengio and colleagues[41]) was used as the training reward. The results showed that under larger library conditions (L and XL, corresponding to 30k and 100k building blocks, respectively), SpaceGFN achieved higher exploration efficiency compared with SynGFN.

Comparison of Policy Network Output Strategies

We further compared the discrete indexing and fingerprint embedding strategies on the sEH target. By evaluating performance metrics including reward trajectories during training, distributions of sampled activity scores, and diversity curves, we found that the two strategies exhibited complementary advantages in different scenarios: the discrete indexing strategy was more efficient in small-scale reactant libraries, whereas the fingerprint embedding strategy performed better in large-scale settings.

# Programmable Chemical Space Framework

In the Discovery mode of SpaceGFN, we provide a DIY chemical space framework. Users can flexibly construct the desired chemical space by leveraging built-in building block libraries (e.g., synthetic building block libraries, natural product–derived building blocks, metabolite-derived building blocks) and reaction libraries (e.g., classical synthetic reactions, enzymatic reactions), or by introducing customized datasets of building blocks and reactions. All inputs can be standardized through customizable preprocessing pipelines (e.g., restrictions on element ranges, ring counts and sizes, and reaction matchability filtering) and then converted into the standard data format supported by SpaceGFN for subsequent model training and space exploration. In the testing phase, after training SpaceGFN separately on the Pseudo-NP, Evo, and various control spaces, we uniformly sampled 10,000 molecules from each space for evaluation.

# Pseudo-NP chemical space

We constructed the Pseudo-NP chemical space using the fragment library provided by Waldmann and colleagues as the building block set, together with 87 classical synthetic reactions curated in our previous SynGFN study as the reaction library. After reaction matchability filtering, 1,666 natural product–derived fragments remained available for use. It should be noted that while companies such as Enamine and Life Chemicals also provide natural product fragment libraries, physicochemical analysis revealed that the fragment library from Waldmann et al. exhibits properties more characteristic of natural products[18]. Specifically, it contains a higher proportion of sp³ carbons, oxygen atoms, and non-aromatic rings, along with lower counts of nitrogen atoms, halogen atoms, and aromatic rings.

To compare the Pseudo-NP space with a commercial chemical space (synthetic chemical space), we constructed a synthetic chemical space using the Enamine building block library (containing 275,951 valid building blocks after data preprocessing). From this set, 1,500 fragments were randomly sampled—matching the scale of the Pseudo-NP building block library—combined with the same 87 synthetic reactions.

For evaluation, we selected four protein targets: epidermal growth factor receptor (EGFR), fibroblast growth factor receptor 1 (FGFR1), proto-oncogene tyrosine-protein kinase Src (SRC), and vascular endothelial growth factor receptor 2 (VEGFR2). The corresponding QSAR activity prediction models for each target were taken from Terayama et al[42]. These targets were chosen because they have relatively abundant activity data and more reliable predictive models.

In the novelty analysis, active compounds for different targets from ChEMBL served as reference molecules; these data were also obtained from Terayama et al. based on ChEMBL version 28. For the NP-likeness evaluation, natural products were used as references, taken from the COCONUT dataset comprising 401,392 natural products. Each molecule was first scored using the QSAR models described above, and for each target, the top 100,000 scoring molecules were selected as reference natural products. We employed AutoDock Vina as the docking tool.

## Evo chemical space

Construction of the Evo Chemical Space

The Evo building block libraries were derived from the Human Metabolome Database (HMDB). Since the concept of Evo spaces is based on combining endogenous molecules with endogenous reactions, we focused on metabolites as the molecular sources. HMDB provides a comprehensive collection of human small molecules, which can be filtered according to four categories: metabolite status (detected and quantified; detected but not quantified; expected but not quantified; predicted), biospecimen (including blood, urine, saliva, cerebrospinal fluid, feces, sweat, breast milk, bile, amniotic fluid, and other biospecimens), origin (exogenous, endogenous, food, plant, microbial, toxin/pollutant, cosmetic, drug, drug metabolite), and cellular location (cell membrane, cytoplasm, nucleus, mitochondria).

We applied two different filtering conditions to construct HMDB-Broadspec and HMDB-Narrowspec libraries. In both cases, all metabolite statuses except predicted were included, to ensure reliability; in future work, predicted metabolites may also be considered to expand the theoretical scale of Evo. For biospecimens, HMDB-Broadspec applied no restrictions, whereas HMDB-Narrowspec only included blood, saliva, cerebrospinal fluid, and breast

milk—chosen to represent samples strongly associated with metabolic activity and relatively safe from toxic accumulation. For origins, HMDB-Broadspec included endogenous, food, plant, microbial, drug, and drug metabolite sources, while HMDB-Narrowspec strictly limited selection to endogenous molecules.

Because HMDB metabolites are generally defined in the range of 50–1500 Da[25], while typical small-molecule drugs have molecular weights around 500 Da, we fragmented the HMDB-Narrowspec and HMDB-Broadspec sets to construct fragment libraries. Following the protocol used by Waldmann for natural product–derived fragments[18], molecules were recursively decomposed using BRICS, generating raw fragment libraries. These were further filtered by physicochemical criteria (AlogP < 3.5; molecular weight 120–350 Da; $\leqslant$3 hydrogen bond donors; $\leqslant$6 hydrogen bond acceptors; $\leqslant$6 rotatable bonds), yielding final fragment sets of 1,622 (HMDB-Narrowspec) and 8,849 (HMDB-Broadspec). After additional preprocessing with the SpaceGFN DIY framework, 1,522 (HMDB-Narrowspec) and 8,197 (HMDB-Broadspec) valid building blocks were obtained.

The Evo reaction library was constructed from RetroBioCat[27], a retrosynthetic design tool for biocatalysis developed by Finnigan and colleagues. RetroBioCat provides 135 generalizable biocatalytic reaction rules. We refined this rule set by correcting and curating templates, ultimately obtaining 125 reaction rules for the Evo reaction library.

Combining the HMDB-Narrowspec or HMDB-Broadspec fragment libraries with the Evo reaction library yielded the Evo-Narrowspec and Evo-Broadspec spaces, respectively.

Control Experiments

To benchmark Evo against synthetic chemical spaces, we constructed Synthetic-S and Synthetic-M by randomly sampling 1,500 and 8,000 building blocks, respectively, from the Enamine library (after preprocessing), matched in size to the Evo fragment libraries, and combining them with 87 classical synthetic reactions. To further test whether the distinctiveness of Evo arises from the "endogenous molecules × endogenous reactions" combination, we designed four additional hybrid spaces: (i) metabolic fragments with synthetic reactions (Evo-N/B+SynT) and (ii) synthetic fragments with enzymatic reactions (Syn-S/M+EvoT). For Evo and all control spaces, the maximum reaction depth was set to two steps.

ADMET Prediction Tools

To evaluate ADMET properties, we used an enhanced version of ADMETlab 2.0, optimized for the DrugFlow drug discovery platform[43]. Compared with the standard ADMETlab 2.0, this version uses updated datasets and retrains models with the MGA algorithm, selecting the best-performing weights for each task based on test set performance.

ADMET Properties

Absorption-related

Caco-2 permeability: predicts intestinal permeability. Reported as log values; > –5.15 indicates good permeability, otherwise poor.

MDCK permeability: predicts in vivo absorption. Reported as log values; > –5.70 indicates good permeability, otherwise poor.

Pgp-inhibitor: whether the molecule is an inhibitor of P-glycoprotein (P-gp), a membrane efflux transporter acting as a detoxification barrier.

Pgp-substrate: whether the molecule is a substrate of P-gp, which may reduce absorption and bioavailability.

HIA (human intestinal absorption): reported as probabilities (0–0.3: good; 0.3–0.7: moderate; 0.7–1.0: poor).

F20% / F30%: oral bioavailability thresholds of 20% and 30%, respectively, reported as probabilities (0–0.3: good; 0.3–0.7: moderate; 0.7–1.0: poor).

Distribution-related

PPB (plasma protein binding): percentage bound to plasma proteins. <90% considered good; >90% considered excessive binding.

VD (volume of distribution): L/kg; 0.04–20 considered good.

Fu (fraction unbound): percentage unbound drug in plasma; ≥5% considered good.

Metabolism-related

Predictions cover CYP450 isoforms (1A2, 3A4, 2C9, 2C19, 2D6), which metabolize over 80% of marketed drugs. Results are probabilities (0–1), with lower values preferred.

Excretion-related

CL (clearance): ml/min/kg; ≥5 good, <5 poor.

T1/2 (half-life): probability values (0–0.3: good; 0.3–0.7: moderate; 0.7–1.0: poor).

Toxicity-related

All predicted as probabilities (0–0.3: good; 0.3–0.7: moderate; 0.7–1.0: poor), including: hERG (cardiotoxicity), hepatotoxicity, mutagenicity, rat oral acute toxicity, carcinogenicity, respiratory toxicity, NR-AR (androgen receptor), NR-ER (estrogen receptor), SR-ARE (oxidative stress), SR-ATAD5 (genotoxicity), SR-HSE (heat shock response), SR-MMP (mitochondrial toxicity), SR-p53 (DNA damage response).

Space Enumeration Experiments

To directly examine the intrinsic properties of Evo, we enumerated the Evo-Narrowspec space with a maximum of two reaction steps. All possible one-step products were generated (2,523,348 molecules), followed by enumeration of second-step products. To reduce computational cost, only 10% of second-step reactions were sampled, yielding 1,869,106,931 molecules in total.

As a control, the Synthetic-S space was enumerated similarly, generating 807,966 one-step products and 624,190,634 two-step products.

For downstream analysis (e.g., ADMET evaluation), we conducted 10 rounds of random sampling, with 8,000 molecules drawn in each round.

To evaluate success rates for ADMET predictions, thresholds were defined as follows: for probability-based indices (0–0.3: good; 0.3–0.7: moderate; 0.7–1.0: poor), success was defined as <0.3; for parameters with specific cutoffs (e.g., PPB <90%), the defined threshold was used; for metabolism-related indices, we also applied a 0.3 cutoff (lower values being preferable).

## Editing Mode

Construction of Edit Rule V1

We collected and curated molecular editing literature according to several editing categories, such as single-atom editing and multi-atom editing, using keyword searches and manual review. Although the concept of molecular editing was formally introduced in 2019[12], related research had already been widely reported earlier; therefore, we included literature from 2015 onward. Based on our background in synthetic chemistry, we selected studies that demonstrated mild reaction conditions, broad substrate scope, single-step transformations, and high selectivity.

Importantly, the presence or absence of explicit drug optimization case studies was not used as a filtering criterion. Given the variability in substrates and combinatorial differences within the Editing mode, we did not require the selected editing strategies to have direct pharmacological effects. Instead, our focus was on identifying novel molecular transformation strategies. Regarding reaction conditions, we did not specifically target photocatalysis, electrocatalysis, or other emerging methodologies, but rather emphasized whether the methods could effectively achieve molecular editing outcomes.

Among the collected editing strategies, C–H activation is particularly noteworthy as a classical approach to drug modification[44]. For this category, we required strong regioselectivity: when functionalization on ring systems could potentially yield multiple products at comparable yields, such methods were excluded due to incompatibility with protecting groups. Only regioselective C–H activation strategies with clearly defined modification sites were retained. Moreover, many early C–H activation methods required directing groups, which are often structurally complex and unsuitable as general substrates for reaction templates. In such cases, we treated directing groups as special protecting groups and tracked the pre-modification structures as substrates. For simple directing groups, however, they were directly incorporated as part of the substrate.

In addition to the four editing categories currently represented, other editing concepts such as functional group exchange, skeletal remodeling, and positional rearrangement also exist. However, due to the limited number of published examples, they were not included at this stage. As these strategies mature, future versions of Edit Rule will expand to include them. We note that stereochemical editing and isotope-based atom–atom exchange are also part of molecular editing, but were excluded here because accurate activity prediction—essential for molecular optimization—is not feasible with current tools for chiral inversions or isotopic substitutions.

After literature collection, we manually compiled molecular editing methods into reaction templates. Although automated tools for reaction template extraction exist, they often perform poorly in atom mapping, particularly for complex editing reactions. To maximize applicability, we manually generalized substrate scopes from reported examples, annotated which structural sites could be modified to broaden template utility, and ensured mechanistic plausibility of transformations. Reaction templates were therefore curated and verified manually to form the Edit Rule V1 dataset.

Editing Mode Workflow

The Editing mode differs from the Discovery mode in that it requires a predefined set of molecules to be optimized. Specifically, the initial state of a molecular generation trajectory is not an empty molecule but a specific substrate drawn from the input molecule set. The reaction library corresponds to Edit Rule V1, rather than the 87 classical synthetic reactions used in Discovery mode. For two-component editing reactions, additional reactants were sourced from the Enamine building block library.

As in Discovery mode, Editing mode requires a reward function, here defined as molecular activity. However, in contrast to Discovery mode, which corresponds to the hit discovery stage of drug development (where the goal is to identify diverse molecules with potential activity), the Editing mode corresponds to lead optimization (where modifications are made to molecules that already possess measurable activity). Since such modifications often involve small structural changes, the activity prediction tool must achieve high accuracy in distinguishing subtle activity differences. To this end, we employed Uni-Dock, a GPU-accelerated docking program that delivers substantial speed improvements over traditional AutoDock Vina while maintaining comparable accuracy.

## Experimental Validation of the Editing Mode

To rigorously evaluate the feasibility and generalizability of integrating molecular editing with generative AI in molecular optimization tasks, we designed a large-scale benchmark. Specifically, we used the comprehensive atlas of approved drug targets constructed by Santos and colleagues as the source of candidate targets. Based on our previously curated high-quality subset of the PDBbind dataset (containing protein–ligand complex structures)[45], which included targets with well-defined binding affinity annotations and resolution $\lesssim$ 2.5 Å, we obtained 129 protein targets with high-resolution crystal structures and reliable affinity labels.

Next, we cross-referenced these targets with activity data from ChEMBL and removed targets without valid active molecules, yielding 116 effective targets. For each target, we then processed the set of known active molecules through the following steps.

(1) remove active molecules that were incompatible with molecular editing templates;

(2) select active molecules with pChEMBL values above a threshold (set to 5 when the remaining actives numbered fewer than 50, otherwise set to 6);

(3) to minimize excessive scoring variation by Uni-Dock[36], retain only active molecules whose Uni-Dock docking score distributions had interquartile ranges < 2.

Through this filtering pipeline, 96 valid targets were retained for the final benchmark. In testing, the maximum reaction depth of SpaceGFN was set to four steps. For evaluation, the top 1,000 optimized molecules ranked by docking score were selected as the assessment set.

## Code Availability

SpaceGFN is available at GitHub (https://github.com/ChemloverYuchen/SpaceGFN).

## Competing Interests Statement

The authors declare no competing interests.

# Figures

**A**

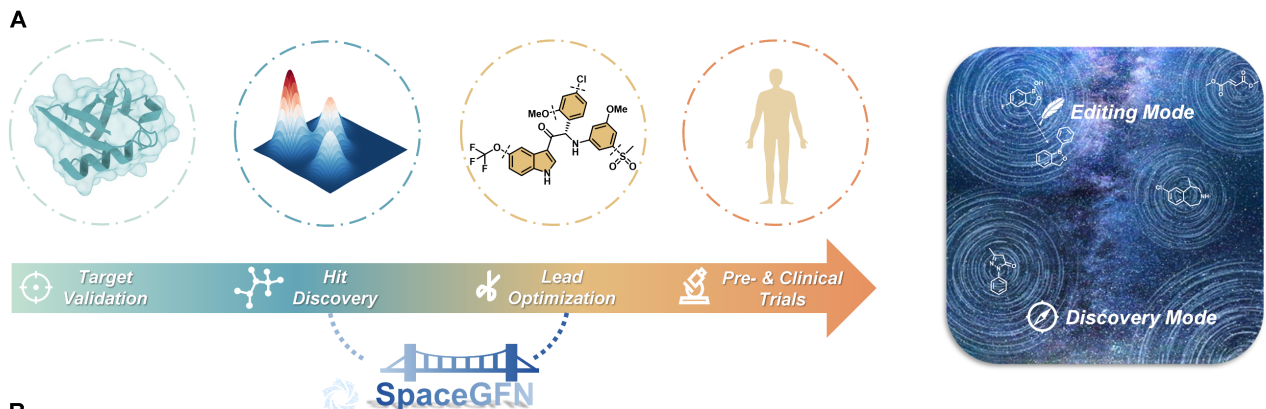

**B**

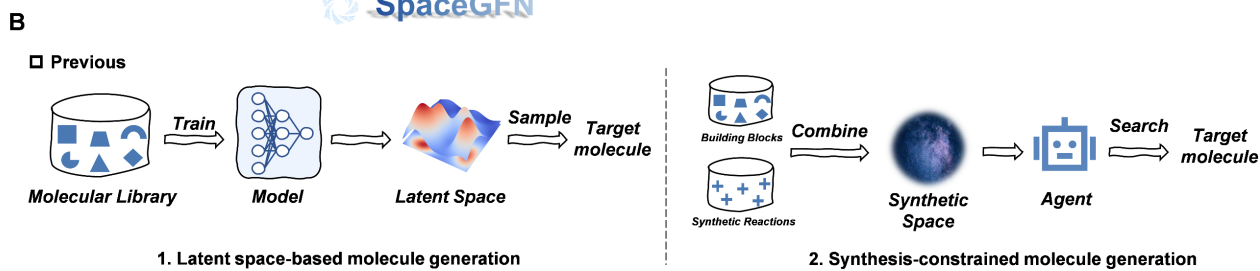

**C**

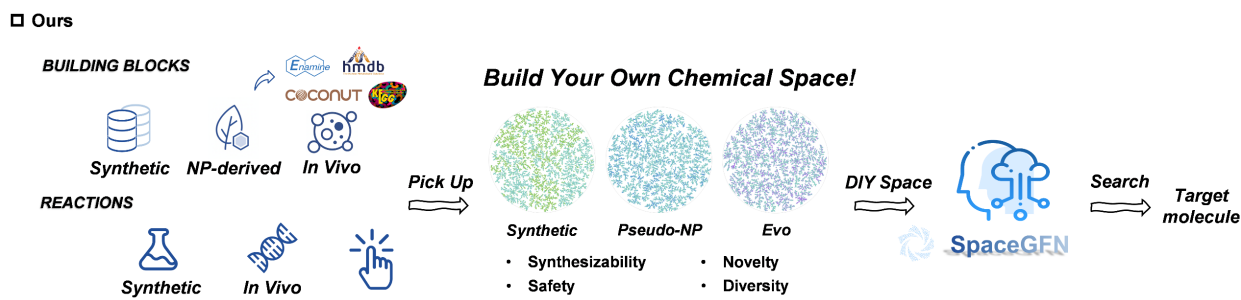

**D**

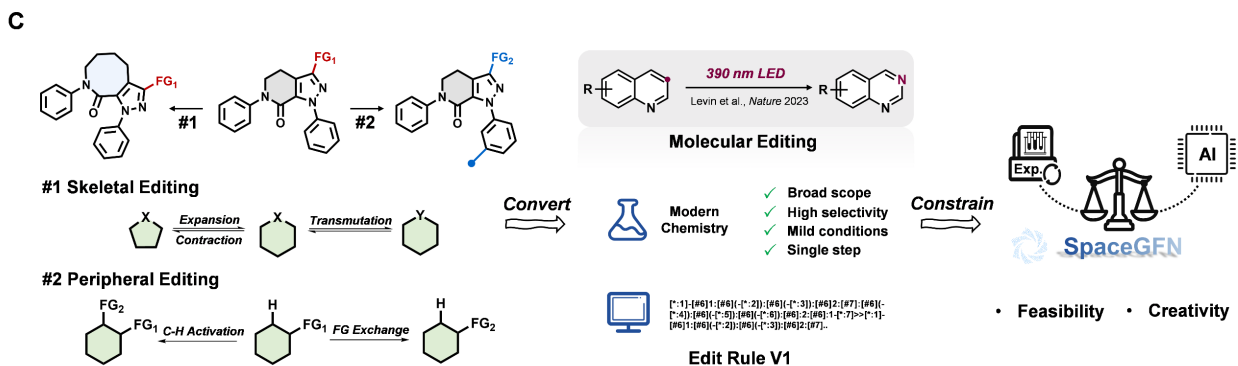
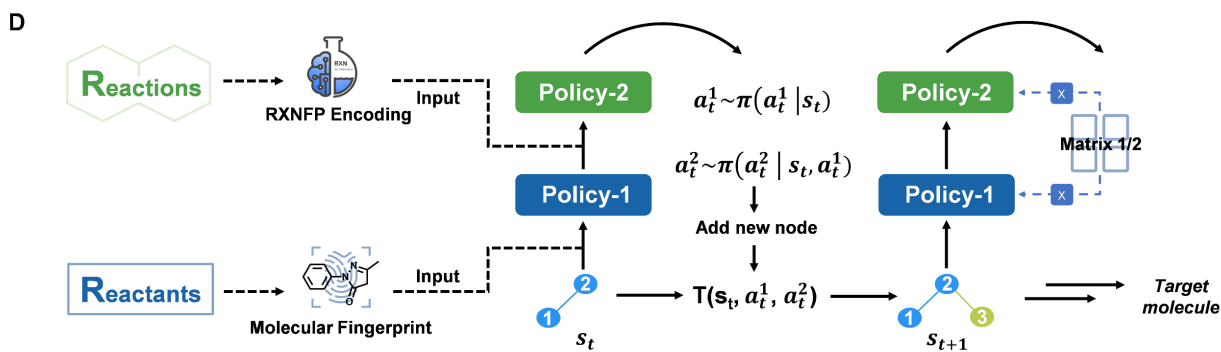

**Fig. 1. Overview of SpaceGFN.**

(A) Overall framework of SpaceGFN. The two modes—Discovery and Editing—are designed for hit discovery and lead optimization in drug development, respectively.

(B) Schematic illustration of the Discovery mode. Unlike conventional latent space–based or synthesis-constrained molecule generation approaches, the Discovery mode of SpaceGFN emphasizes the deliberate design of chemical space. Through a DIY chemical space framework, researchers can freely construct novel spaces of interest and efficiently search for target molecules using SpaceGFN.

(C) Schematic illustration of the Editing mode. The concept of molecular editing from synthetic chemistry is incorporated into SpaceGFN, where modern chemical methodologies are compiled into SMARTS-based reaction templates to form the Edit Rule V1 dataset. These templates serve as synthetic constraints in the generation process, overcoming the limitations of traditional experimental approaches and previous GenAI methods, and establishing a new balance between creativity and feasibility.

(D) Algorithmic schematic of SpaceGFN. The molecular generation process is formulated as a Markov decision process in which reactants are sequentially combined through predefined reaction rules. Two policy models (Policy-1 and Policy-2) govern the hierarchical actions of reaction selection and reactant selection (optionally guided by fingerprint embeddings), guided by rewards such as molecular activity.

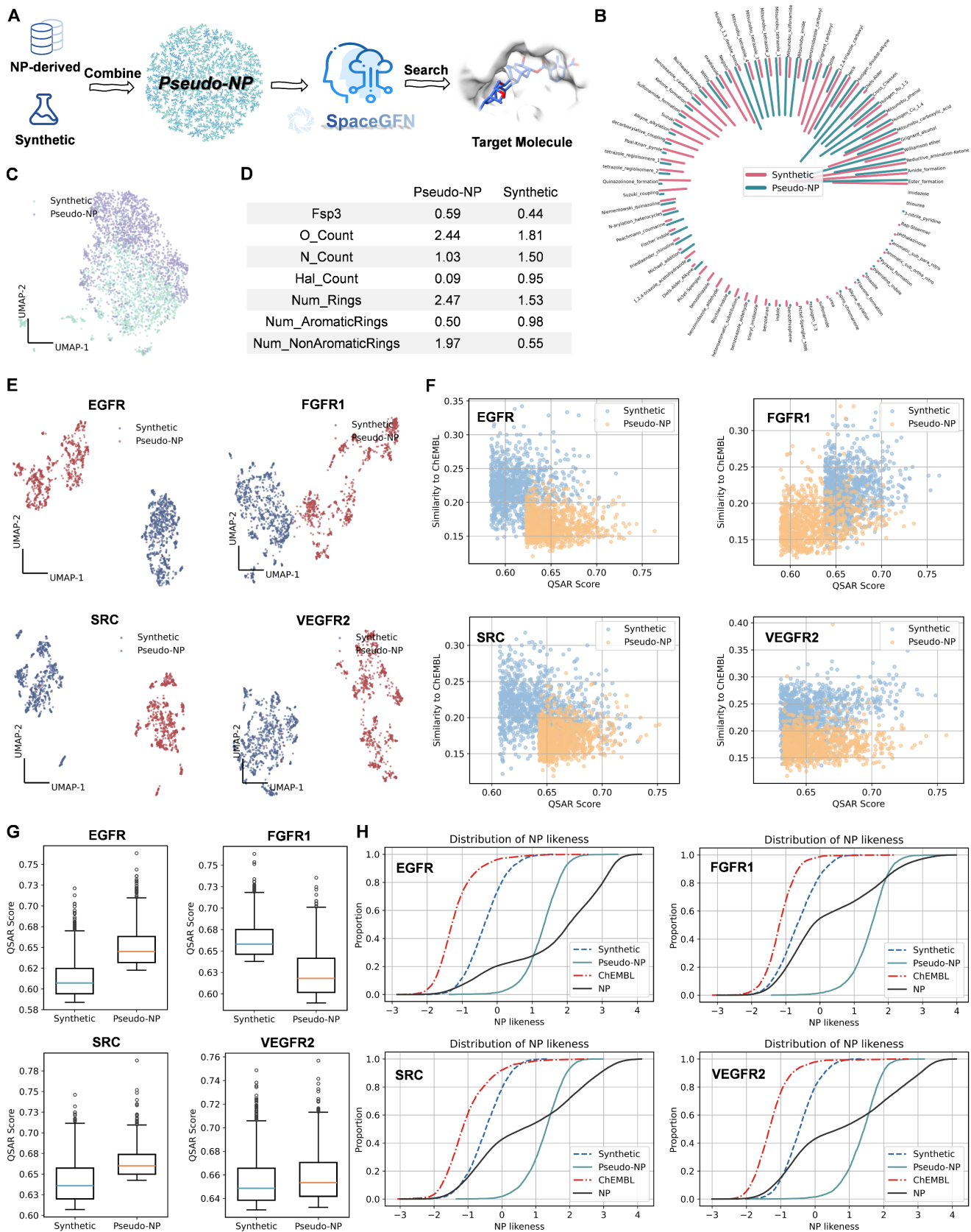

Fig. 2. Case study of the Discovery mode: construction and validation of the pseudo-natural product (Pseudo-NP) chemical space.

(A) Schematic illustration of Pseudo-NPchemical space construction and application. The Pseudo-NPspace is built by combining NP-derived building blocks with synthetic reactions, and SpaceGFN is used to efficiently explore this space to generate potential drug-like molecules.

(B) Comparison of synthetic reaction preferences (number of matchable building blocks) between NP-derived and synthetic building block libraries.

(C) UMAP projection comparing NP-derived and synthetic building block libraries.

(D) Comparison of NP-related properties between NP-derived and synthetic building block libraries.

(E) UMAP distribution of sampled molecules from Pseudo-NP and Synthetic spaces across different targets.

(F) Novelty comparison of molecules sampled from Pseudo-NP and Synthetic spaces, measured by Tanimoto similarity to known active compounds from ChEMBL corresponding to the same targets.

(G) QSAR score distributions of molecules sampled from Pseudo-NP and Synthetic spaces across different targets.

(H) NP-likeness distributions of molecules sampled from Pseudo-NP and Synthetic spaces, known active compounds from ChEMBL, and natural products from COCONUT (top 100,000 molecules ranked by predicted QSAR activity).

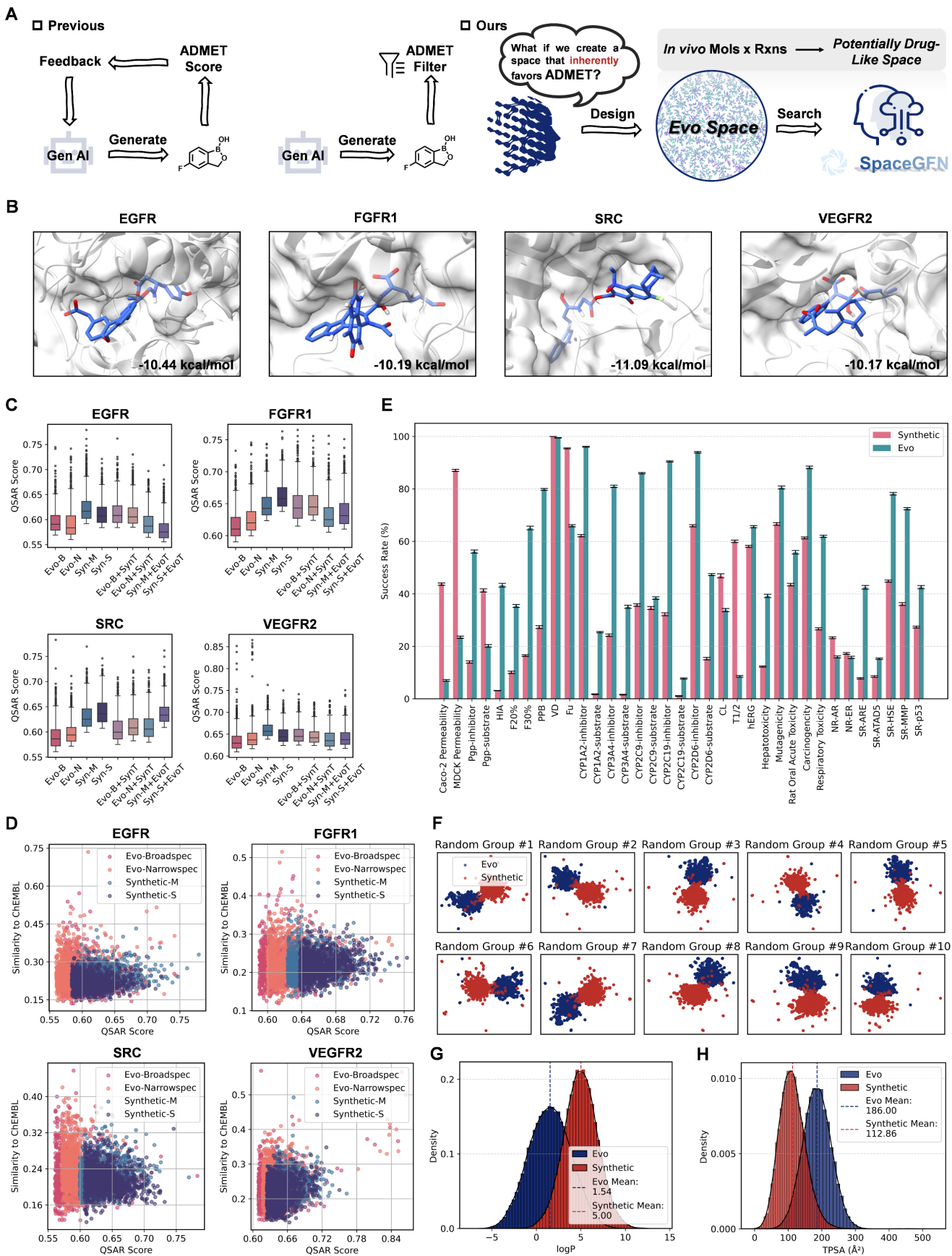

Fig. 3. Case study of the Discovery mode: construction and validation of the Evo chemical space.

(A) Concept of the Evo space. Conventional approaches rely on ADMET prediction tools, which are limited by sparse data and cannot fully resolve ADMET risks in drug discovery. We propose an alternative strategy: Evo space, constructed from endogenous molecules used as building blocks and endogenous reactions, is expected to inherently reduce ADMET risks from the outset.

(B) Representative docking examples of molecules generated by SpaceGFN within the Evo space across different targets.

(C) QSAR score distributions of sampled molecules from Evo and control spaces across different targets. Evo-B and Evo-N represent Evo-Broadspec and Evo-Narrowspec, which differ in the selection range of endogenous building blocks. Syn-S and Syn-M represent two Synthetic chemical spaces, Synthetic-S and Synthetic-M. Evo-B+SynT and Evo-N+SynT denote combinations of endogenous building block libraries with synthetic reaction libraries, while Syn-S+EvoT and Syn-M+EvoT denote combinations of synthetic building block libraries with endogenous reaction libraries. Further details are provided in Methods.

(D) Novelty comparison of sampled molecules from Evo and Synthetic spaces across different targets, measured by Tanimoto similarity to known active compounds from ChEMBL.

(E) Success rates of Evo and Synthetic spaces across 35 key ADMET properties predicted by the enhanced version of ADMETlab 2.0. Success rate is defined as the proportion of randomly sampled molecules meeting predefined thresholds.

(F) UMAP projections of molecules randomly sampled (10 runs, 8000 molecules per run) from Evo and Synthetic spaces.

(G) LogP distributions of molecules randomly sampled from Evo and Synthetic spaces.

(H) TPSA (topological polar surface area) distributions of molecules randomly sampled from Evo and Synthetic spaces.

**Fig. 4. Illustration and case studies of the Editing mode.**

(A) Schematic of molecular editing. Molecular editing can be classified into scaffold editing and peripheral editing: the former includes representative types such as single-atom and multi-atom editing, while the latter includes C–H activation and functional group–functional group exchange.

(B) Construction of the Edit Rule V1 dataset of editing-style reaction templates. Through literature collection, data processing, and SMARTS compilation, we curated a dataset comprising 300 key molecular editing reactions.

(C) Case study of molecular optimization using the Editing mode. This mode not only achieves molecular optimization (with docking scores as metric) but also provides explicit synthetic routes, with each reaction step supported by corresponding literature references.

(D) Demonstration of the plug-and-play feature of Editing mode. Newly developed editing methods can be rapidly incorporated into SpaceGFN and readily applied.

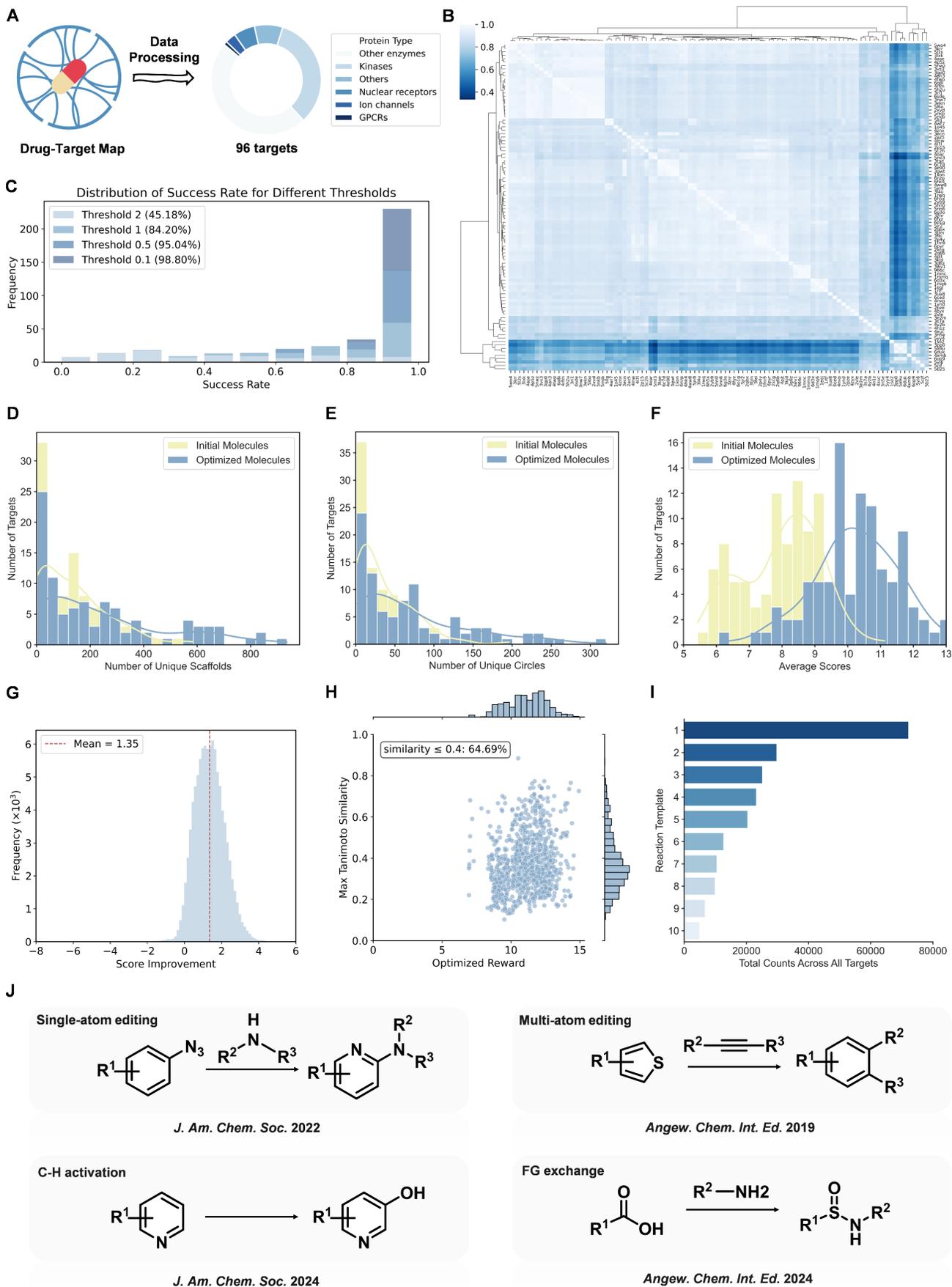

**Fig. 5. Validation of the Editing mode.**

(A) Illustration of the 96-target test set. Based on the target landscape of approved drugs and subsequent data processing, 96 representative protein targets were selected for molecular optimization tests.

(B) Similarity analysis of the 96 targets. Target feature vectors were obtained using the protein language model ESM2, and cosine similarity was used to construct a similarity matrix, which was visualized as a hierarchical clustering heatmap.

(C) Success rate across the 96 targets. Success rate is defined as the proportion of targets for which the difference between pre- and post-optimization docking scores evaluated by Uni-Dock, exceeds a specified threshold.

(D) Distribution of Murcko scaffolds before and after optimization for each target. Analysis was performed on the Top K molecules ranked by docking score after optimization, where K corresponds to the number of initial molecules per target.

(E) Distribution of Circles counts before and after optimization for each target, based on the Top K molecules as defined above.

(F) Distribution of mean docking scores before and after optimization across the 96 targets, as evaluated by Uni-Dock. Higher scores indicate stronger predicted activity.

(G) Overall distribution of docking score changes across the 96 targets.

(H) Distribution of Tanimoto similarity between the Top 10 optimized molecules per target and their corresponding initial molecules.

(I) Top 10 most frequent molecular editing reactions in optimization tasks. Frequency was measured among molecules with docking score improvements greater than 0.5 kcal/mol. Detailed reaction templates are provided in Fig. S54.

(J) Representative high-frequency molecular editing templates. One example is shown for each type of editing reaction.

# References

1. Beck, H., Härter, M., Haß, B., Schmeck, C. & Baerfacker, L. Small molecules and their impact in drug discovery: A perspective on the occasion of the 125th anniversary of the Bayer Chemical Research Laboratory. *Drug Discov. Today* **27**, 1560–1574 (2022).
2. Reymond, J.-L. & Awale, M. Exploring Chemical Space for Drug Discovery Using the Chemical Universe Database. *ACS Chem. Neurosci.* **3**, 649–657 (2012).
3. Zhang, K. *et al.* Artificial intelligence in drug development. *Nat. Med.* **31**, 45–59 (2025).
4. Lavecchia, A. Navigating the frontier of drug-like chemical space with cutting-edge generative AI models. *Drug Discov. Today* **29**, 104133 (2024).
5. Brown, D. G. & Boström, J. Analysis of Past and Present Synthetic Methodologies on Medicinal Chemistry: Where Have All the New Reactions Gone?: Miniperspective. *J. Med. Chem.* **59**, 4443–4458 (2016).
6. Sadybekov, A. V. & Katritch, V. Computational approaches streamlining drug discovery. *Nature* **616**, 673–685 (2023).
7. Gómez-Bombarelli, R. *et al.* Automatic Chemical Design Using a Data-Driven Continuous Representation of Molecules. *ACS Cent. Sci.* **4**, 268–276 (2018).
8. Alakhdar, A., Poczos, B. & Washburn, N. Diffusion Models in De Novo Drug Design. *J. Chem. Inf. Model.* **64**, 7238–7256 (2024).
9. Wang, J. *et al.* Multi-constraint molecular generation based on conditional transformer, knowledge distillation and reinforcement learning. *Nat. Mach. Intell.* **3**, 914–922 (2021).
10. Loeffler, H. H. *et al.* Reinvent 4: Modern AI–driven generative molecule design. *J. Cheminformatics* **16**, 20 (2024).
11. Karageorgis, G., Foley, D. J., Laraia, L. & Waldmann, H. Principle and design of pseudo-natural products. *Nat. Chem.* **12**, 227–235 (2020).
12. Campos, K. R. *et al.* The importance of synthetic chemistry in the pharmaceutical industry. *Science* **363**, eaat0805 (2019).
13. Zhu, Y. *et al.* SynGFN: learning across chemical space with generative flow-based molecular discovery. *Nat. Comput. Sci.* **6**, 29–38 (2025).
14. Heinzke, A. L. *et al.* Occurrence of "Natural Selection" in Successful Small Molecule Drug Discovery. *J. Med. Chem.* **67**, 11226–11241 (2024).
15. Grigalunas, M., Brakmann, S. & Waldmann, H. Chemical Evolution of Natural Product Structure. *J. Am. Chem. Soc.* **144**, 3314–3329 (2022).
16. Grigalunas, M. *et al.* Natural product fragment combination to performance-diverse pseudo-natural products. *Nat. Commun.* **12**, 1883 (2021).
17. Bag, S. *et al.* A divergent intermediate strategy yields biologically diverse pseudo-natural products. *Nat. Chem.* **16**, 945–958 (2024).
18. Over, B. *et al.* Natural-product-derived fragments for fragment-based ligand discovery. *Nat. Chem.* **5**, 21–28 (2013).
19. Gaulton, A. *et al.* ChEMBL: a large-scale bioactivity database for drug discovery. *Nucleic Acids Res.* **40**, D1100–D1107 (2012).
20. Ertl, P., Roggo, S. & Schuffenhauer, A. Natural Product-likeness Score and Its Application for Prioritization of Compound Libraries. *J. Chem. Inf. Model.* **48**, 68–74 (2008).
21. Zhang, K. *et al.* Artificial intelligence in drug development. *Nat. Med.* **31**, 45–59 (2025).
22. Huang, D. Z., Baber, J. C. & Bahmanyar, S. S. The challenges of generalizability in artificial intelligence for ADME/Tox endpoint and activity prediction. *Expert Opin. Drug Discov.* **16**, 1045–1056 (2021).
23. Wu, Z. *et al.* Leveraging language model for advanced multiproperty molecular optimization via prompt engineering. *Nat. Mach. Intell.* **6**, 1359–1369 (2024).
24. Yang, J. *et al.* Structure-guided discovery of bile acid derivatives for treating liver diseases without causing itch. *Cell* **187**, 7164-7182.e18 (2024).
25. Wishart, D. S. *et al.* HMDB: the Human Metabolome Database. *Nucleic Acids Res.* **35**, D521–D526 (2007).
26. Wishart, D. S. *et al.* HMDB 5.0: the Human Metabolome Database for 2022. *Nucleic Acids Res.* **50**, D622–D631 (2022).
27. Finnigan, W., Hepworth, L. J., Flitsch, S. L. & Turner, N. J. RetroBioCat as a computer-aided synthesis planning tool for biocatalytic reactions and cascades. *Nat. Catal.* **4**, 98–104 (2021).

28. Luo, W. *et al.* Bridging Machine Learning and Thermodynamics for Accurate p$K_a$ Prediction. *JACS Au* **4**, 3451–3465 (2024).
29. Ma, C., Lindsley, C. W., Chang, J. & Yu, B. Rational Molecular Editing: A New Paradigm in Drug Discovery. *J. Med. Chem.* **67**, 11459–11466 (2024).
30. Li, E.-Q., Lindsley, C. W., Chang, J. & Yu, B. Molecular Skeleton Editing for New Drug Discovery. *J. Med. Chem.* **67**, 13509–13511 (2024).
31. Coley, C. W., Green, W. H. & Jensen, K. F. RDChiral: An RDKit Wrapper for Handling Stereochemistry in Retrosynthetic Template Extraction and Application. *J. Chem. Inf. Model.* **59**, 2529–2537 (2019).
32. Chen, S. *et al.* Deep lead optimization enveloped in protein pocket and its application in designing potent and selective ligands targeting LTK protein. *Nat. Mach. Intell.* **7**, 448–458 (2025).
33. Paschke, A.-S. K. *et al.* Carbon-to-nitrogen atom swap enables direct access to benzimidazoles from drug-like indoles. *Nat. Chem.* https://doi.org/10.1038/s41557-025-01904-x (2025) doi:10.1038/s41557-025-01904-x.
34. Li, J., Tang, P., Fan, Y. & Lu, H. Skeletal editing of pyrrolidines by nitrogen-atom insertion. *Science* **389**, 275–281 (2025).
35. Santos, R. *et al.* A comprehensive map of molecular drug targets. *Nat. Rev. Drug Discov.* **16**, 19–34 (2017).
36. Yu, Y. *et al.* Uni-Dock: GPU-Accelerated Docking Enables Ultralarge Virtual Screening. *J. Chem. Theory Comput.* **19**, 3336–3345 (2023).
37. Xie, Y., Xu, Z., Ma, J. & Mei, Q. How Much Space Has Been Explored? Measuring the Chemical Space Covered by Databases and Machine-Generated Molecules. Preprint at https://doi.org/10.48550/arXiv.2112.12542 (2023).
38. Jain, M. *et al.* GFlowNets for AI-Driven Scientific Discovery. *Digit. Discov.* **2**, 557–577 (2023).
39. Malkin, N., Jain, M., Bengio, E., Sun, C. & Bengio, Y. Trajectory balance: Improved credit assignment in GFlowNets. Preprint at https://doi.org/10.48550/arXiv.2201.13259 (2023).
40. Schwaller, P. *et al.* Mapping the space of chemical reactions using attention-based neural networks. *Nat. Mach. Intell.* **3**, 144–152 (2021).
41. Bengio, E., Jain, M., Korablyov, M., Precup, D. & Bengio, Y. Flow Network based Generative Models for Non-Iterative Diverse Candidate Generation. Preprint at https://doi.org/10.48550/arXiv.2106.04399 (2021).
42. Yoshizawa, T. *et al.* A data-driven generative strategy to avoid reward hacking in multi-objective molecular design. *Nat. Commun.* **16**, 2409 (2025).
43. Shen, C. *et al.* DrugFlow: An AI-Driven One-Stop Platform for Innovative Drug Discovery. *J. Chem. Inf. Model.* **64**, 5381–5391 (2024).
44. Huo, T. *et al.* Late-stage modification of bioactive compounds: Improving druggability through efficient molecular editing. *Acta Pharm. Sin. B* **14**, 1030–1076 (2024).
45. Shen, C. *et al.* Boosting Protein–Ligand Binding Pose Prediction and Virtual Screening Based on Residue–Atom Distance Likelihood Potential and Graph Transformer. *J. Med. Chem.* **65**, 10691–10706 (2022).

# Supporting Information

## Designing the Haystack:
## Programmable Chemical Space for Generative Drug Discovery

Yuchen Zhu[1], Donghai Zhao[1], Yangyang Zhang[1], Yitong Li[1], Xiaorui Wang[1], Shuwang Li[2], Yue Kong[2], Beichen Zhang[2], Chang Liu[2], Xingcai Zhang[3*], Tingjun Hou[1*], Chang-Yu Hsieh[1*]

[1] College of Pharmaceutical Sciences, Zhejiang University, Hangzhou, 310058

[2] LEPU MEDICAL TECHNOLOGY (BEIJING) CO., Ltd, Beijing, 102200

[3] Department of Chemical and Nano Engineering, University of California, San Diego, La Jolla, CA, USA

Corresponding authors

    Chang-Yu Hsieh

    E-mail: kimhsieh@zju.edu.cn

    Tingjun Hou

    E-mail: tingjunhou@zju.edu.cn

    Xingcai Zhang

    E-mail: xiz292@ucsd.edu

# Supplementary Figures

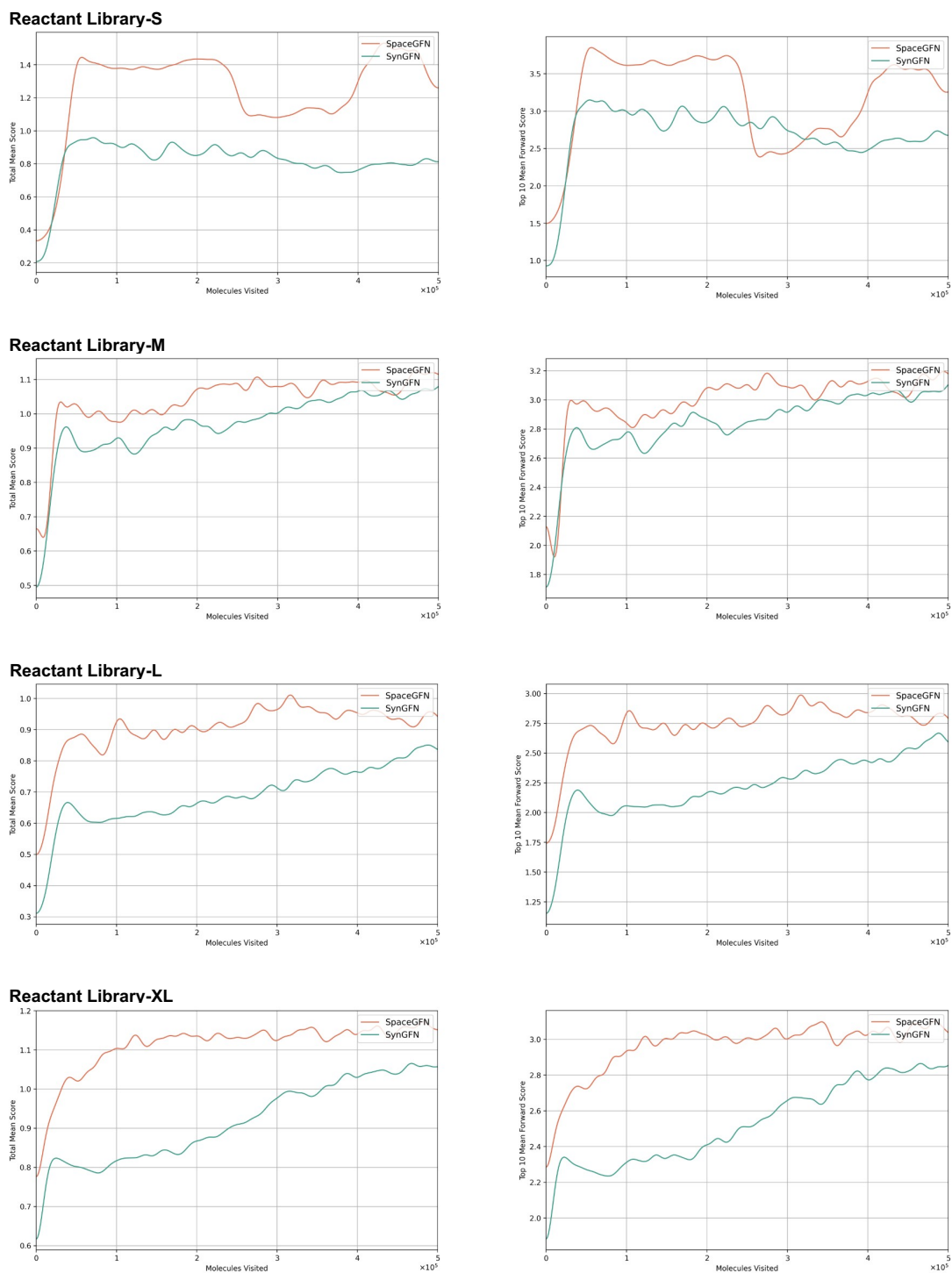

Supplementary Figure 1. Comparison of training performance between SpaceGFN and SynGFN under different reactant library sizes. The QSAR-predicted activity score of sEH was used as the training objective. SynGFN employed its best-performing pretrained version. Reactant libraries S, M, L, and XL contained 1,879, 7,958, 26,509, and 99,886 building blocks, respectively.

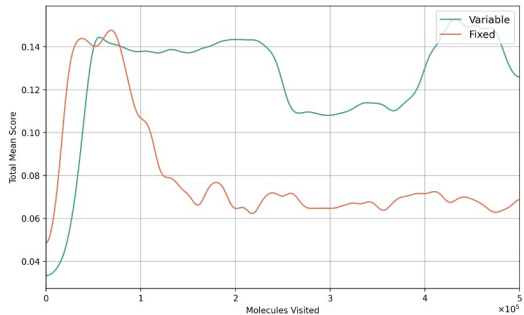
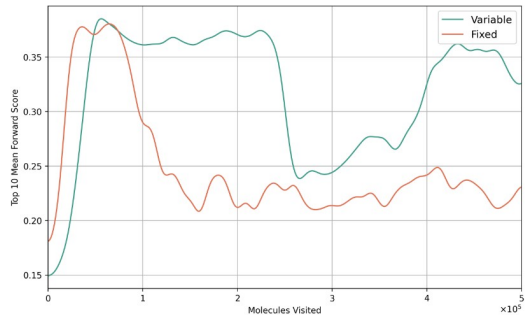
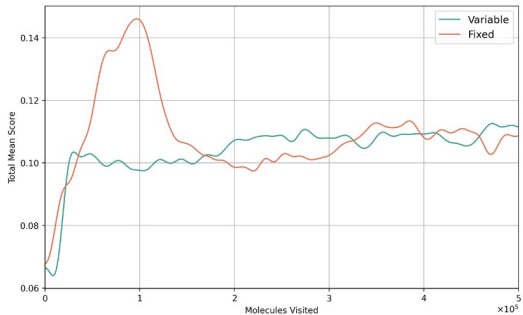
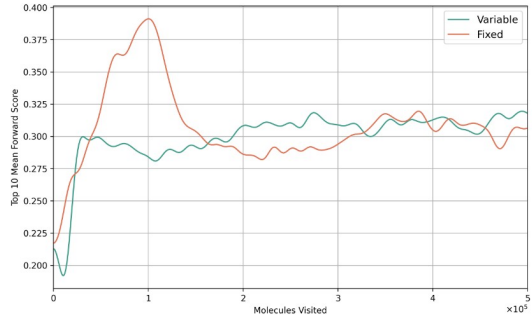
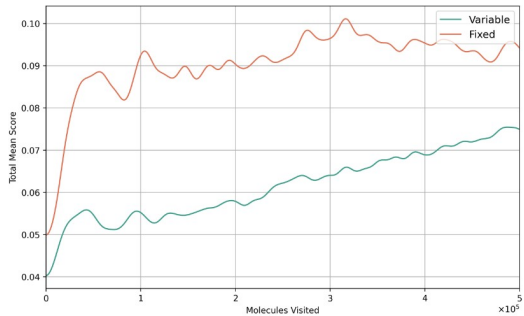
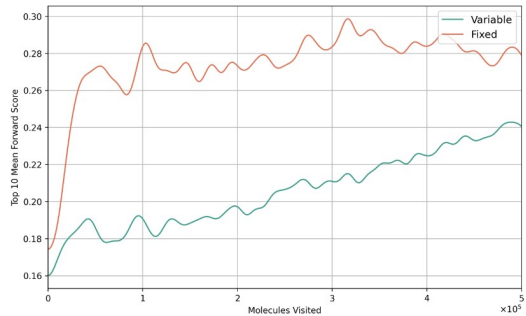
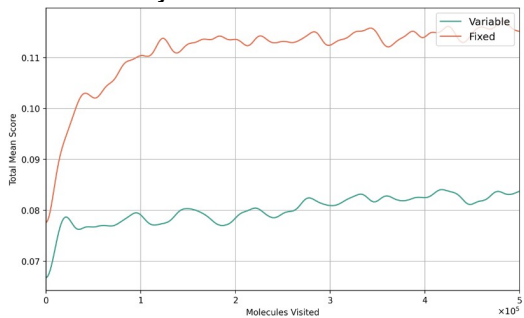
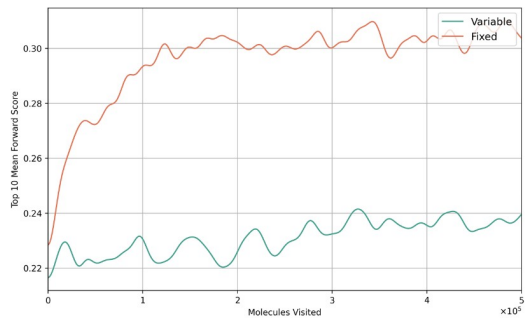

Supplementary Figure 2. Comparison of training performance of SpaceGFN on the sEH target under different reactant library sizes using two policy designs. The policie networks included a discrete-index policy (Variable, with output dimensions varying with the number of reactants) and a fingerprint-embedding policy (Fixed, with output dimensions fixed). Reactant libraries S, M, L, and XL contained 1,879, 7,958, 26,509, and 99,886 building blocks, respectively.

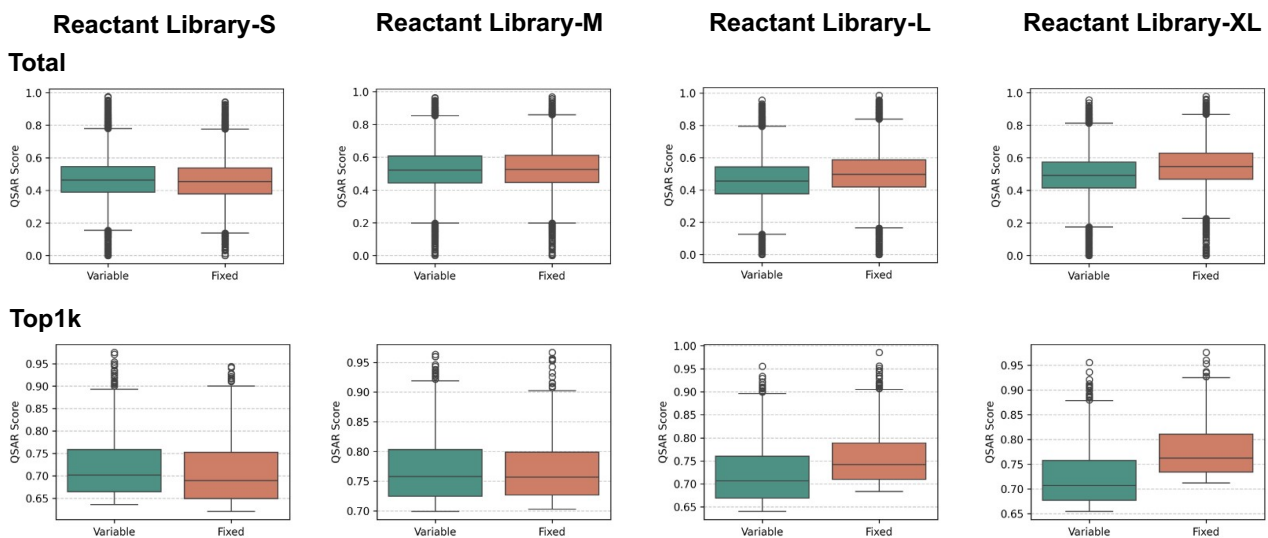

Supplementary Figure 3. Comparison of QSAR scores of molecules sampled by SpaceGFN on the sEH target under different reactant library sizes using two policy designs. The policy networks included a discrete-index policy (Variable, with output dimensions varying with the number of reactants) and a fingerprint-embedding policy (Fixed, with output dimensions fixed). Reactant libraries S, M, L, and XL contained 1,879, 7,958, 26,509, and 99,886 building blocks, respectively.

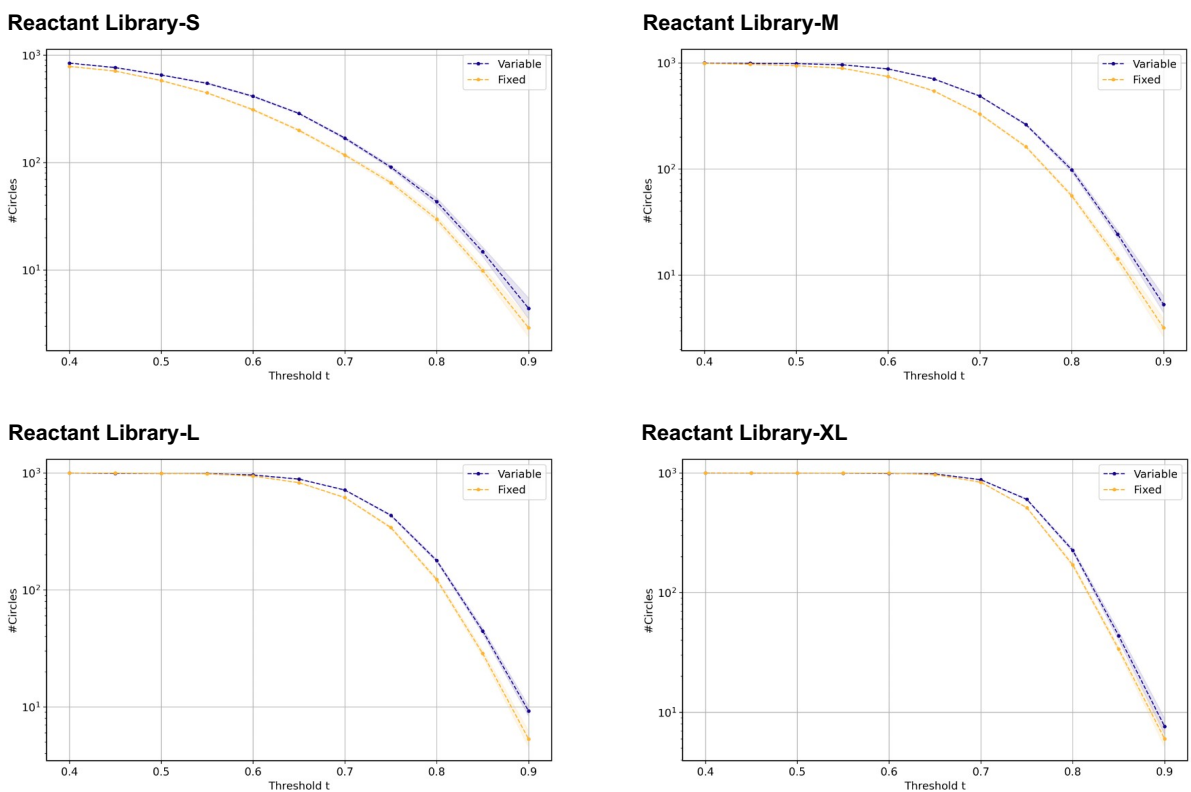

Supplementary Figure 4. Comparison of #Circles variation curves for molecules sampled by SpaceGFN on the sEH target under different reactant library sizes. #Circles is a representative metric for evaluating molecular set

diversity, and the curves show its variation across different thresholds (t). Reactant libraries S, M, L, and XL contained 1,879, 7,958, 26,509, and 99,886 building blocks, respectively.

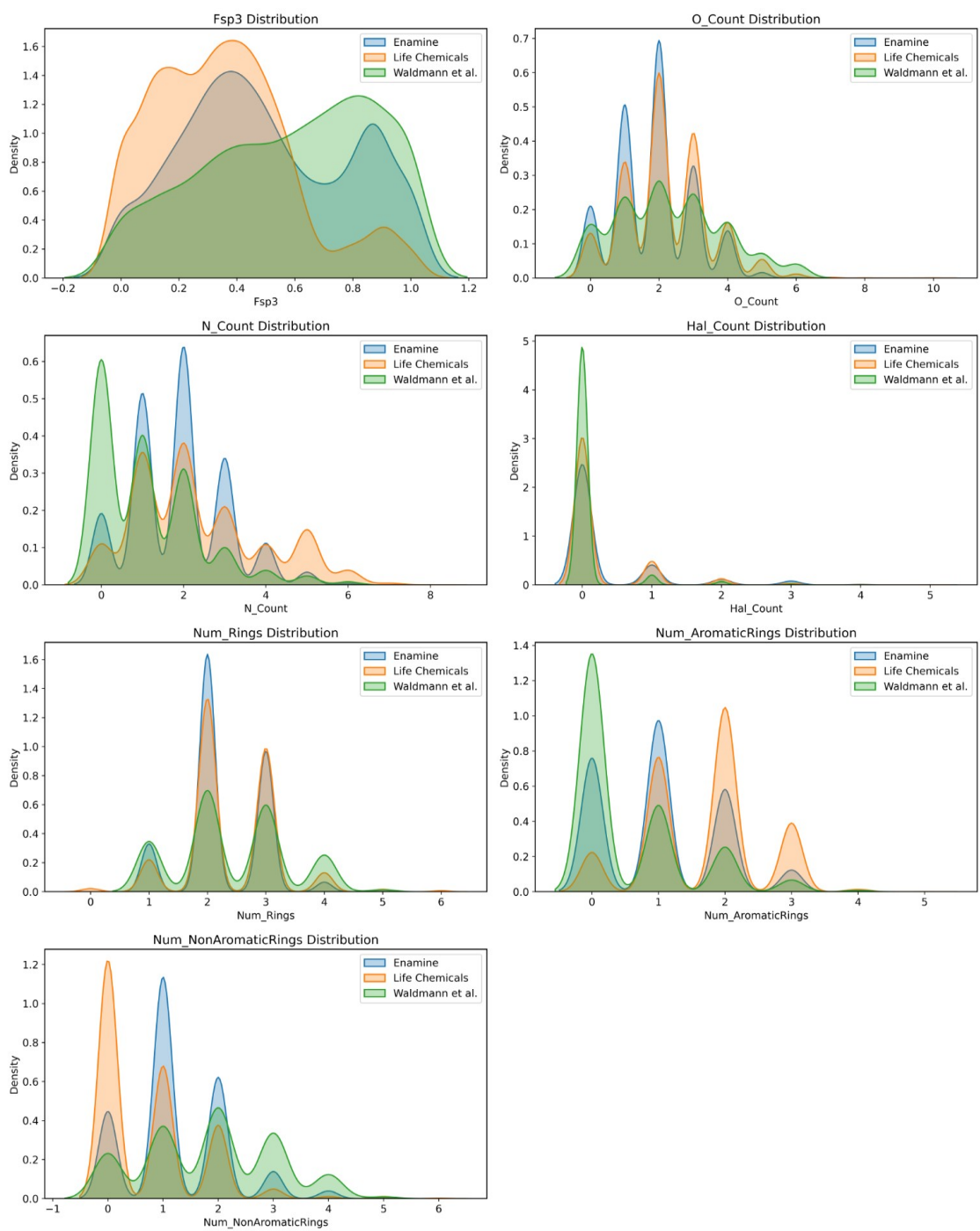

Supplementary Figure 5. Distribution of NP-related physicochemical properties across different natural product fragment libraries. We compared the NP-derived fragment library collected by Waldmann et al. with NP fragment libraries provided by Enamine and Life Chemicals. The analysis focused on physicochemical properties that are known to differ markedly between natural products and general synthetic molecules, including fraction of sp³ car-

bons (Fsp³), number of oxygen atoms (O_Count), number of nitrogen atoms (N_Count), number of halogen atoms (Hal_Count), total ring count (Num_Rings), number of aromatic rings (Num_AromaticRings), and number of non-aromatic rings (Num_NonAromaticRings).

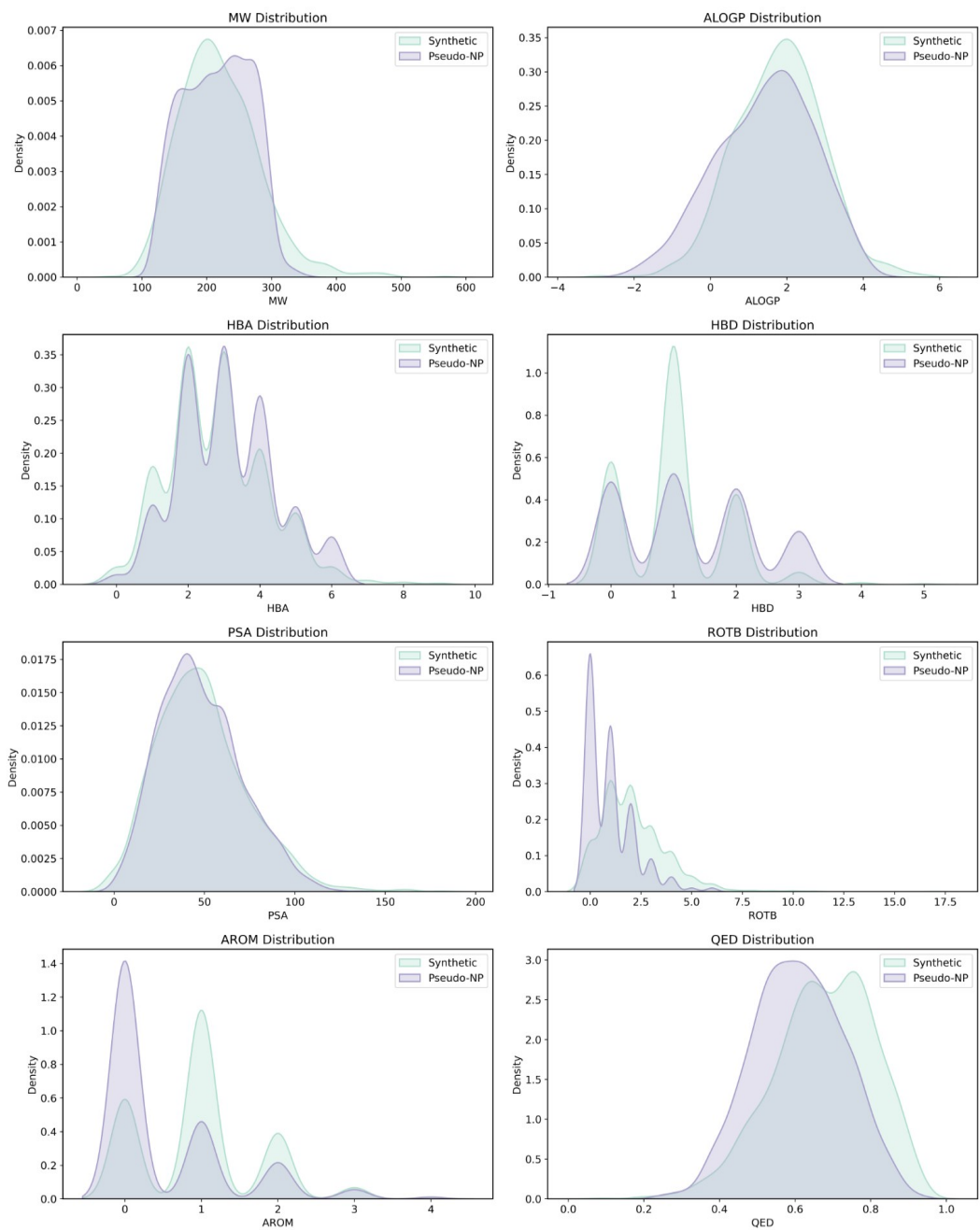

Supplementary Figure 6. Comparison of general physicochemical property distributions between the NP-derived fragment library used for Pseudo-NP construction and the synthetic fragment library used as control. The analyzed properties include molecular weight (MW), octanol–water partition coefficient (ALOGP), number of hydrogen bond acceptors (HBA), number of hydrogen bond donors (HBD), polar surface area (PSA), number of rotatable bonds (ROTB), number of aromatic rings (AROM), and quantitative estimate of drug-likeness (QED).

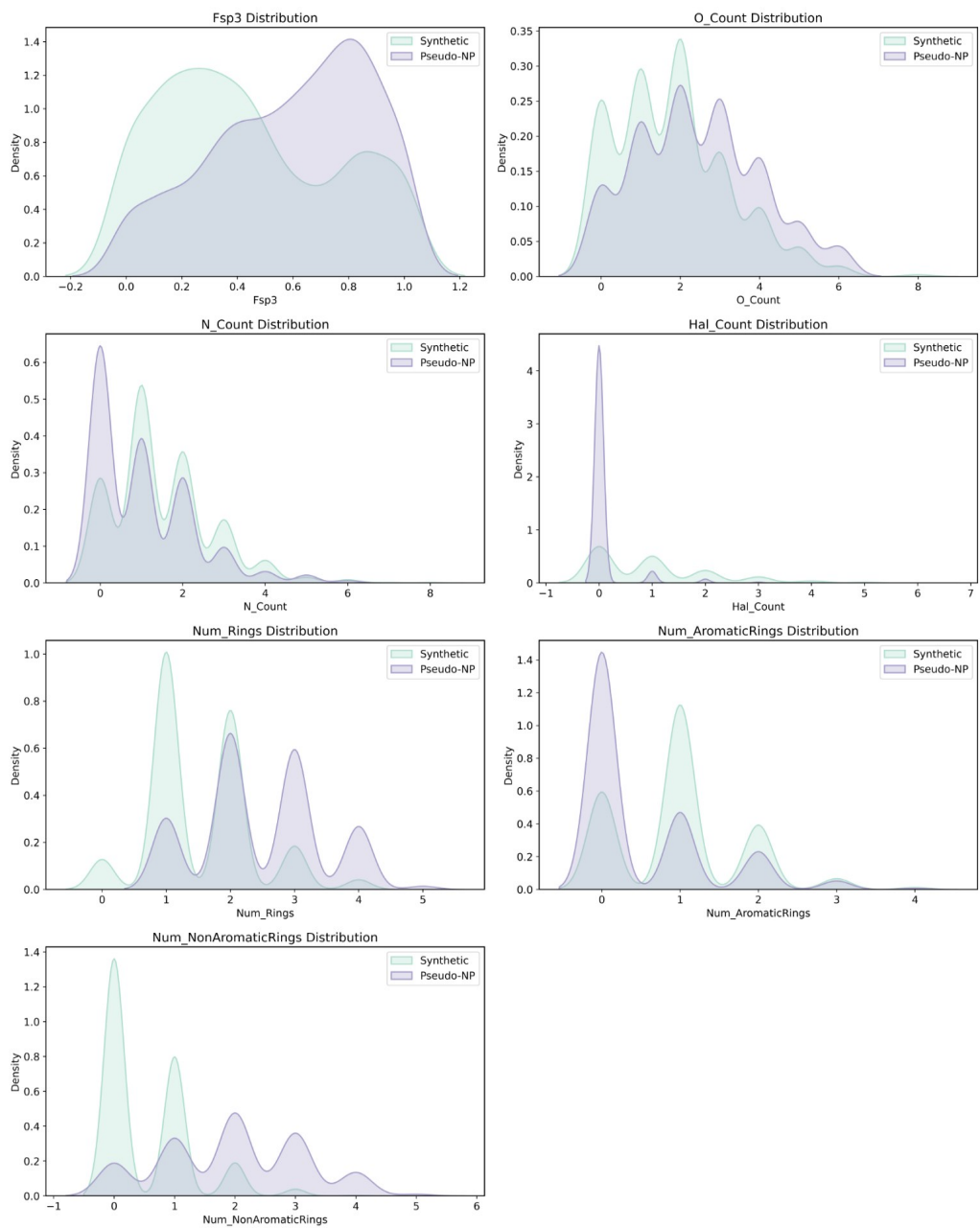

Supplementary Figure 7. Comparison of NP-related physicochemical property distributions between the NP-derived fragment library used for Pseudo-NP construction and the synthetic fragment library used as control. The analyzed properties are those that show significant differences between natural products and general chemical compounds, including fraction of sp3 carbons (Fsp3), number of oxygen atoms (O_Count), number of nitrogen atoms (N_Count), number of halogen atoms (Hal_Count), number of rings (Num_Rings), number of aromatic rings (Num_AromaticRings), and number of non-aromatic rings (Num_NonAromaticRings).

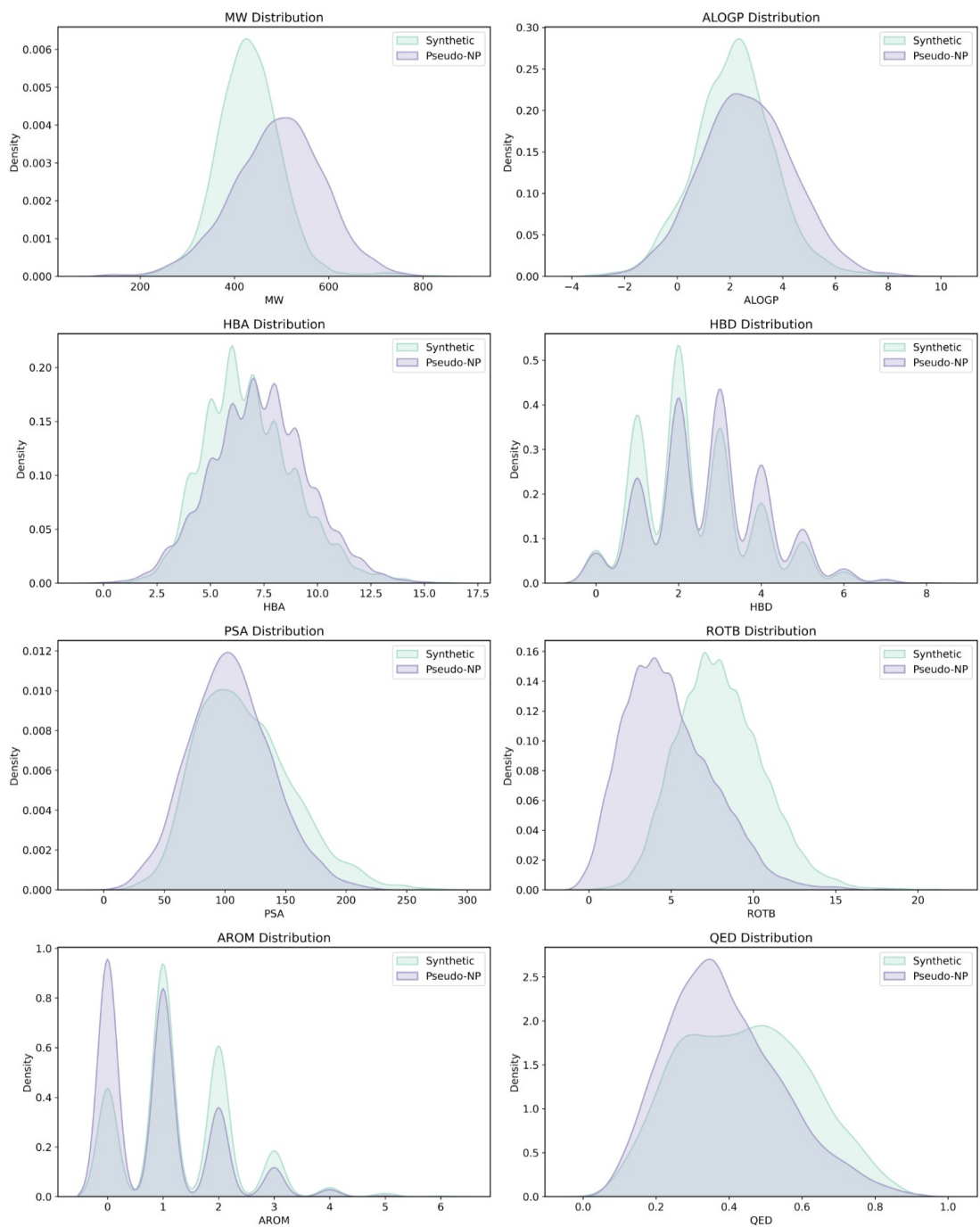

Supplementary Figure 8. Comparison of general physicochemical property distributions of sampled molecules from the Pseudo-NP space and the control Synthetic space under the EGFR target. The analyzed properties include molecular weight (MW), partition coefficient logP (ALOGP), number of hydrogen bond acceptors (HBA), number of hydrogen bond donors (HBD), polar surface area (PSA), number of rotatable bonds (ROTB), number of aromatic rings (AROM), and quantitative estimate of drug-likeness (QED).

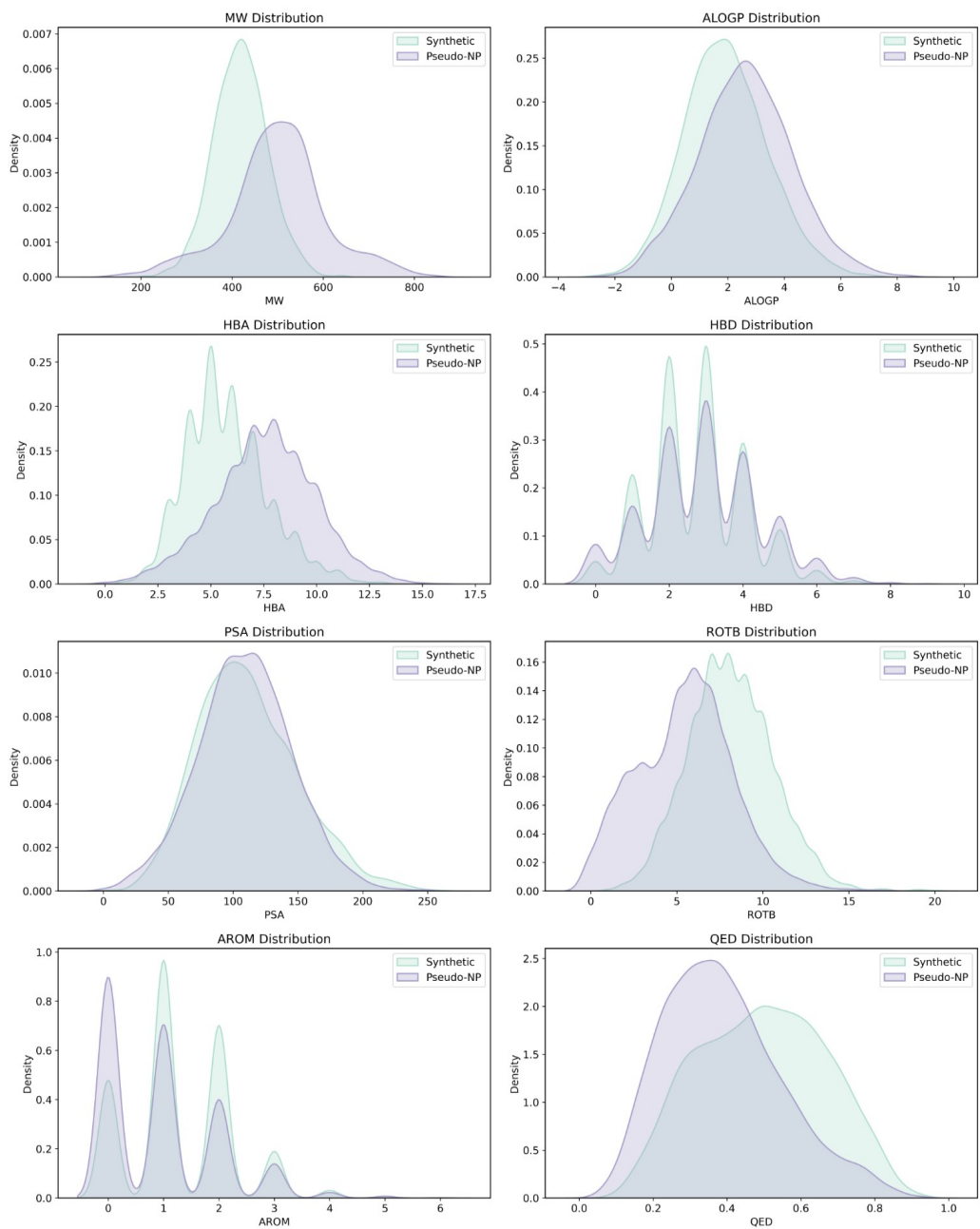

Supplementary Figure 9. Comparison of general physicochemical property distributions of sampled molecules from the Pseudo-NP space and the control Synthetic space under the FGFR1 target. The analyzed properties include molecular weight (MW), partition coefficient logP (ALOGP), number of hydrogen bond acceptors (HBA), number of hydrogen bond donors (HBD), polar surface area (PSA), number of rotatable bonds (ROTB), number of aromatic rings (AROM), and quantitative estimate of drug-likeness (QED).

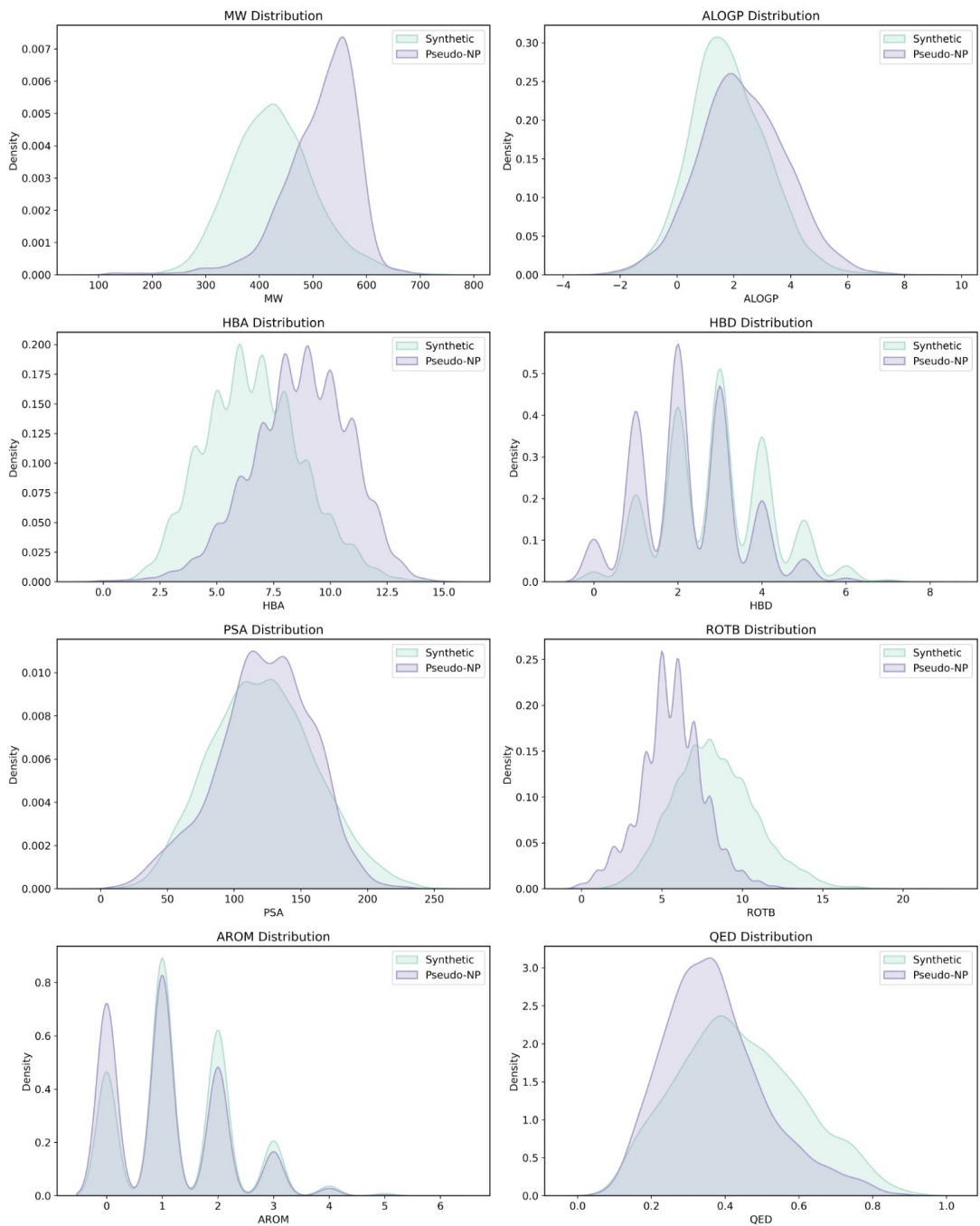

Supplementary Figure 10. Comparison of general physicochemical property distributions of sampled molecules from the Pseudo-NP space and the control Synthetic space under the SRC target. The analyzed properties include molecular weight (MW), partition coefficient logP (ALOGP), number of hydrogen bond acceptors (HBA), number of hydrogen bond donors (HBD), polar surface area (PSA), number of rotatable bonds (ROTB), number of aromatic rings (AROM), and quantitative estimate of drug-likeness (QED).

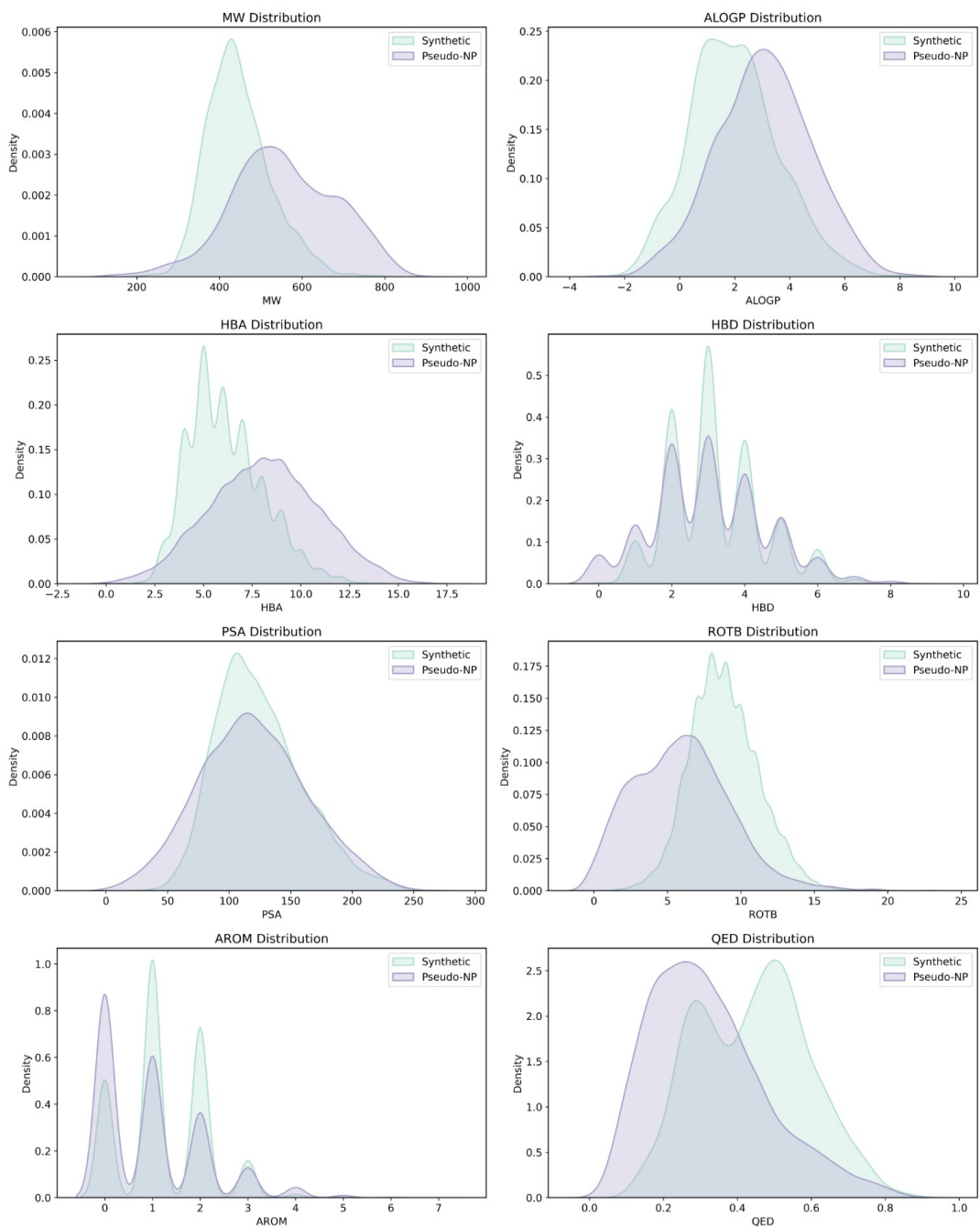

Supplementary Figure 11. Comparison of general physicochemical property distributions of sampled molecules from the Pseudo-NP space and the control Synthetic space under the VEGFR2 target. The analyzed properties include molecular weight (MW), partition coefficient logP (ALOGP), number of hydrogen bond acceptors (HBA), number of hydrogen bond donors (HBD), polar surface area (PSA), number of rotatable bonds (ROTB), number of aromatic rings (AROM), and quantitative estimate of drug-likeness (QED).

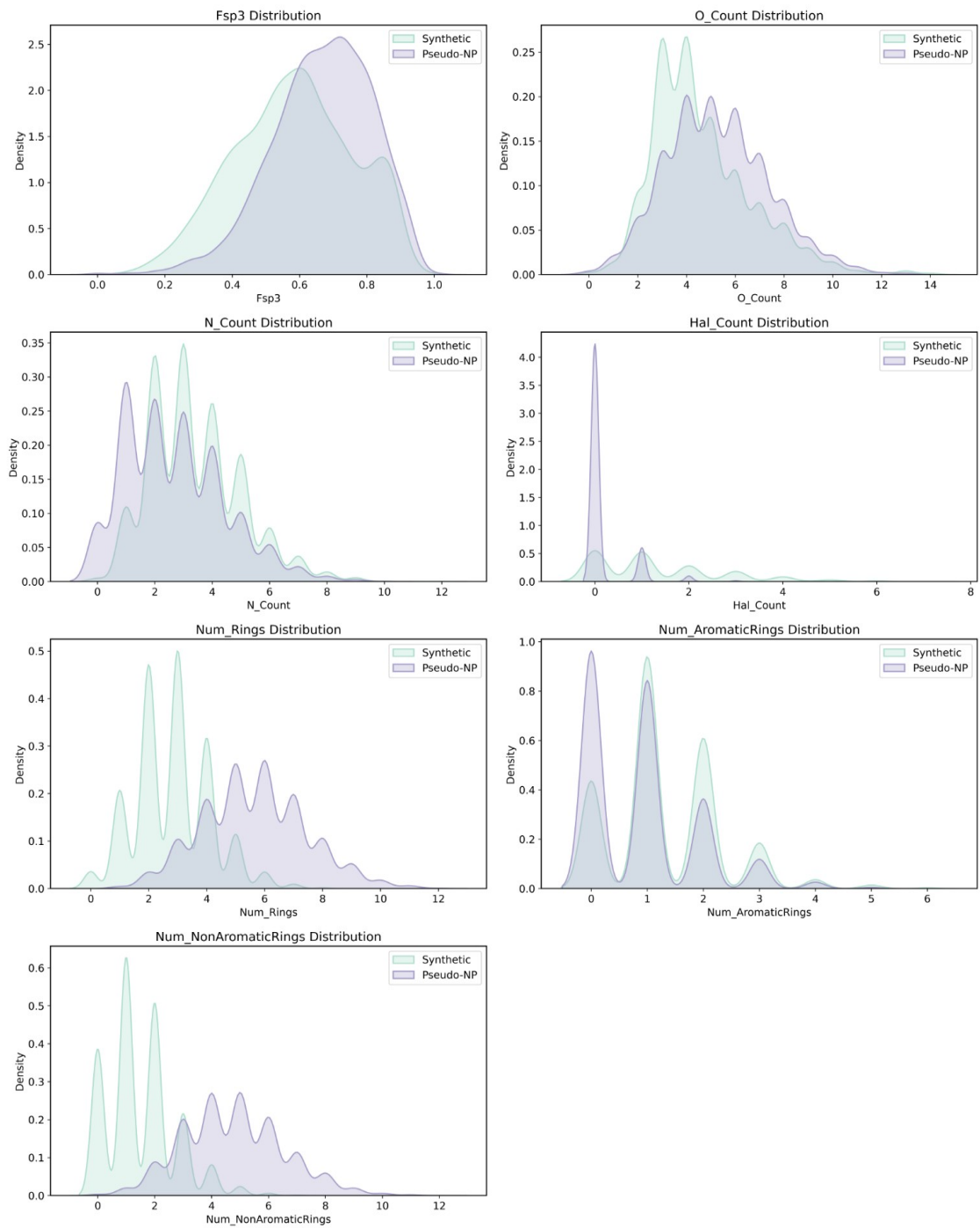

Supplementary Figure 12. Comparison of NP-related physicochemical property distributions of sampled molecules from the Pseudo-NP space and the control Synthetic space under the EGFR target. The analyzed properties include fraction of sp³ carbons (Fsp3), number of oxygen atoms (O_Count), number of nitrogen atoms (N_Count), number of halogen atoms (Hal_Count), number of rings (Num_Rings), number of aromatic rings (Num_AromaticRings), and number of non-aromatic rings (Num_NonAromaticRings).

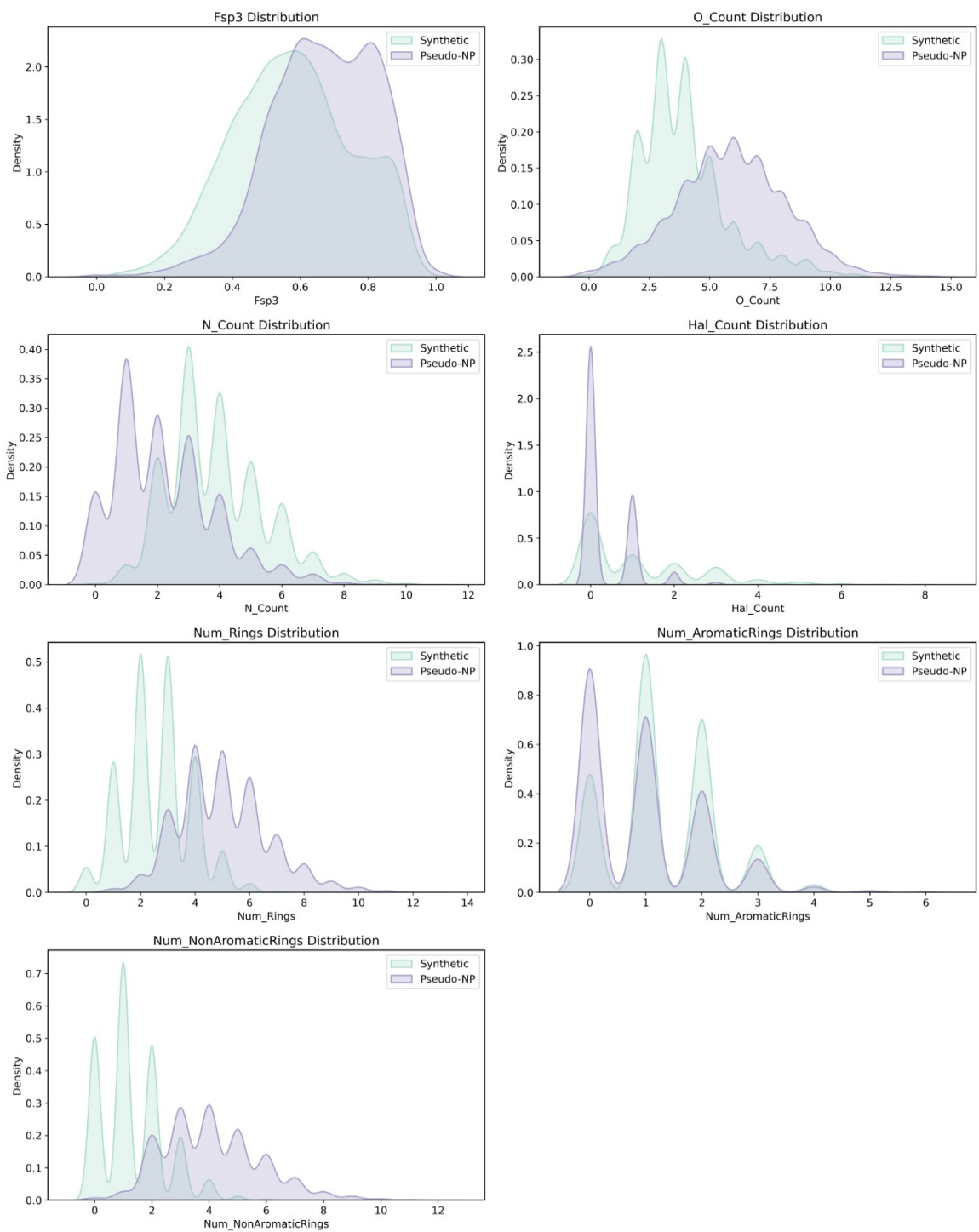

Supplementary Figure 13. Comparison of NP-related physicochemical property distributions of sampled molecules from the Pseudo-NP space and the control Synthetic space under the FGFR1 target. The analyzed properties include fraction of sp³ carbons (Fsp3), number of oxygen atoms (O_Count), number of nitrogen atoms (N_Count), number of halogen atoms (Hal_Count), number of rings (Num_Rings), number of aromatic rings (Num_AromaticRings), and number of non-aromatic rings (Num_NonAromaticRings).

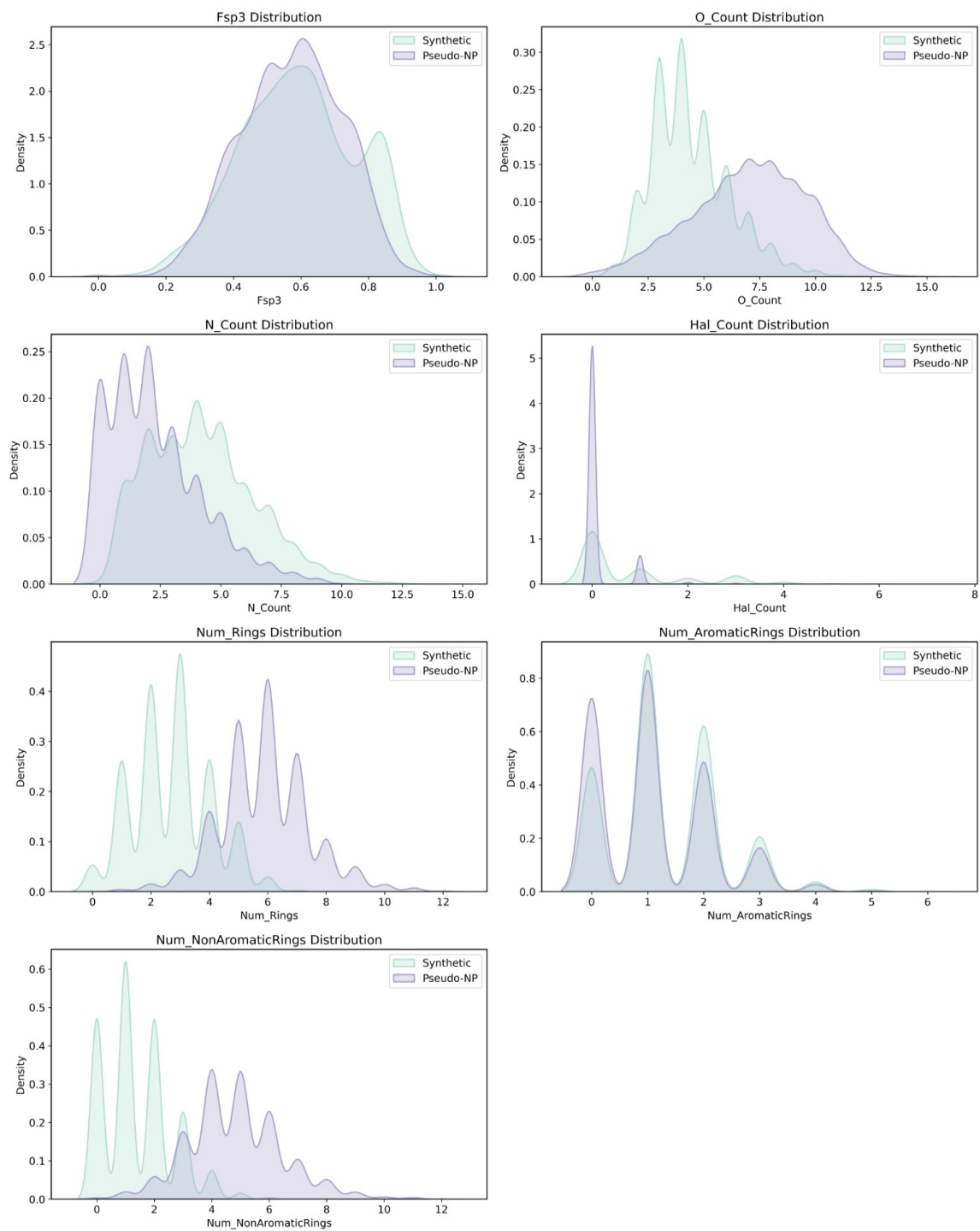

Supplementary Figure 14. Comparison of NP-related physicochemical property distributions of sampled molecules from the Pseudo-NP space and the control Synthetic space under the SRC target. The analyzed properties include fraction of sp³ carbons (Fsp3), number of oxygen atoms (O_Count), number of nitrogen atoms (N_Count), number of halogen atoms (Hal_Count), number of rings (Num_Rings), number of aromatic rings (Num_AromaticRings), and number of non-aromatic rings (Num_NonAromaticRings).

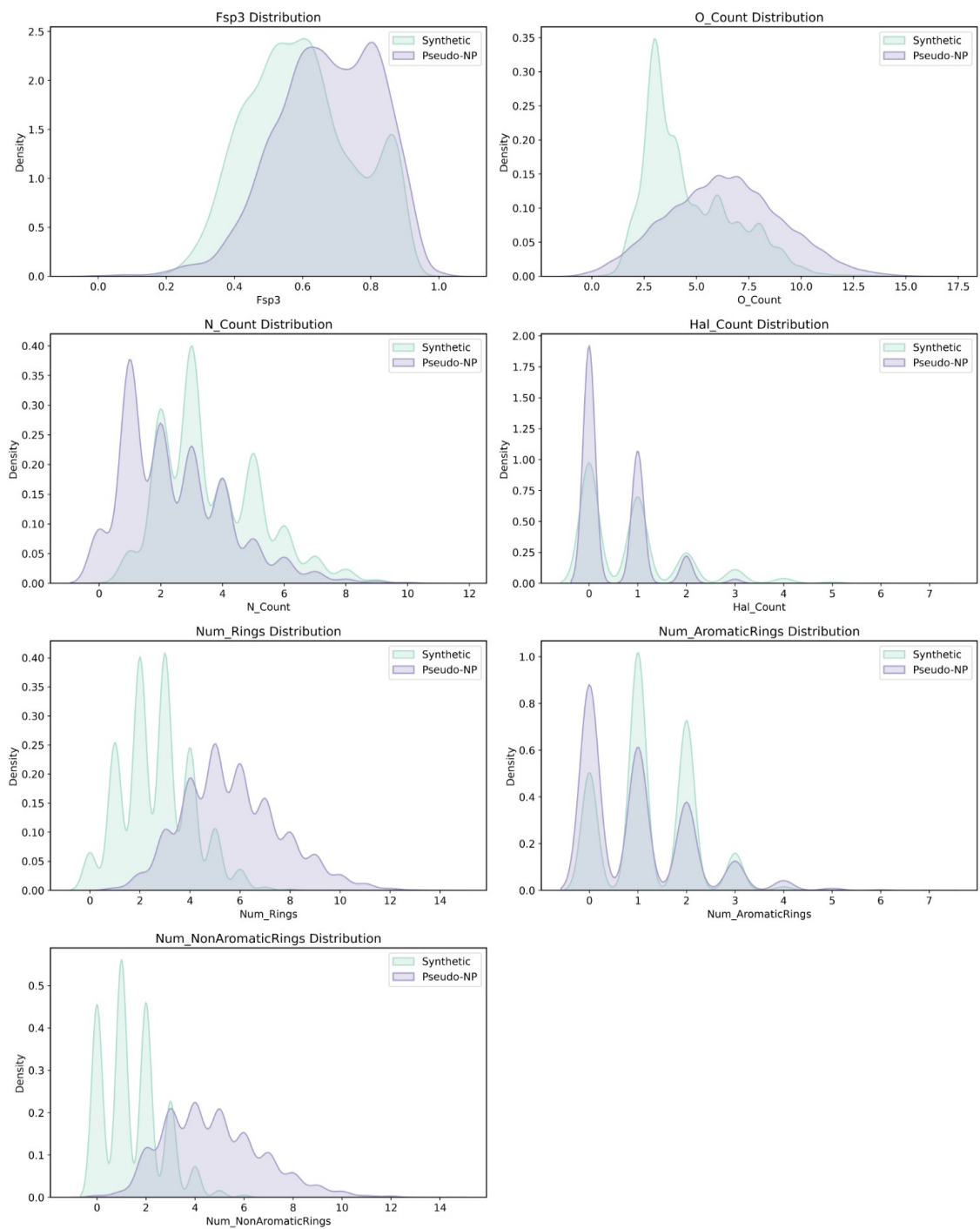

Supplementary Figure 15. Comparison of NP-related physicochemical property distributions of sampled molecules from the Pseudo-NP space and the control Synthetic space under the VEGFR2 target. The analyzed properties include fraction of sp³ carbons (Fsp3), number of oxygen atoms (O_Count), number of nitrogen atoms (N_Count), number of halogen atoms (Hal_Count), number of rings (Num_Rings), number of aromatic rings (Num_AromaticRings), and number of non-aromatic rings (Num_NonAromaticRings).

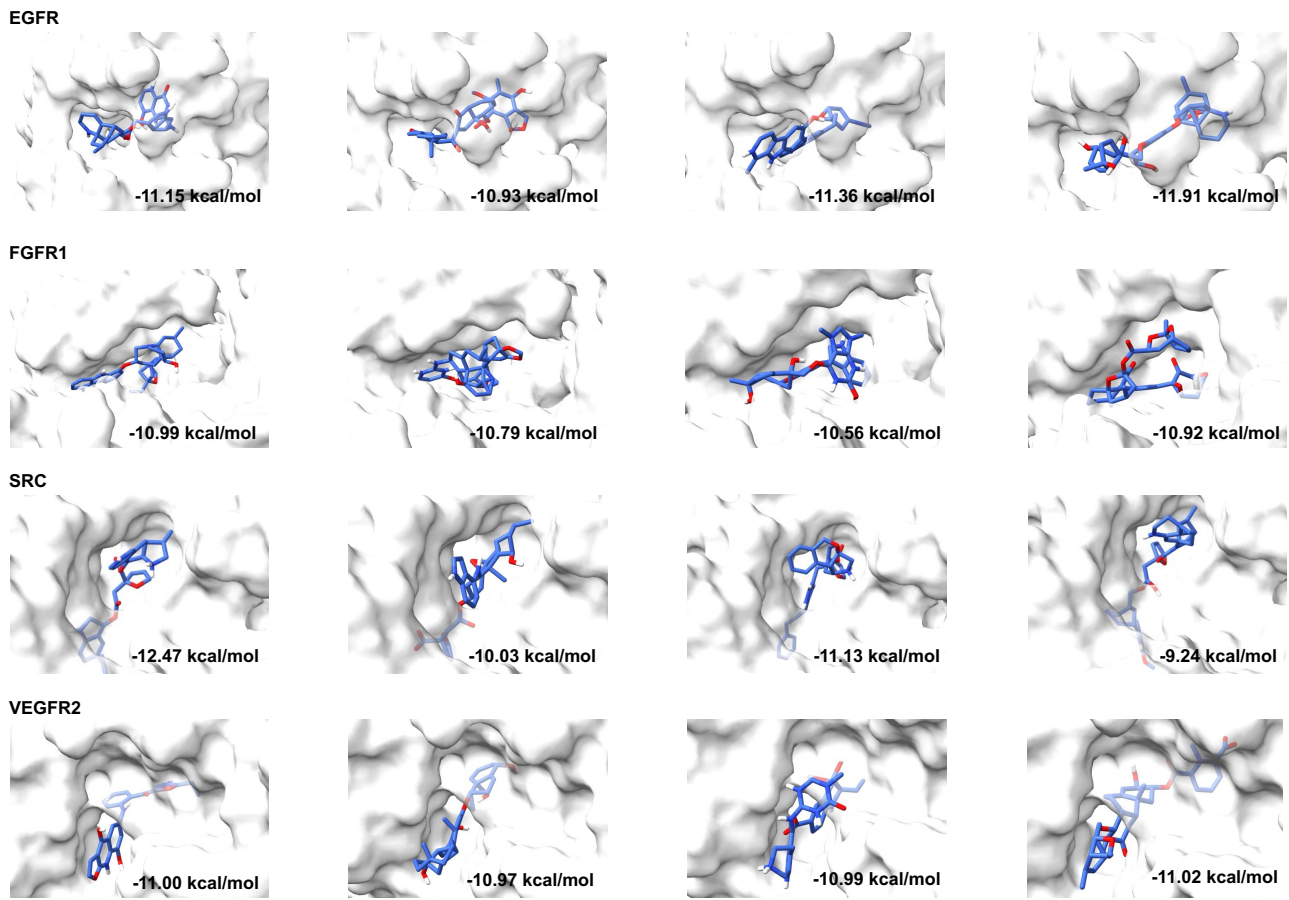

Supplementary Figure 16. Representative docking cases of sampled molecules from the Pseudo-NP space against different targets. For each target, four generated molecules with their docking poses are shown. Docking was performed using AutoDock Vina.

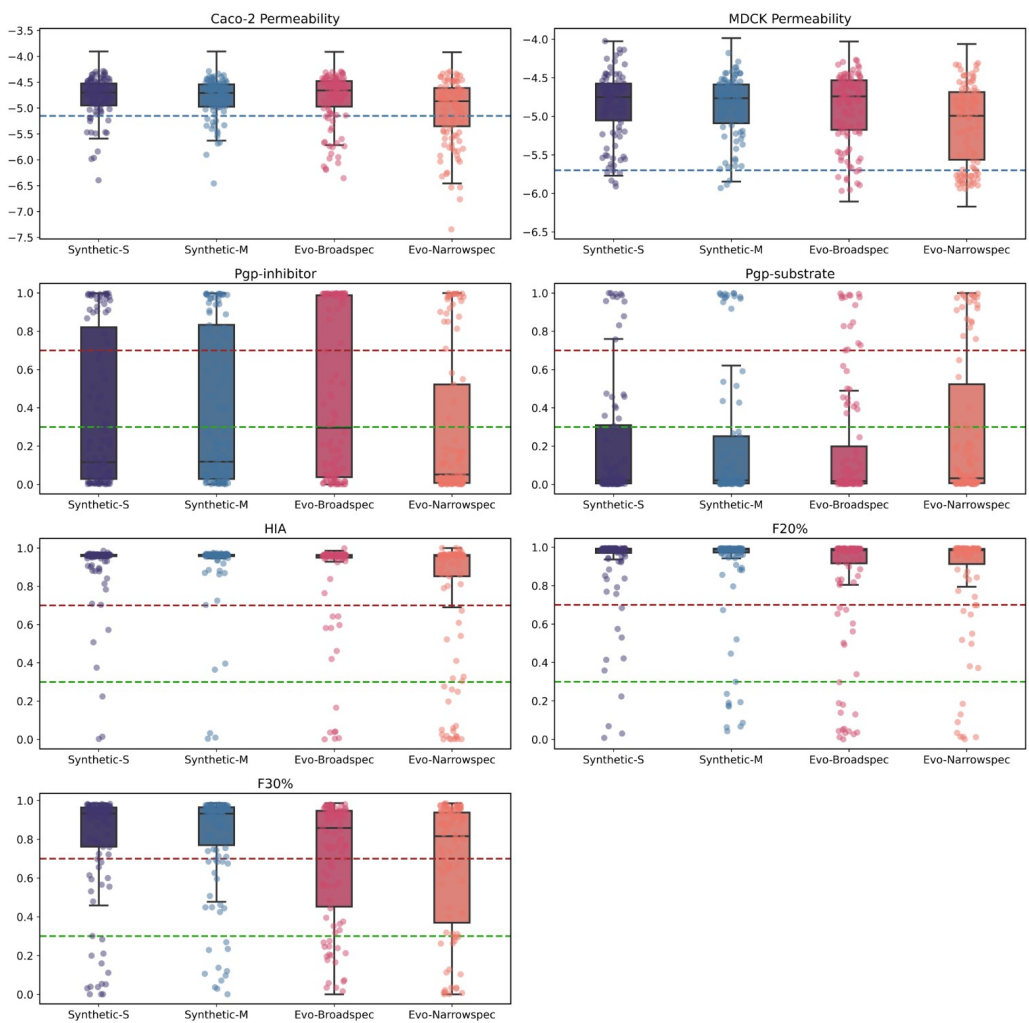

Supplementary Figure 17. Distribution of absorption-related properties for the building block libraries of Evo and control Synthetic spaces, predicted using the enhanced ADMETlab 2.0. Boxplots show distributions based on all data points, while overlaid dots represent 100 randomly sampled points for clarity.

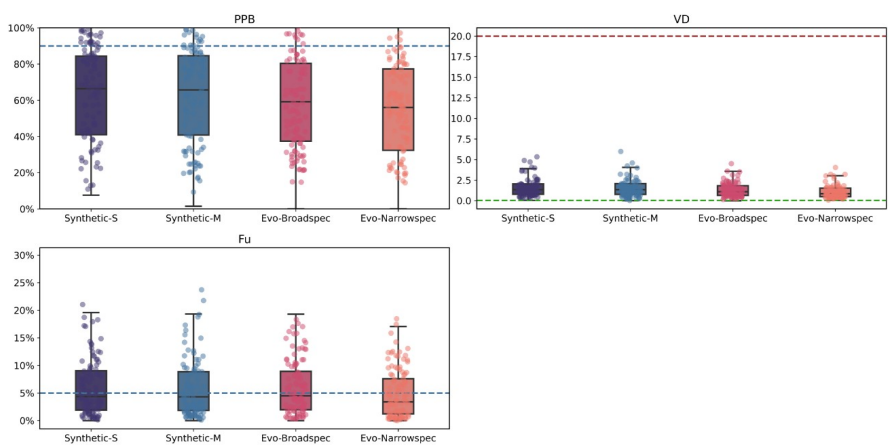

Supplementary Figure 18. Distribution-related properties of the building block libraries used in Evo and the control Synthetic chemical spaces. Properties were predicted using the enhanced ADMETlab 2.0. Boxplots show distributions based on all data points; overlaid dots represent 100 randomly sampled points for clarity.

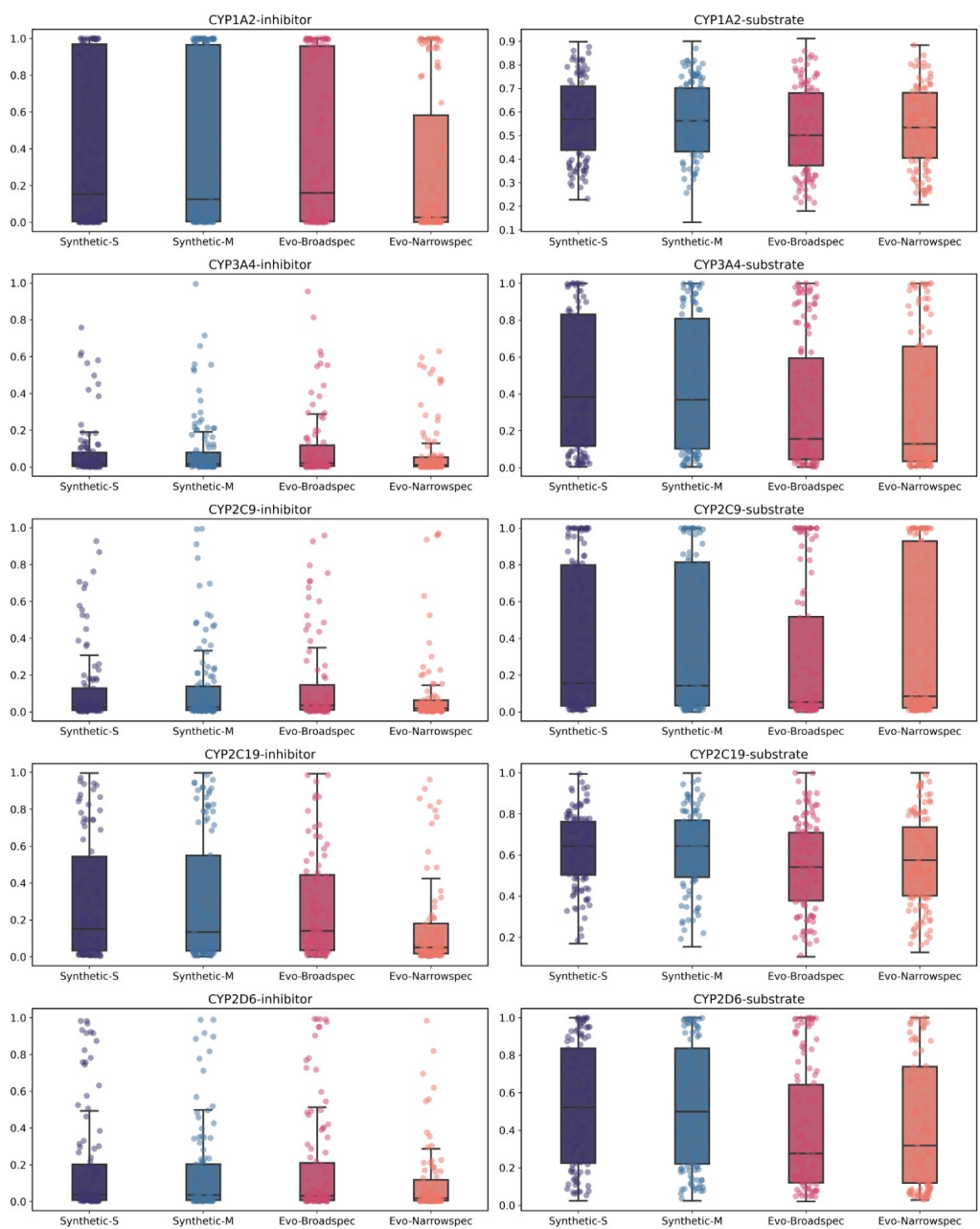

Supplementary Figure 19. Metabolism-related properties of the building block libraries used in Evo and the control Synthetic chemical spaces. Properties were predicted using the enhanced ADMETlab 2.0. Boxplots show distributions based on all data points; overlaid dots represent 100 randomly sampled points for clarity.

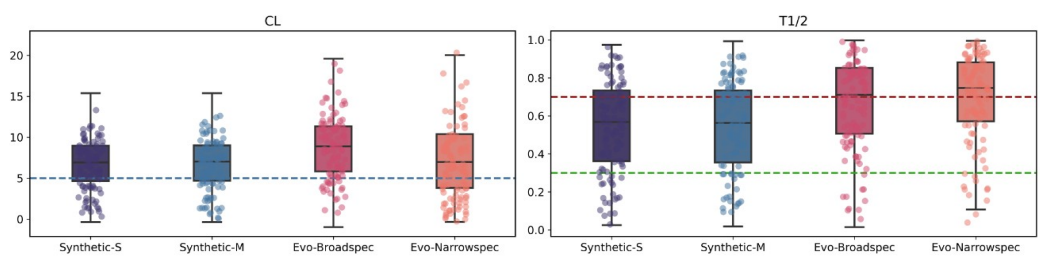

Supplementary Figure 20. Excretion-related properties of the building block libraries used in Evo and the control Synthetic chemical spaces. Properties were predicted using the enhanced ADMETlab 2.0. Boxplots show distributions based on all data points; overlaid dots represent 100 randomly sampled points for clarity.

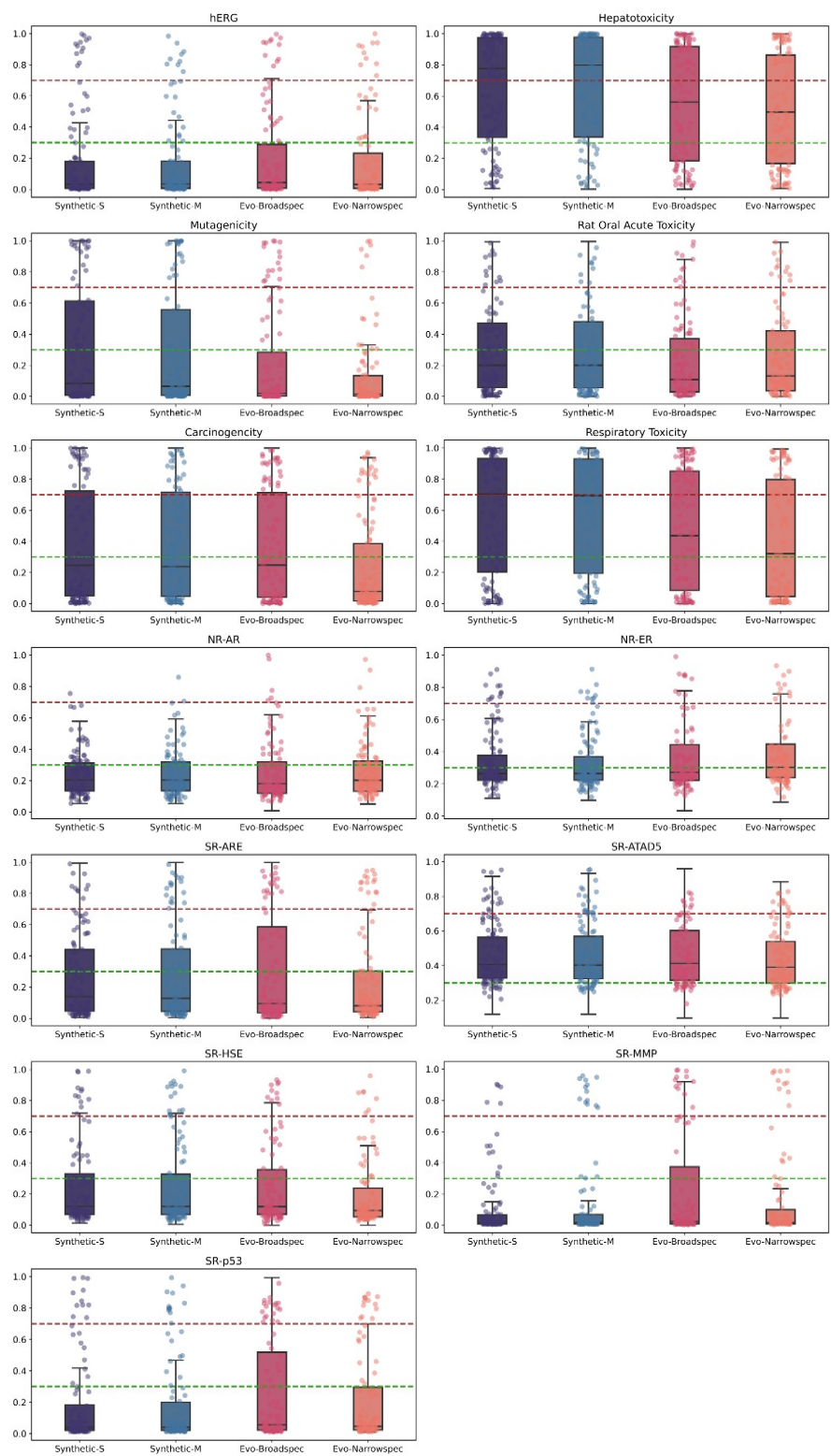

Supplementary Figure 21. Toxicity-related properties of the building block libraries used in Evo and the control Synthetic chemical spaces. Properties were predicted using the enhanced ADMETlab 2.0. Boxplots show distributions based on all data points; overlaid dots represent 100 randomly sampled points for clarity.

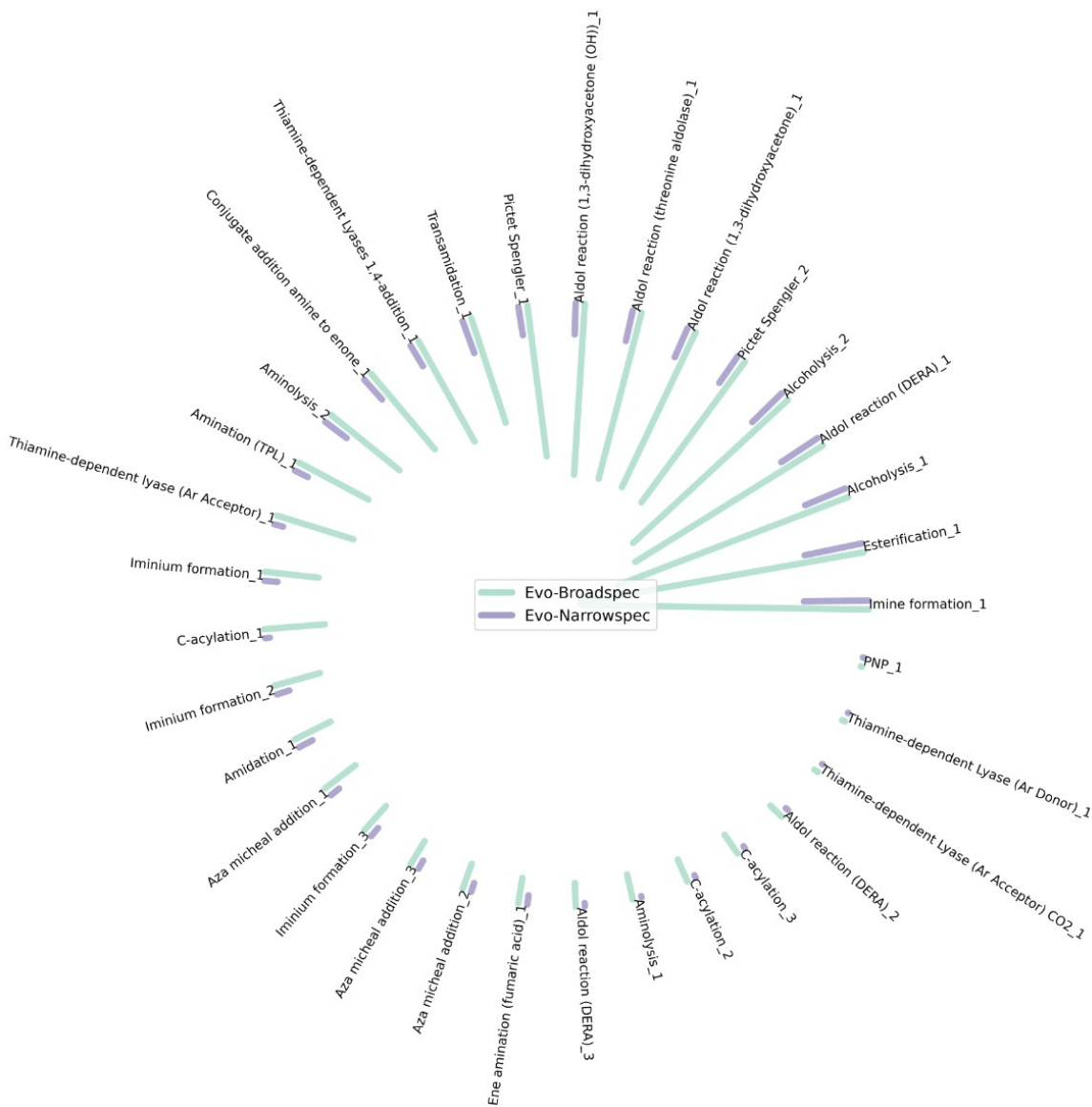

Supplementary Figure 22. Comparison of reaction preferences (number of matched building blocks) between the two Evo spaces. Reactions are derived from the Evo reaction library consisting of 125 endogenous reactions.

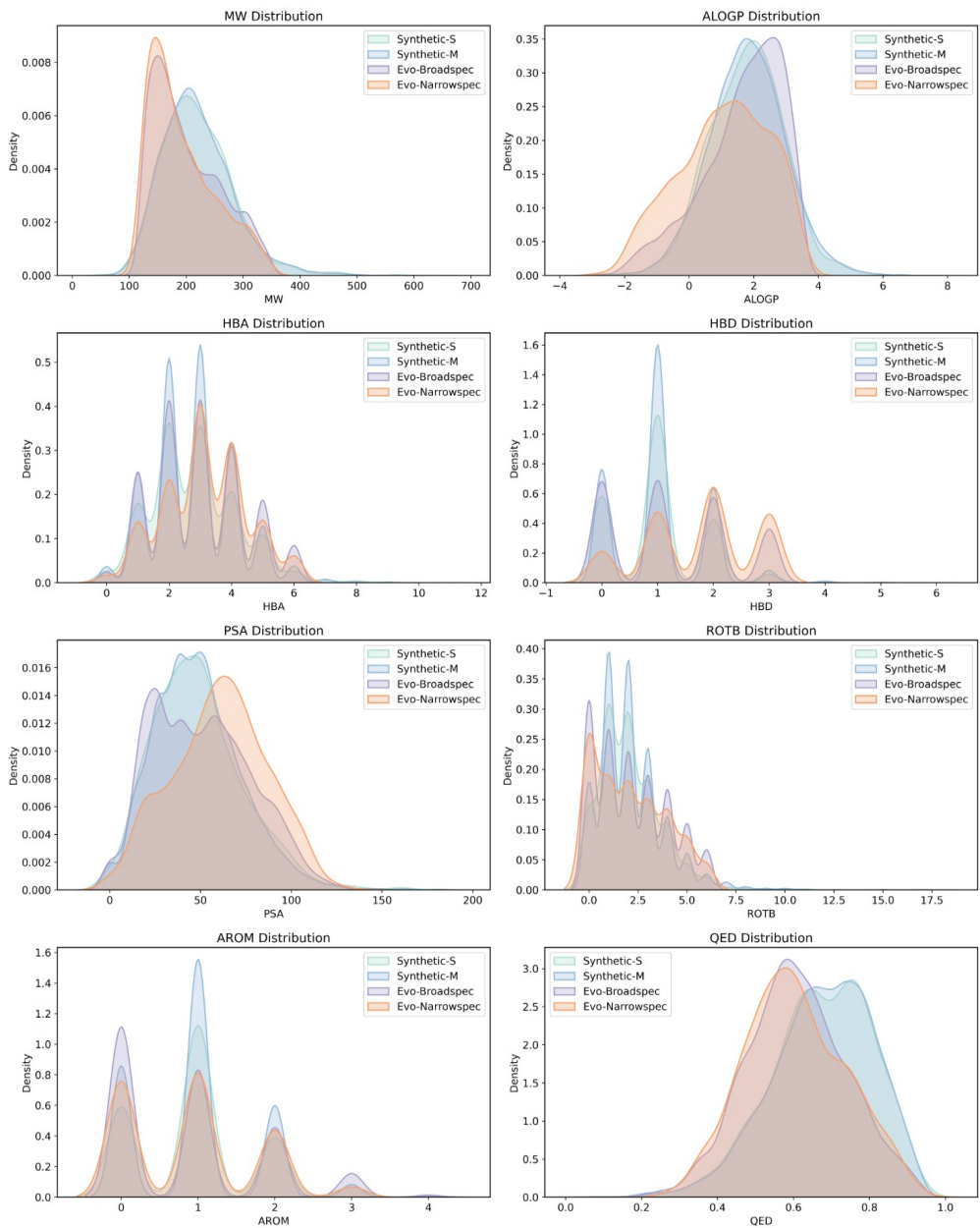

Supplementary Figure 23. Comparison of general physicochemical property distributions of the building block libraries used in Evo and Synthetic spaces. Considered properties include molecular weight (MW), octanol–water partition coefficient (ALOGP), number of hydrogen bond acceptors (HBA), number of hydrogen bond donors (HBD), polar surface area (PSA), number of rotatable bonds (ROTB), number of aromatic rings (AROM), and quantitative estimate of drug-likeness (QED).

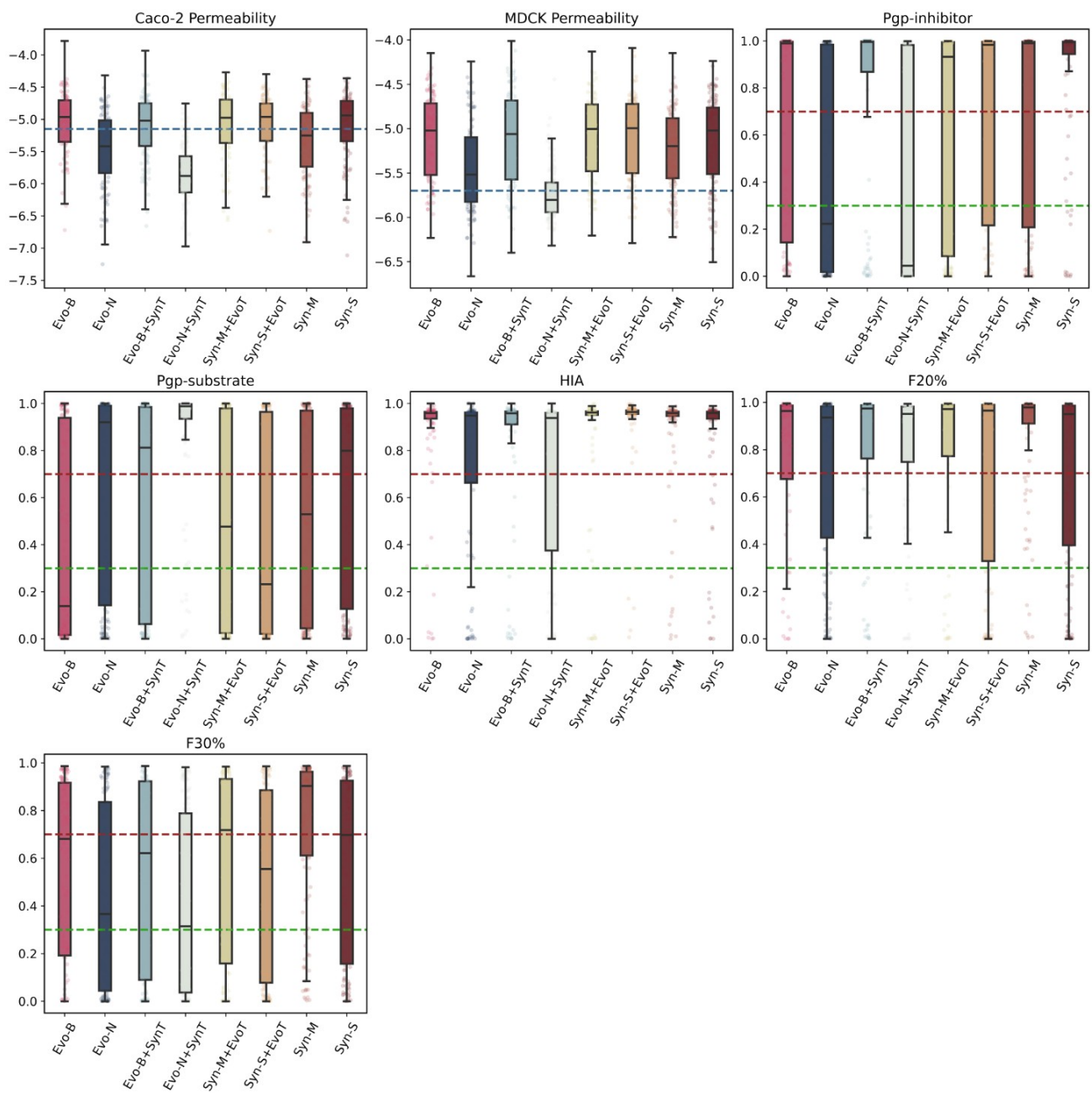

Supplementary Figure 24. Distribution of Absorption-related properties of the top 1000 QSAR-ranked molecules sampled in Evo and all control spaces under the EGFR target. Properties were predicted using the enhanced version of ADMETlab 2.0. Evo-B and Evo-N represent Evo-Broadspec and Evo-Narrowspec, respectively, which differ in the selection range of building blocks (see Methods for details). Syn-S and Syn-M denote the two control Synthetic spaces, Synthetic-S and Synthetic-M. Evo-B+SynT and Evo-N+SynT represent two mixed spaces constructed by combining the metabolic building blocks from Evo with the synthetic reaction set. Syn-S+EvoT and Syn-M+EvoT represent two mixed spaces constructed by combining the enzymatic reactions from Evo with the synthetic building block library.

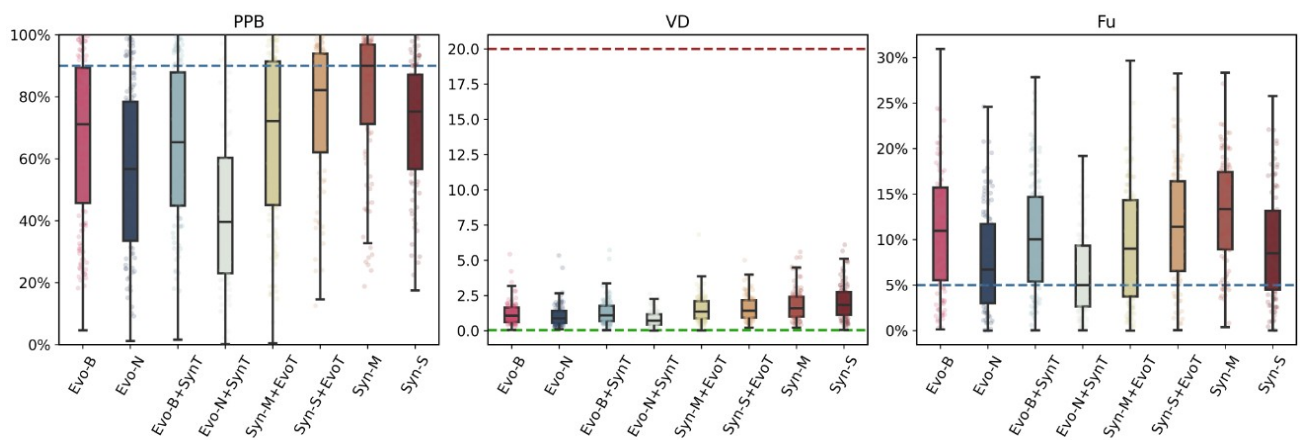

Supplementary Figure 25. Distribution of Distribution-related properties of the top 1000 QSAR-ranked molecules sampled in Evo and all control spaces under the EGFR target. Properties were predicted using the enhanced version of ADMETlab 2.0. Evo-B and Evo-N represent Evo-Broadspec and Evo-Narrowspec, respectively, which differ in the selection range of building blocks (see Methods for details). Syn-S and Syn-M denote the two control Synthetic spaces, Synthetic-S and Synthetic-M. Evo-B+SynT and Evo-N+SynT represent two mixed spaces constructed by combining the metabolic building blocks from Evo with the synthetic reaction set. Syn-S+EvoT and Syn-M+EvoT represent two mixed spaces constructed by combining the enzymatic reactions from Evo with the synthetic building block library.

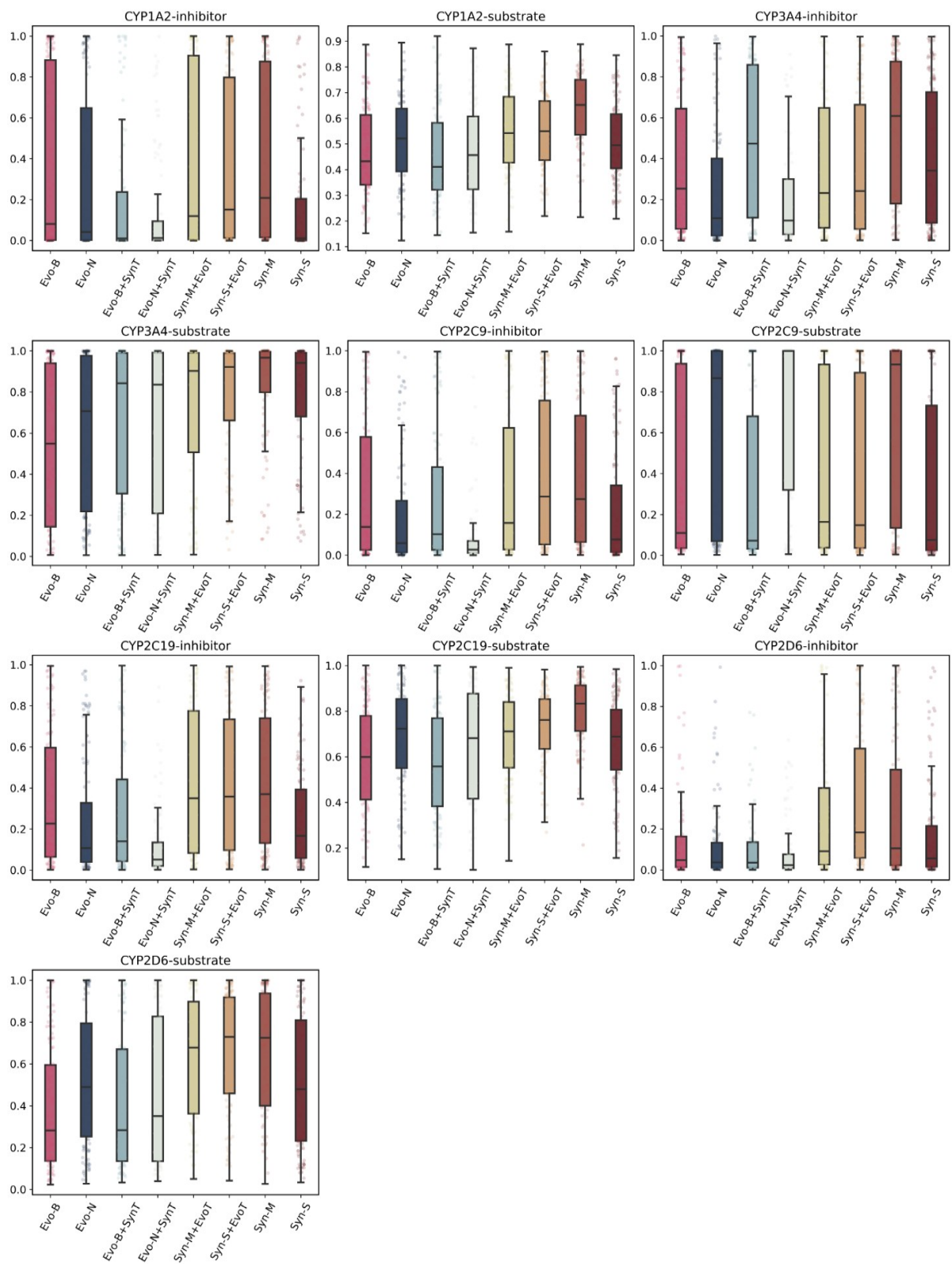

Supplementary Figure 26. Distribution of Metabolism-related properties of the top 1000 QSAR-ranked molecules sampled in Evo and all control spaces under the EGFR target. Properties were predicted using the enhanced version of ADMETlab 2.0. Evo-B and Evo-N represent Evo-Broadspec and Evo-Narrowspec, respectively, which differ in the selection range of building blocks (see Methods for details). Syn-S and Syn-M denote

the two control Synthetic spaces, Synthetic-S and Synthetic-M. Evo-B+SynT and Evo-N+SynT represent two mixed spaces constructed by combining the metabolic building blocks from Evo with the synthetic reaction set. Syn-S+EvoT and Syn-M+EvoT represent two mixed spaces constructed by combining the enzymatic reactions from Evo with the synthetic building block library.

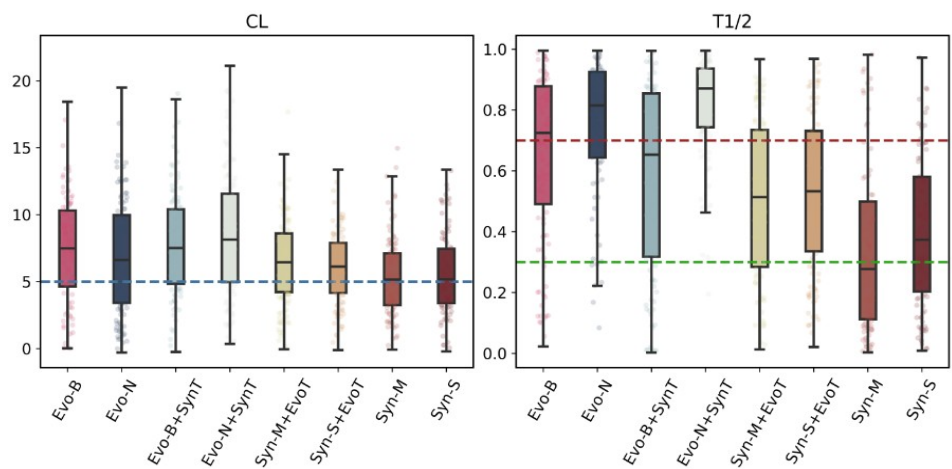

Supplementary Figure 27. Distribution of Excretion-related properties of the top 1000 QSAR-ranked molecules sampled in Evo and all control spaces under the EGFR target. Properties were predicted using the enhanced version of ADMETlab 2.0. Evo-B and Evo-N represent Evo-Broadspec and Evo-Narrowspec, respectively, which differ in the selection range of building blocks (see Methods for details). Syn-S and Syn-M denote the two control Synthetic spaces, Synthetic-S and Synthetic-M. Evo-B+SynT and Evo-N+SynT represent two mixed spaces constructed by combining the metabolic building blocks from Evo with the synthetic reaction set. Syn-S+EvoT and Syn-M+EvoT represent two mixed spaces constructed by combining the enzymatic reactions from Evo with the synthetic building block library.

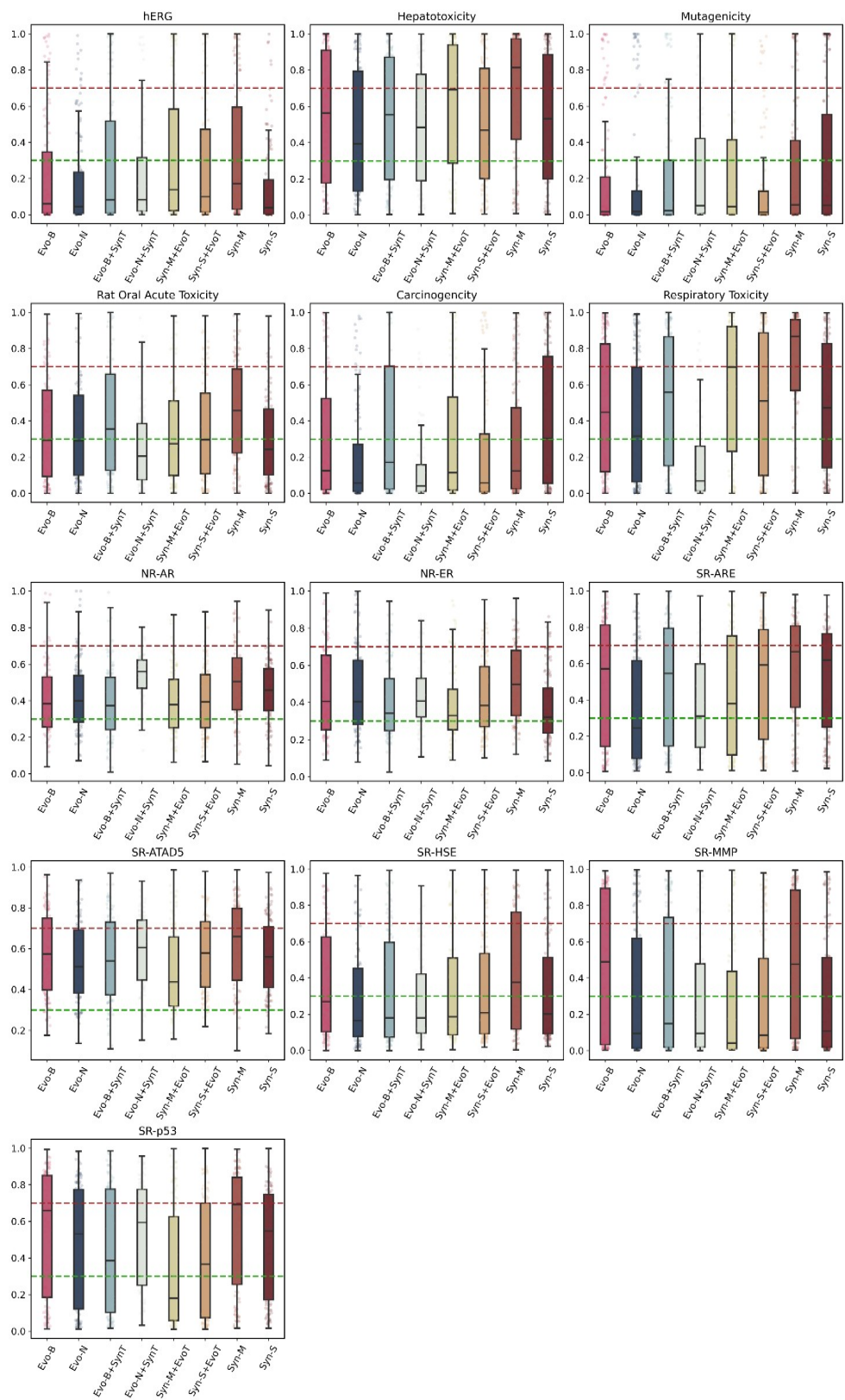

Supplementary Figure 28. Distribution of Toxicity-related properties of the top 1000 QSAR-ranked molecules sampled in Evo and all control spaces under the EGFR target. Properties were predicted using the enhanced version of ADMETlab 2.0. Evo-B and Evo-N represent Evo-Broadspec and Evo-Narrowspec, respectively, which differ in the selection range of building blocks (see Methods for details). Syn-S and Syn-M denote the two control

Synthetic spaces, Synthetic-S and Synthetic-M. Evo-B+SynT and Evo-N+SynT represent two mixed spaces constructed by combining the metabolic building blocks from Evo with the synthetic reaction set. Syn-S+EvoT and Syn-M+EvoT represent two mixed spaces constructed by combining the enzymatic reactions from Evo with the synthetic building block library.

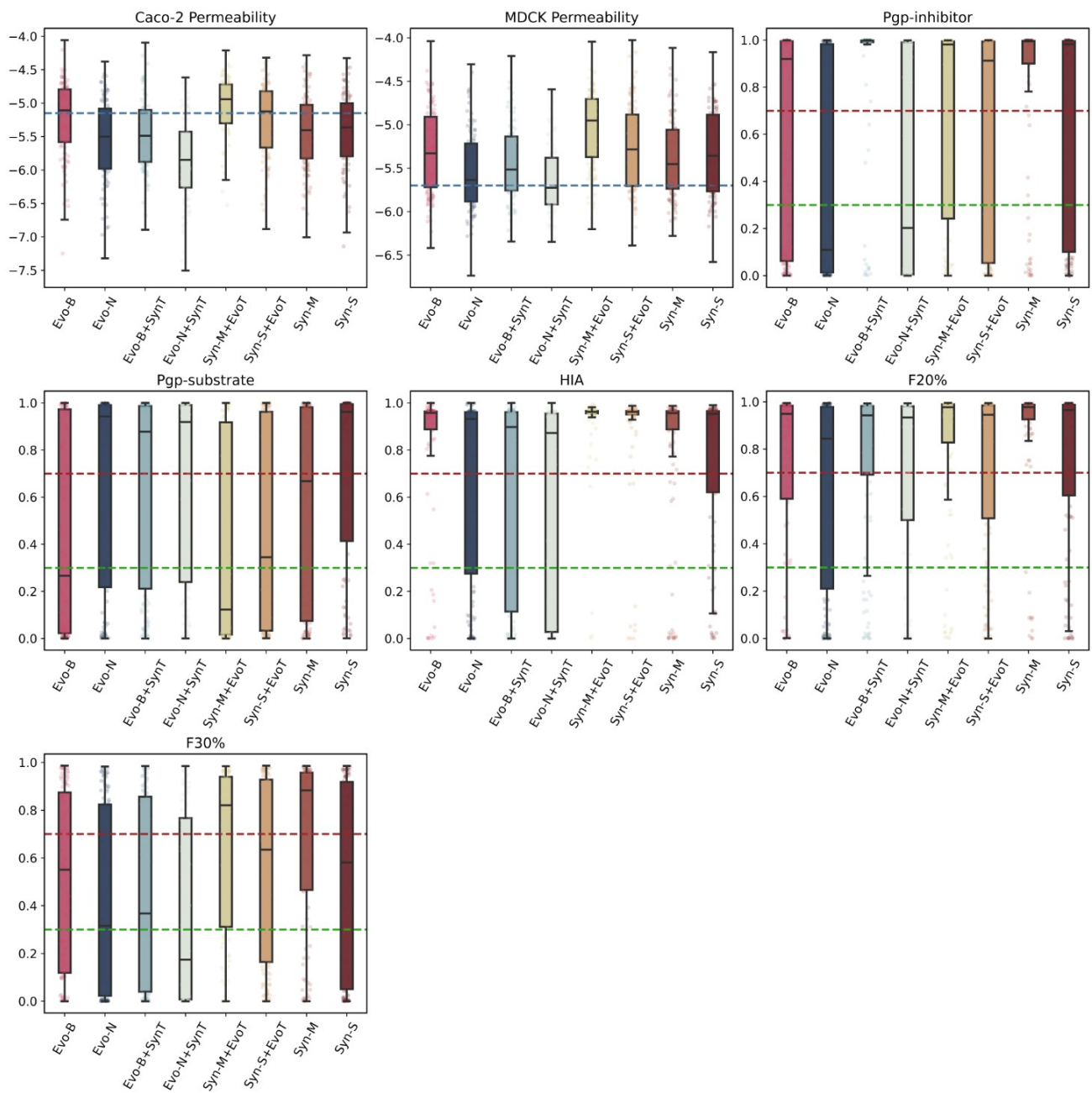

Supplementary Figure 29. Distribution of Absorption-related properties of the top 1000 QSAR-ranked molecules sampled in Evo and all control spaces under the FGFR1 target. Properties were predicted using the enhanced version of ADMETlab 2.0. Evo-B and Evo-N represent Evo-Broadspec and Evo-Narrowspec, respectively, which differ in the selection range of building blocks (see Methods for details). Syn-S and Syn-M denote the two control Synthetic spaces, Synthetic-S and Synthetic-M. Evo-B+SynT and Evo-N+SynT represent two

mixed spaces constructed by combining the metabolic building blocks from Evo with the synthetic reaction set. Syn-S+EvoT and Syn-M+EvoT represent two mixed spaces constructed by combining the enzymatic reactions from Evo with the synthetic building block library.

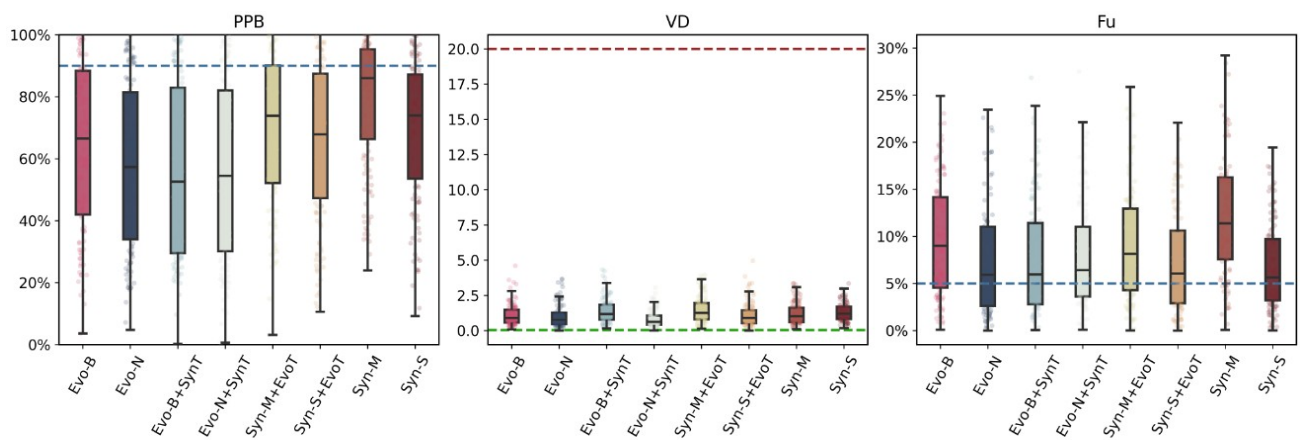

Supplementary Figure 30. Distribution of Distribution-related properties of the top 1000 QSAR-ranked molecules sampled in Evo and all control spaces under the FGFR1 target. Properties were predicted using the enhanced version of ADMETlab 2.0. Evo-B and Evo-N represent Evo-Broadspec and Evo-Narrowspec, respectively, which differ in the selection range of building blocks (see Methods for details). Syn-S and Syn-M denote the two control Synthetic spaces, Synthetic-S and Synthetic-M. Evo-B+SynT and Evo-N+SynT represent two mixed spaces constructed by combining the metabolic building blocks from Evo with the synthetic reaction set. Syn-S+EvoT and Syn-M+EvoT represent two mixed spaces constructed by combining the enzymatic reactions from Evo with the synthetic building block library.

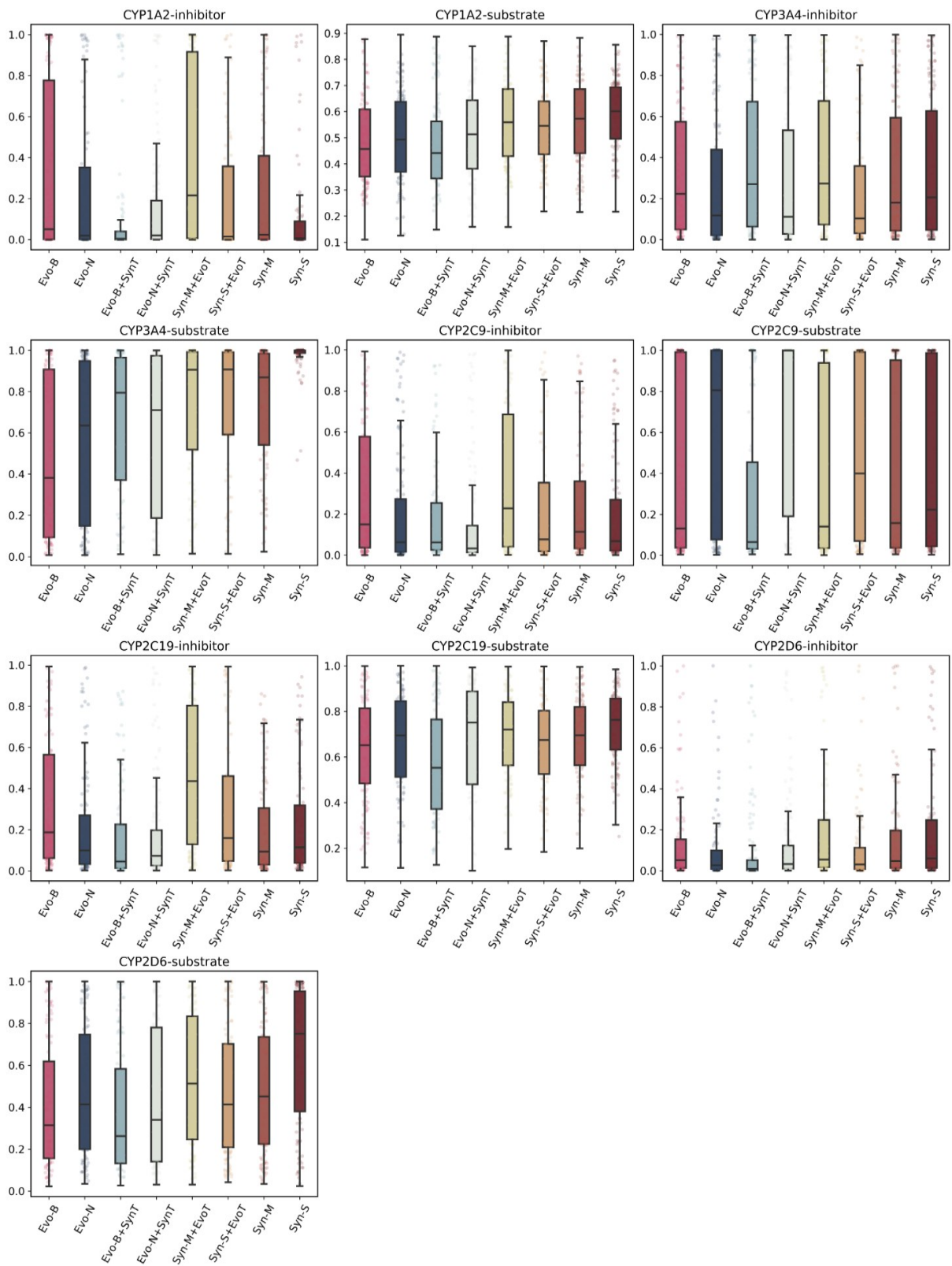

Supplementary Figure 31. Distribution of Metabolism-related properties of the top 1000 QSAR-ranked molecules sampled in Evo and all control spaces under the FGFR1 target. Properties were predicted using the enhanced version of ADMETlab 2.0. Evo-B and Evo-N represent Evo-Broadspec and Evo-Narrowspec, respectively, which differ in the selection range of building blocks (see Methods for details). Syn-S and Syn-M denote

the two control Synthetic spaces, Synthetic-S and Synthetic-M. Evo-B+SynT and Evo-N+SynT represent two mixed spaces constructed by combining the metabolic building blocks from Evo with the synthetic reaction set. Syn-S+EvoT and Syn-M+EvoT represent two mixed spaces constructed by combining the enzymatic reactions from Evo with the synthetic building block library.

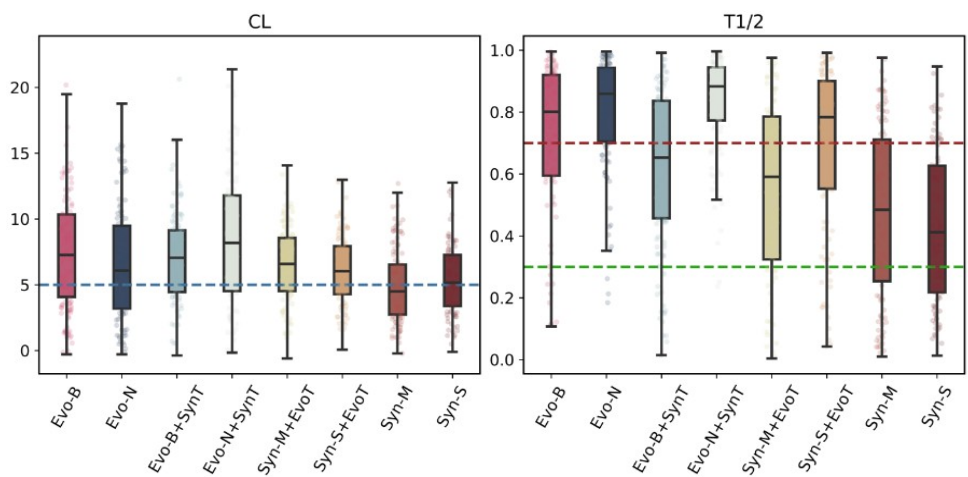

Supplementary Figure 32. Distribution of Excretion-related properties of the top 1000 QSAR-ranked molecules sampled in Evo and all control spaces under the FGFR1 target. Properties were predicted using the enhanced version of ADMETlab 2.0. Evo-B and Evo-N represent Evo-Broadspec and Evo-Narrowspec, respectively, which differ in the selection range of building blocks (see Methods for details). Syn-S and Syn-M denote the two control Synthetic spaces, Synthetic-S and Synthetic-M. Evo-B+SynT and Evo-N+SynT represent two mixed spaces constructed by combining the metabolic building blocks from Evo with the synthetic reaction set. Syn-S+EvoT and Syn-M+EvoT represent two mixed spaces constructed by combining the enzymatic reactions from Evo with the synthetic building block library.

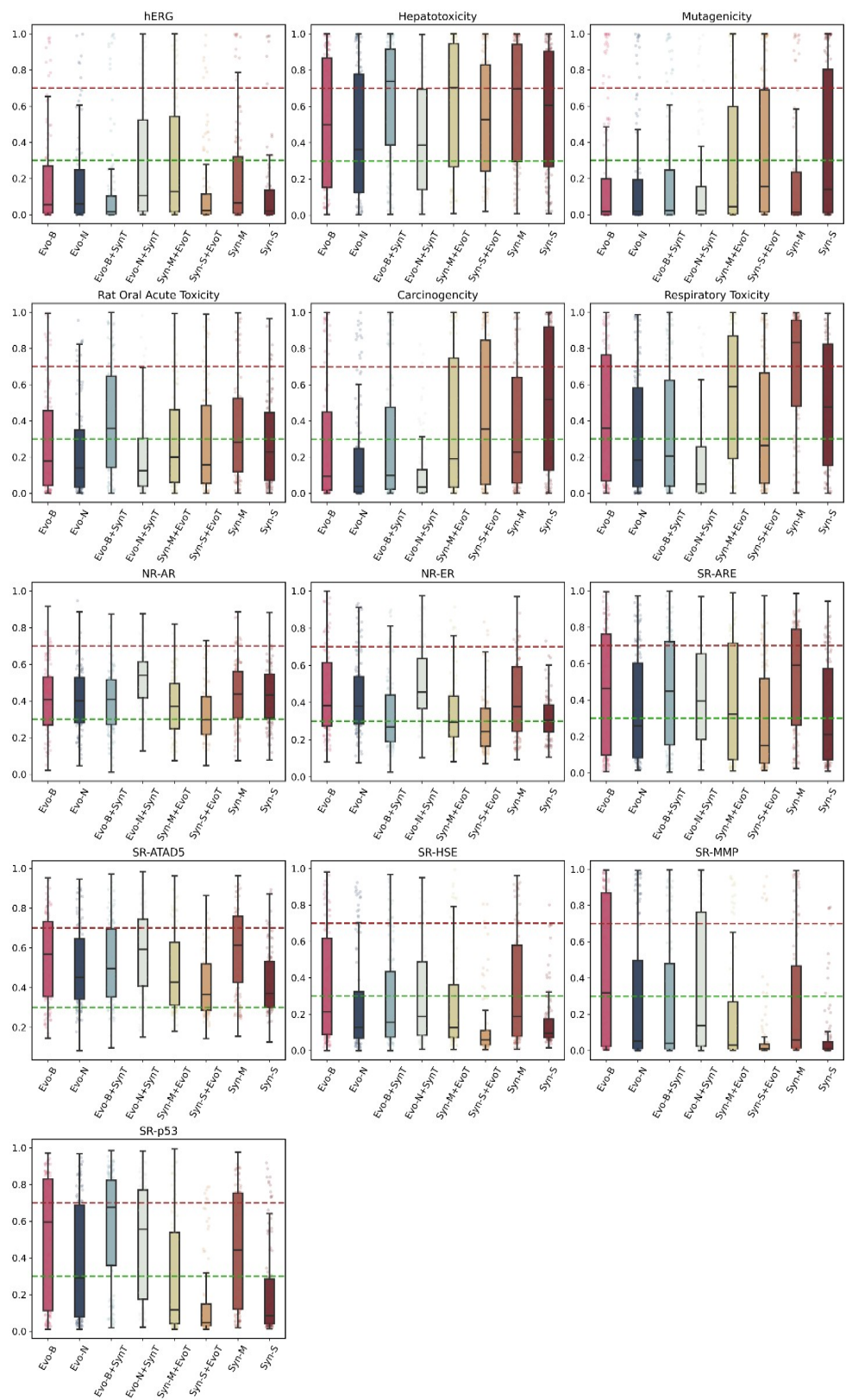

Supplementary Figure 33. Distribution of Toxicity-related properties of the top 1000 QSAR-ranked molecules sampled in Evo and all control spaces under the FGFR1 target. Properties were predicted using the enhanced version of ADMETlab 2.0. Evo-B and Evo-N represent Evo-Broadspec and Evo-Narrowspec, respectively, which differ in the selection range of building blocks (see Methods for details). Syn-S and Syn-M denote the two control

Synthetic spaces, Synthetic-S and Synthetic-M. Evo-B+SynT and Evo-N+SynT represent two mixed spaces constructed by combining the metabolic building blocks from Evo with the synthetic reaction set. Syn-S+EvoT and Syn-M+EvoT represent two mixed spaces constructed by combining the enzymatic reactions from Evo with the synthetic building block library.

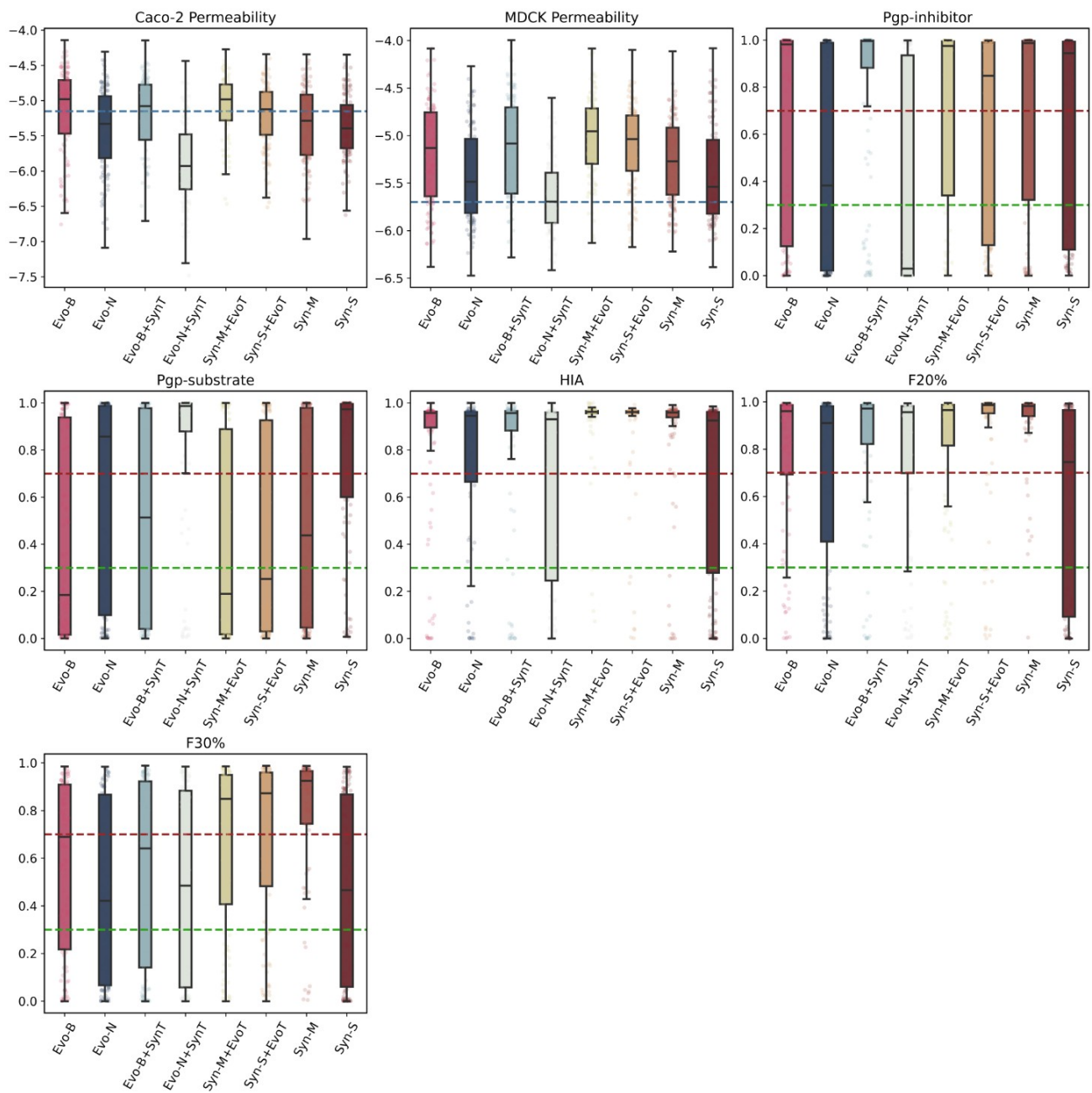

Supplementary Figure 34. Distribution of Absorption-related properties of the top 1000 QSAR-ranked molecules sampled in Evo and all control spaces under the SRC target. Properties were predicted using the enhanced version of ADMETlab 2.0. Evo-B and Evo-N represent Evo-Broadspec and Evo-Narrowspec, respectively, which differ in the selection range of building blocks (see Methods for details). Syn-S and Syn-M denote the two control

Synthetic spaces, Synthetic-S and Synthetic-M. Evo-B+SynT and Evo-N+SynT represent two mixed spaces constructed by combining the metabolic building blocks from Evo with the synthetic reaction set. Syn-S+EvoT and Syn-M+EvoT represent two mixed spaces constructed by combining the enzymatic reactions from Evo with the synthetic building block library.

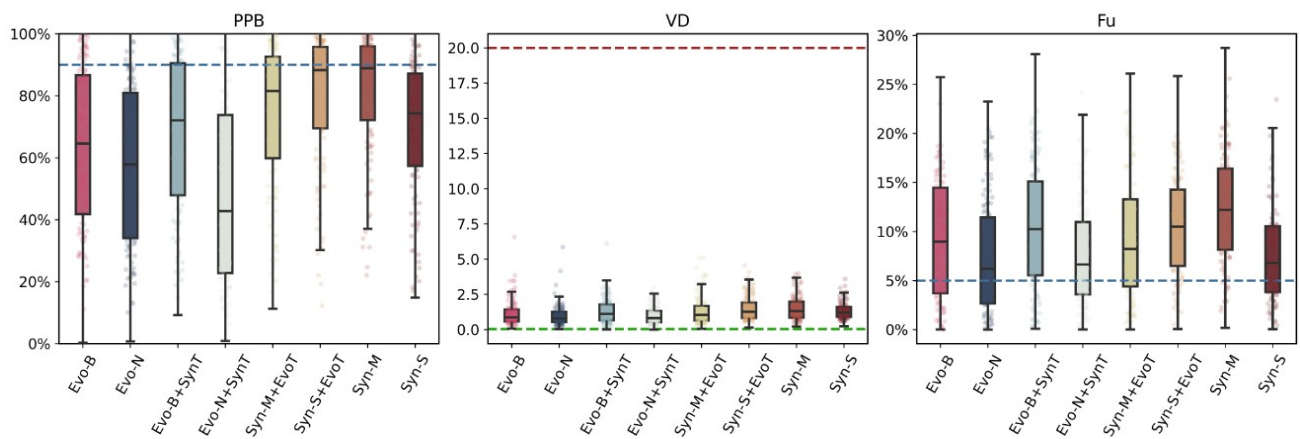

Supplementary Figure 35. Distribution of Distribution-related properties of the top 1000 QSAR-ranked molecules sampled in Evo and all control spaces under the SRC target. Properties were predicted using the enhanced version of ADMETlab 2.0. Evo-B and Evo-N represent Evo-Broadspec and Evo-Narrowspec, respectively, which differ in the selection range of building blocks (see Methods for details). Syn-S and Syn-M denote the two control Synthetic spaces, Synthetic-S and Synthetic-M. Evo-B+SynT and Evo-N+SynT represent two mixed spaces constructed by combining the metabolic building blocks from Evo with the synthetic reaction set. Syn-S+EvoT and Syn-M+EvoT represent two mixed spaces constructed by combining the enzymatic reactions from Evo with the synthetic building block library.

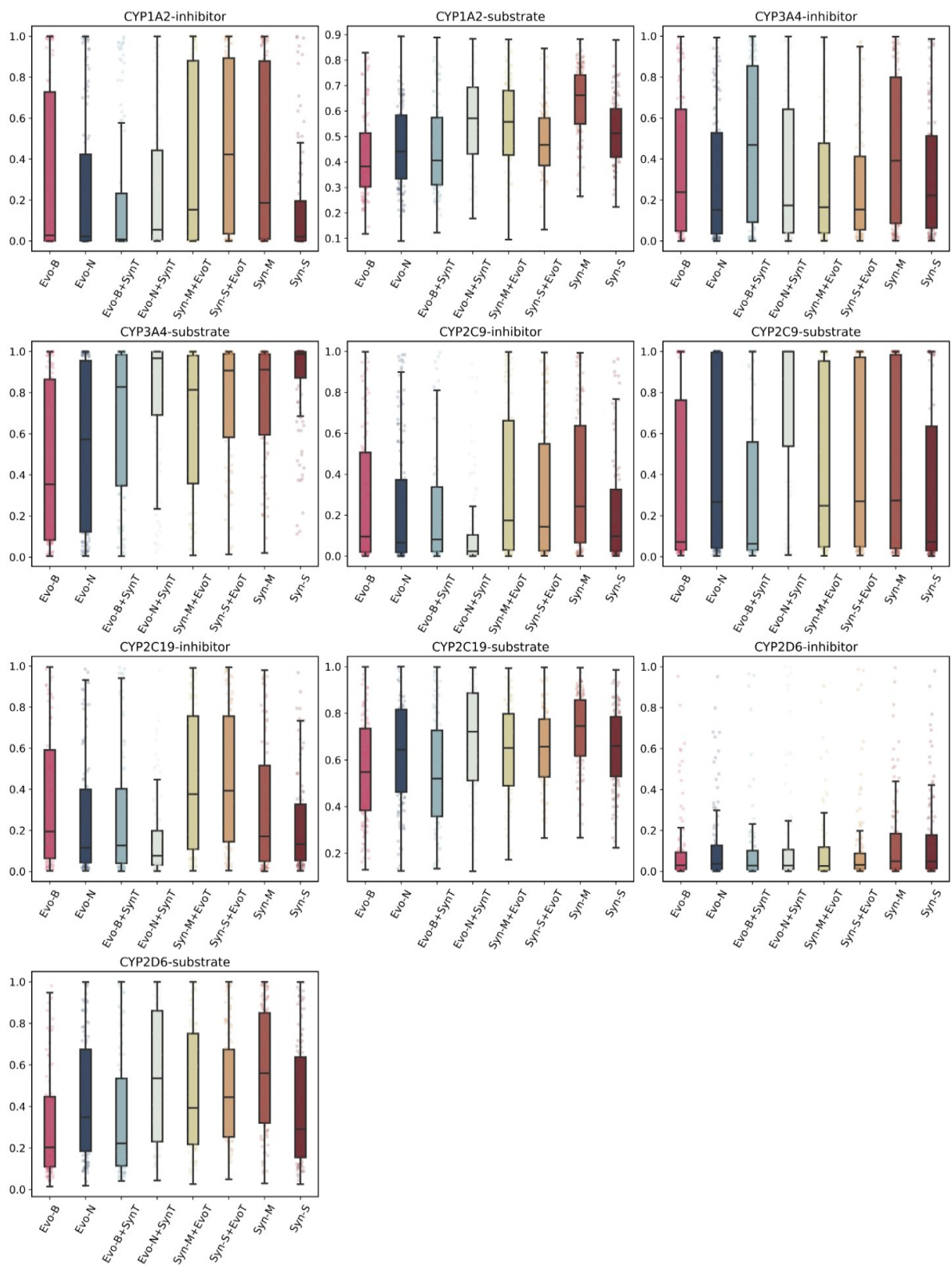

Supplementary Figure 36. Distribution of Metabolism-related properties of the top 1000 QSAR-ranked molecules sampled in Evo and all control spaces under the SRC target. Properties were predicted using the enhanced version of ADMETlab 2.0. Evo-B and Evo-N represent Evo-Broadspec and Evo-Narrowspec, respectively, which differ in the selection range of building blocks (see Methods for details). Syn-S and Syn-M denote the two control

Synthetic spaces, Synthetic-S and Synthetic-M. Evo-B+SynT and Evo-N+SynT represent two mixed spaces constructed by combining the metabolic building blocks from Evo with the synthetic reaction set. Syn-S+EvoT and Syn-M+EvoT represent two mixed spaces constructed by combining the enzymatic reactions from Evo with the synthetic building block library.

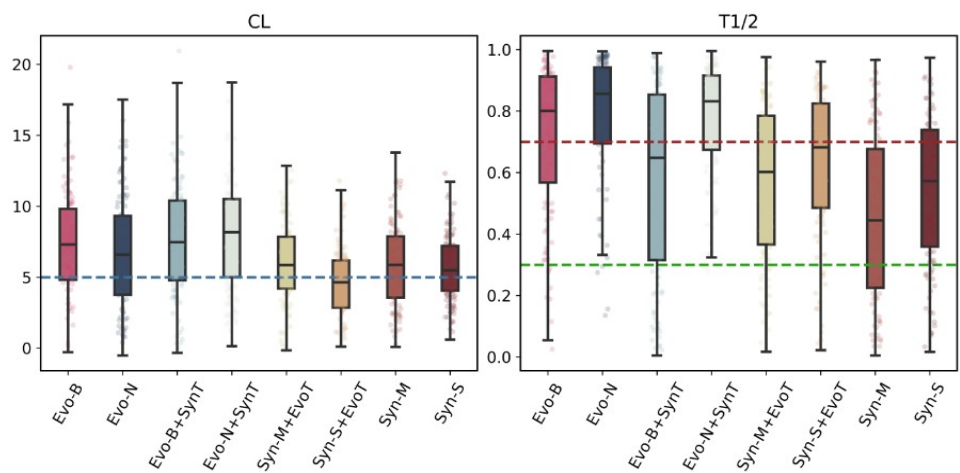

Supplementary Figure 37. Distribution of Excretion-related properties of the top 1000 QSAR-ranked molecules sampled in Evo and all control spaces under the SRC target. Properties were predicted using the enhanced version of ADMETlab 2.0. Evo-B and Evo-N represent Evo-Broadspec and Evo-Narrowspec, respectively, which differ in the selection range of building blocks (see Methods for details). Syn-S and Syn-M denote the two control Synthetic spaces, Synthetic-S and Synthetic-M. Evo-B+SynT and Evo-N+SynT represent two mixed spaces constructed by combining the metabolic building blocks from Evo with the synthetic reaction set. Syn-S+EvoT and Syn-M+EvoT represent two mixed spaces constructed by combining the enzymatic reactions from Evo with the synthetic building block library.

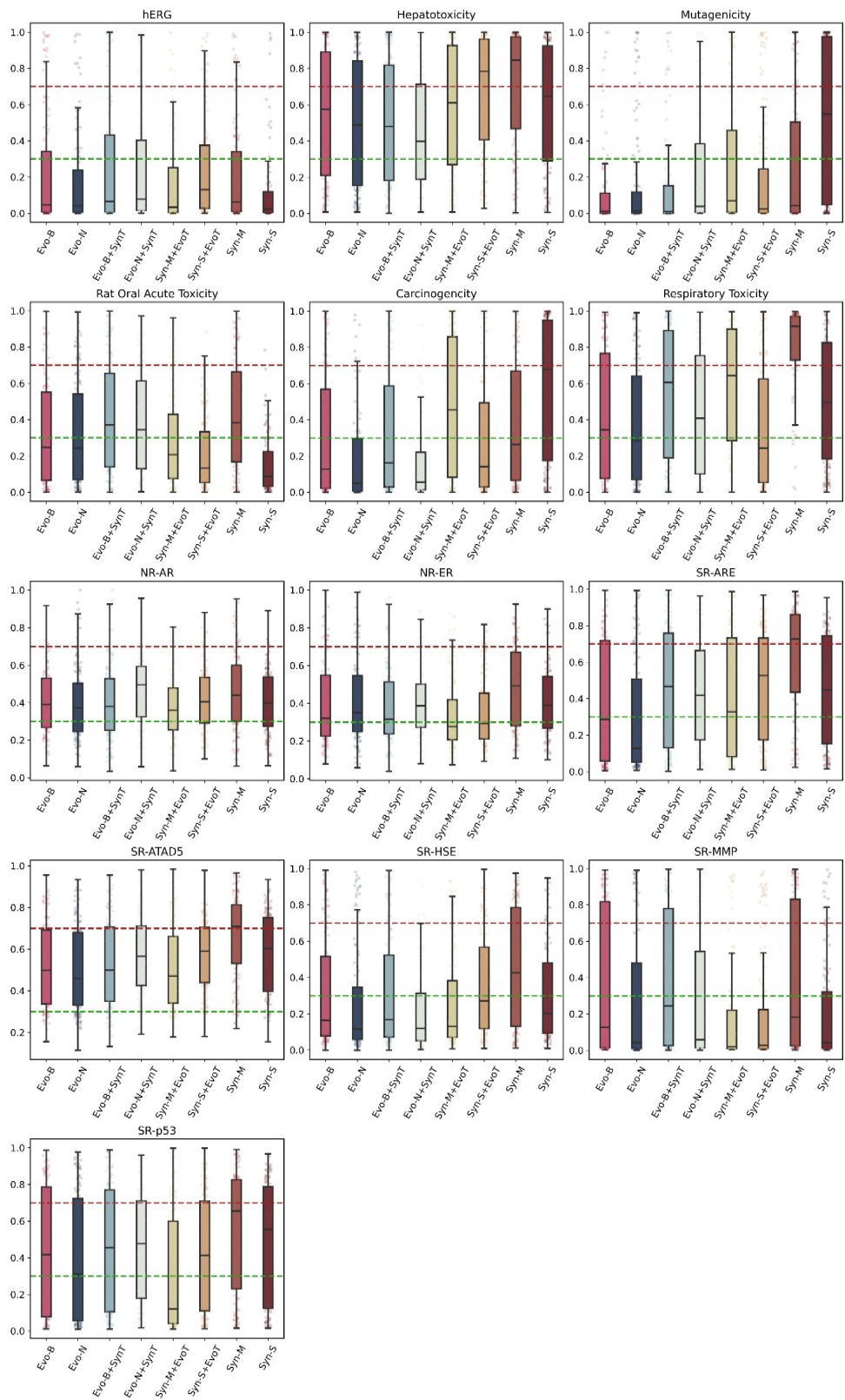

Supplementary Figure 38. Distribution of Toxicity-related properties of the top 1000 QSAR-ranked molecules sampled in Evo and all control spaces under the SRC target. Properties were predicted using the enhanced version of ADMETlab 2.0. Evo-B and Evo-N represent Evo-Broadspec and Evo-Narrowspec, respectively, which differ in the selection range of building blocks (see Methods for details). Syn-S and Syn-M denote the two control

Synthetic spaces, Synthetic-S and Synthetic-M. Evo-B+SynT and Evo-N+SynT represent two mixed spaces constructed by combining the metabolic building blocks from Evo with the synthetic reaction set. Syn-S+EvoT and Syn-M+EvoT represent two mixed spaces constructed by combining the enzymatic reactions from Evo with the synthetic building block library.

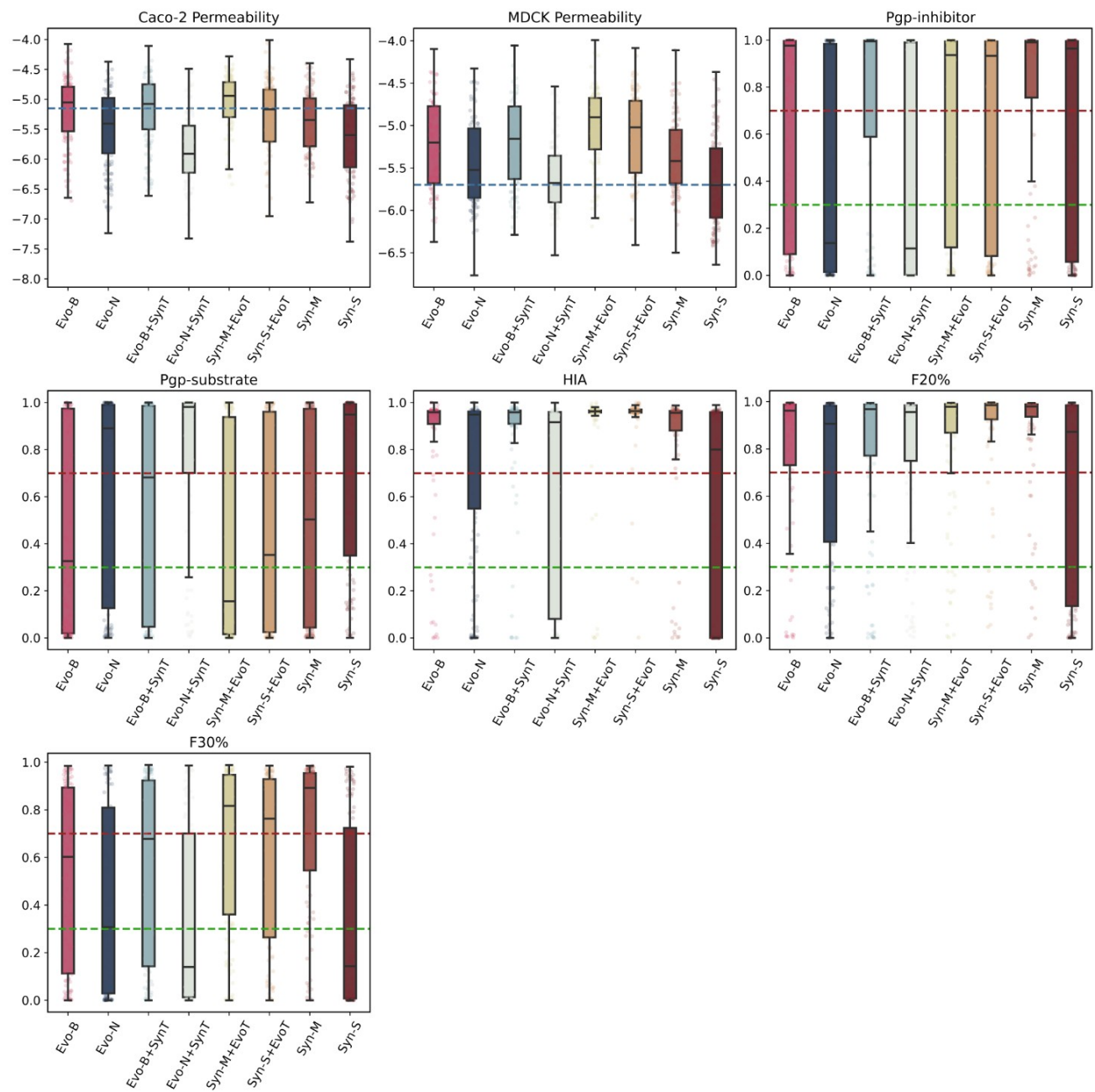

Supplementary Figure 39. Distribution of Absorption-related properties of the top 1000 QSAR-ranked molecules sampled in Evo and all control spaces under the VEGFR2 target. Properties were predicted using the enhanced version of ADMETlab 2.0. Evo-B and Evo-N represent Evo-Broadspec and Evo-Narrowspec, respectively, which differ in the selection range of building blocks (see Methods for details). Syn-S and Syn-M denote the two control Synthetic spaces, Synthetic-S and Synthetic-M. Evo-B+SynT and Evo-N+SynT represent two

mixed spaces constructed by combining the metabolic building blocks from Evo with the synthetic reaction set. Syn-S+EvoT and Syn-M+EvoT represent two mixed spaces constructed by combining the enzymatic reactions from Evo with the synthetic building block library.

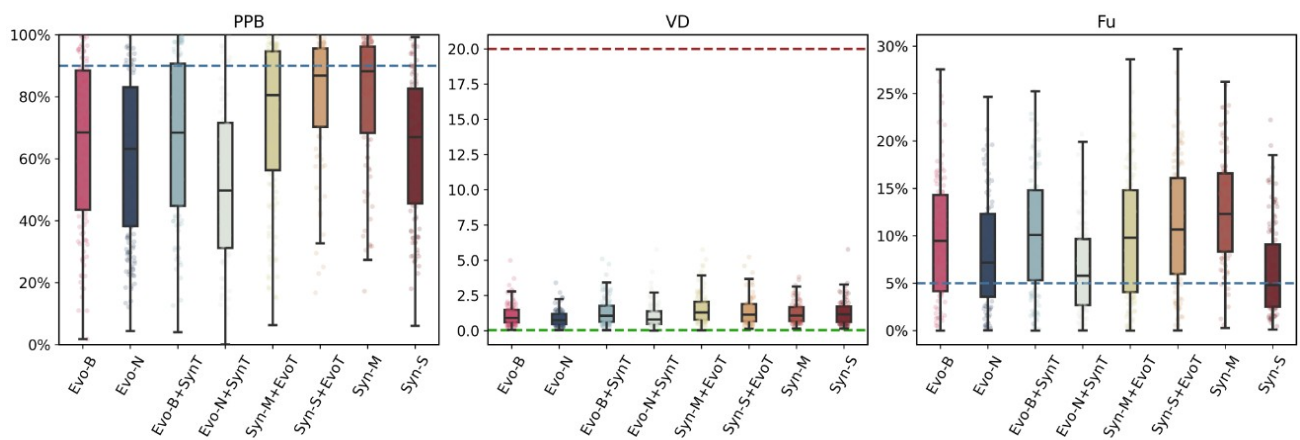

Supplementary Figure 40. Distribution of Distribution-related properties of the top 1000 QSAR-ranked molecules sampled in Evo and all control spaces under the VEGFR2 target. Properties were predicted using the enhanced version of ADMETlab 2.0. Evo-B and Evo-N represent Evo-Broadspec and Evo-Narrowspec, respectively, which differ in the selection range of building blocks (see Methods for details). Syn-S and Syn-M denote the two control Synthetic spaces, Synthetic-S and Synthetic-M. Evo-B+SynT and Evo-N+SynT represent two mixed spaces constructed by combining the metabolic building blocks from Evo with the synthetic reaction set. Syn-S+EvoT and Syn-M+EvoT represent two mixed spaces constructed by combining the enzymatic reactions from Evo with the synthetic building block library.

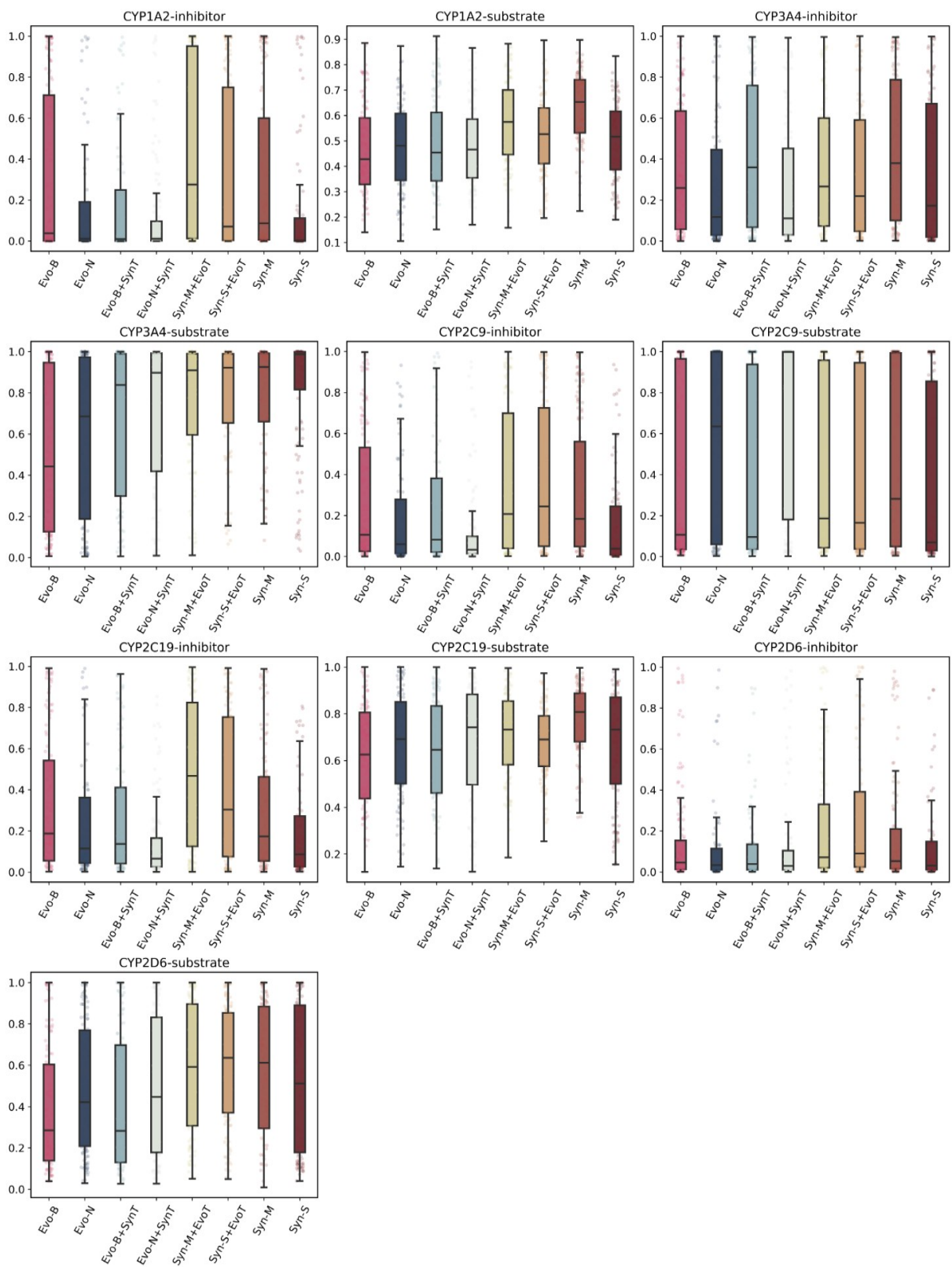

Supplementary Figure 41. Distribution of Metabolism-related properties of the top 1000 QSAR-ranked molecules sampled in Evo and all control spaces under the VEGFR2 target. Properties were predicted using the enhanced version of ADMETlab 2.0. Evo-B and Evo-N represent Evo-Broadspec and Evo-Narrowspec, respectively, which differ in the selection range of building blocks (see Methods for details). Syn-S and Syn-M denote

the two control Synthetic spaces, Synthetic-S and Synthetic-M. Evo-B+SynT and Evo-N+SynT represent two mixed spaces constructed by combining the metabolic building blocks from Evo with the synthetic reaction set. Syn-S+EvoT and Syn-M+EvoT represent two mixed spaces constructed by combining the enzymatic reactions from Evo with the synthetic building block library.

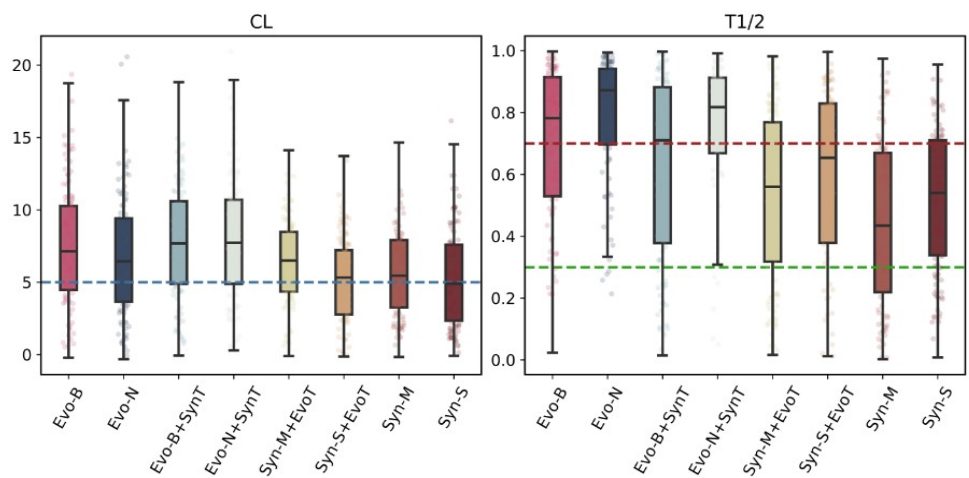

Supplementary Figure 42. Distribution of Excretion-related properties of the top 1000 QSAR-ranked molecules sampled in Evo and all control spaces under the VEGFR2 target. Properties were predicted using the enhanced version of ADMETlab 2.0. Evo-B and Evo-N represent Evo-Broadspec and Evo-Narrowspec, respectively, which differ in the selection range of building blocks (see Methods for details). Syn-S and Syn-M denote the two control Synthetic spaces, Synthetic-S and Synthetic-M. Evo-B+SynT and Evo-N+SynT represent two mixed spaces constructed by combining the metabolic building blocks from Evo with the synthetic reaction set. Syn-S+EvoT and Syn-M+EvoT represent two mixed spaces constructed by combining the enzymatic reactions from Evo with the synthetic building block library.

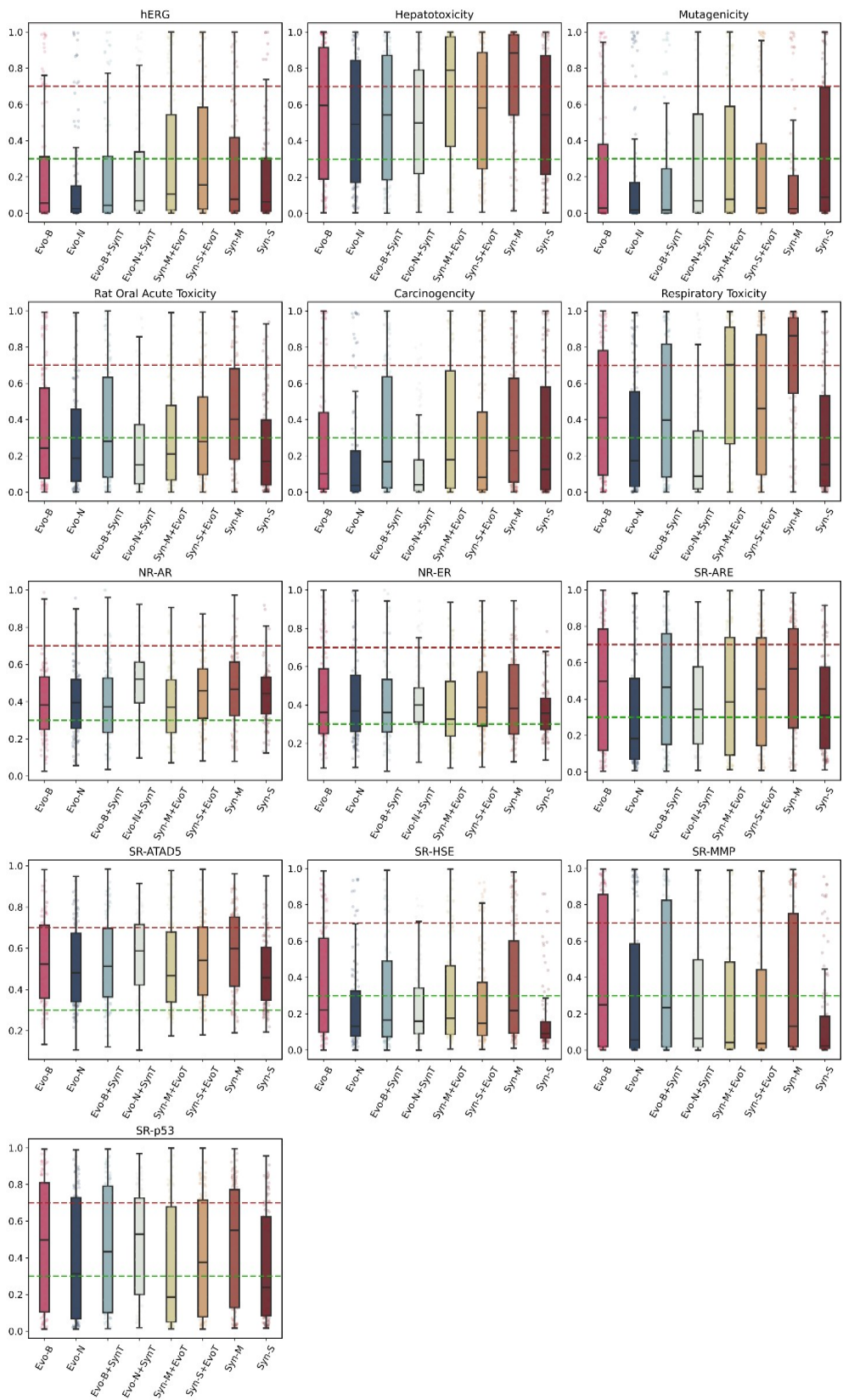

Supplementary Figure 43. Distribution of Toxicity-related properties of the top 1000 QSAR-ranked molecules sampled in Evo and all control spaces under the VEGFR2 target. Properties were predicted using the enhanced version of ADMETlab 2.0. Evo-B and Evo-N represent Evo-Broadspec and Evo-Narrowspec, respectively, which differ in the selection range of building blocks (see Methods for details). Syn-S and Syn-M denote the two control Synthetic spaces, Synthetic-S and Synthetic-M. Evo-B+SynT and Evo-N+SynT represent two mixed spaces constructed by combining the metabolic building blocks from Evo with the synthetic reaction set. Syn-S+EvoT and Syn-M+EvoT represent two mixed spaces constructed by combining the enzymatic reactions from Evo with the synthetic building block library.

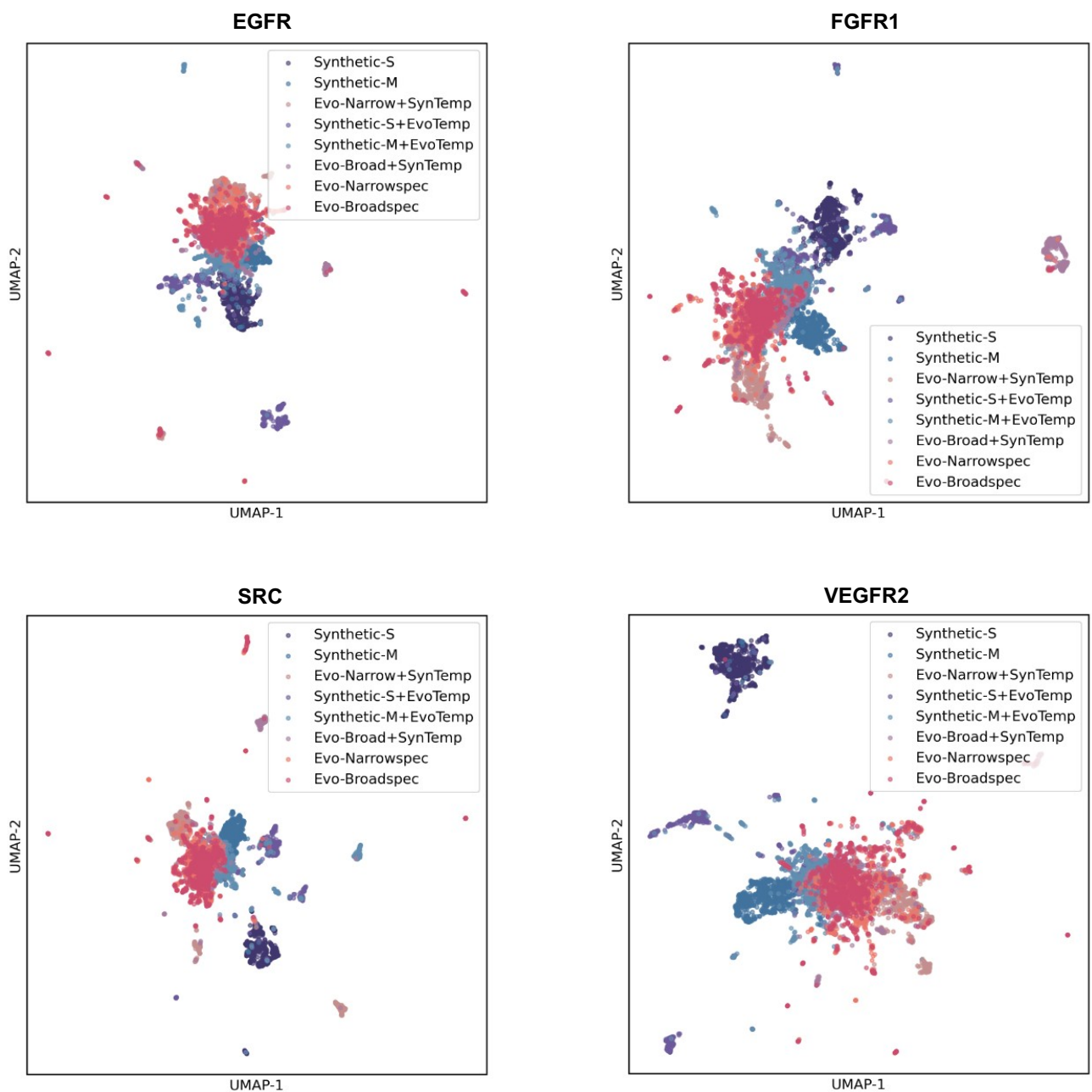

Supplementary Figure 44. UMAP distributions of the top 1000 QSAR-ranked molecules sampled from Evo and all control spaces across different targets. Evo-Narrow+SynTemp and Evo-Broad+SynTemp represent two

mixed spaces constructed by combining the metabolic building block libraries from Evo with the synthetic reaction set. Synthetic-S+EvoTemp and Synthetic-M+EvoTemp represent two mixed spaces constructed by combining the enzymatic reactions from Evo with the synthetic building block libraries.

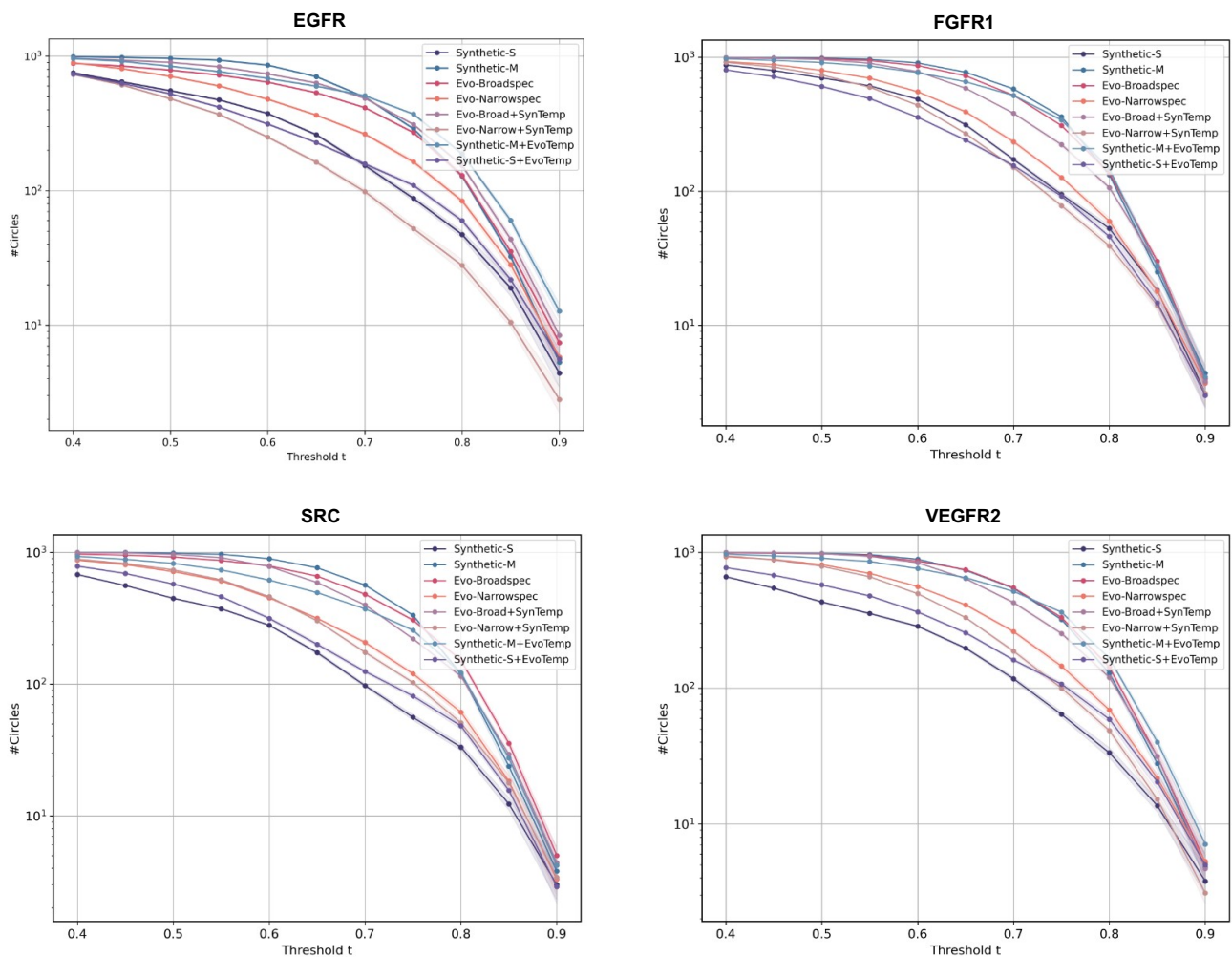

Supplementary Figure 45. Comparison of #Circles curves for the top 1000 QSAR-ranked molecules sampled from Evo and all control spaces across different targets. #Circles is a representative metric for evaluating molecular set diversity. The curves show the variation of #Circles at different threshold values t. Evo-Narrow+SynTemp and Evo-Broad+SynTemp represent two mixed spaces constructed by combining the metabolic building block libraries from Evo with the synthetic reaction set. Synthetic-S+EvoTemp and Synthetic-M+EvoTemp represent two mixed spaces constructed by combining the enzymatic reactions from Evo with the synthetic building block libraries.

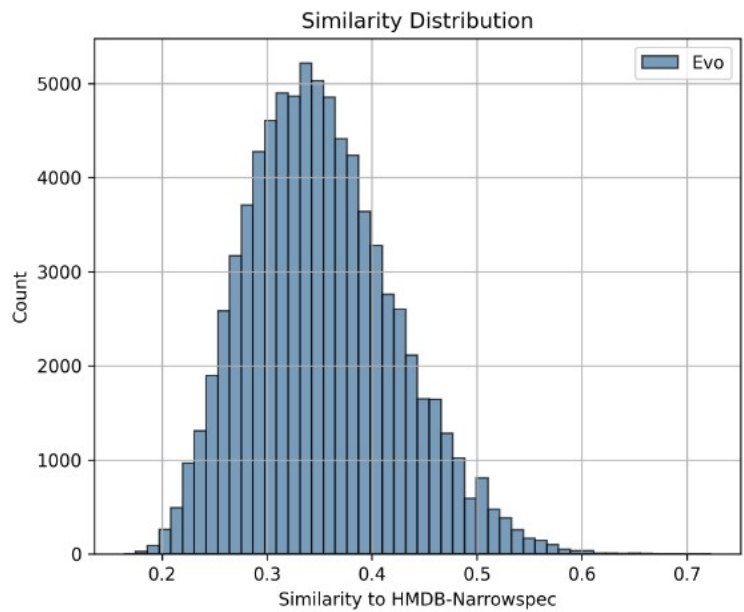

Supplementary Figure 46. Tanimoto similarity distribution between sampled molecules from the enumerated Evo space and the corresponding HMDB-Narrowspec molecules. HMDB-Narrowspec refers to the block library source of Evo-Narrowspec, as detailed in Methods.

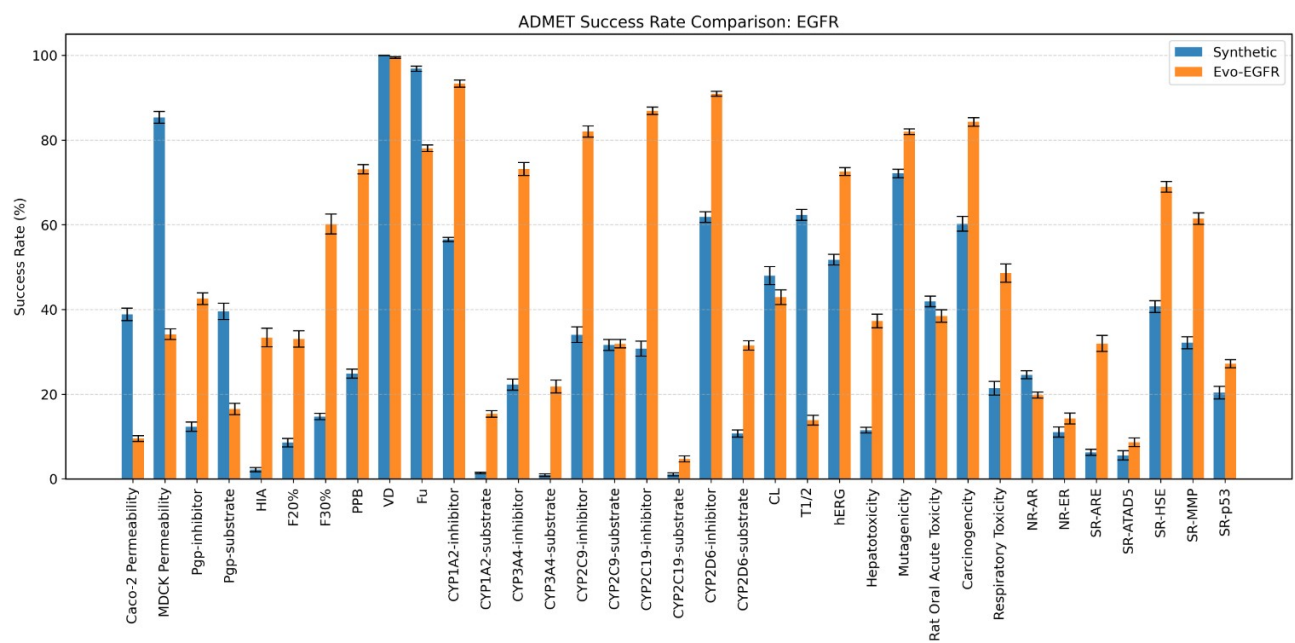

Supplementary Figure 47. Success rate statistics of 35 key ADMET properties (predicted using the enhanced ADMETlab 2.0) for the top 1k molecules ranked by QSAR scores under the EGFR target. Molecules were obtained from random sampling (10 runs, 8000 molecules each) of the enumerated Evo space and the control Synthetic space. Success rate is defined as the proportion of sampled molecules that satisfy the predefined property thresholds within a given space.

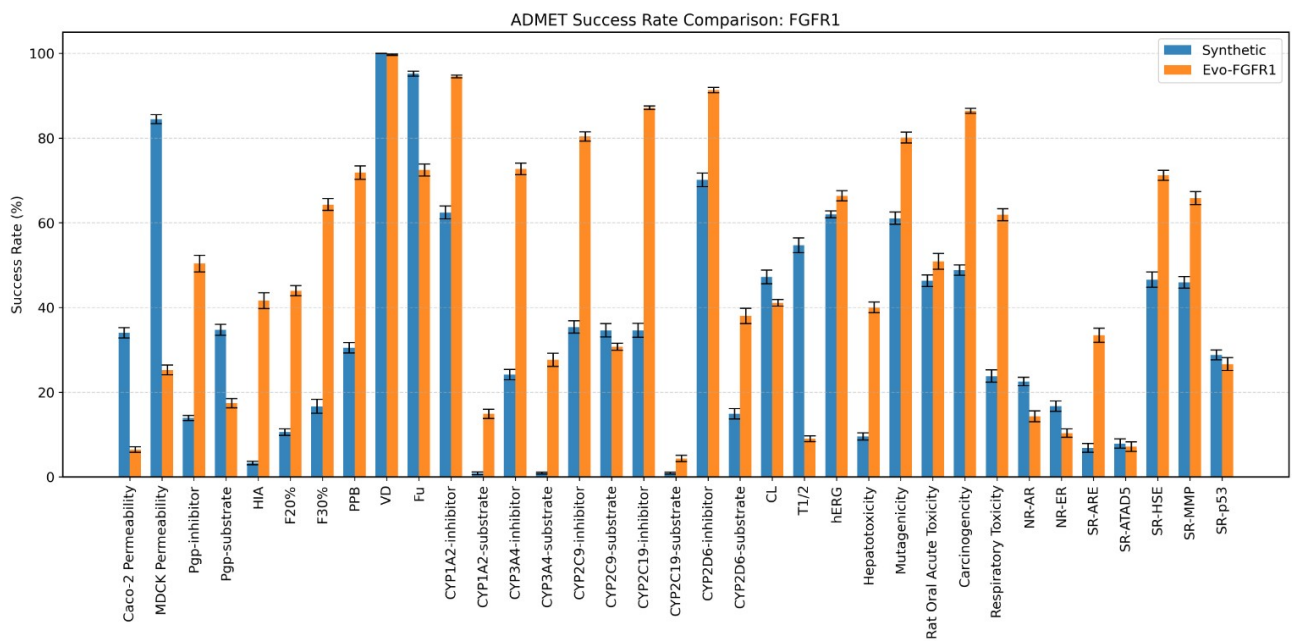

Supplementary Figure 48. Success rate statistics of 35 key ADMET properties (predicted using the enhanced ADMETlab 2.0) for the top 1k molecules ranked by QSAR scores under the FGFR1 target. Molecules were obtained from random sampling (10 runs, 8000 molecules each) of the enumerated Evo space and the control Synthetic space. Success rate is defined as the proportion of sampled molecules that satisfy the predefined property thresholds within a given space.

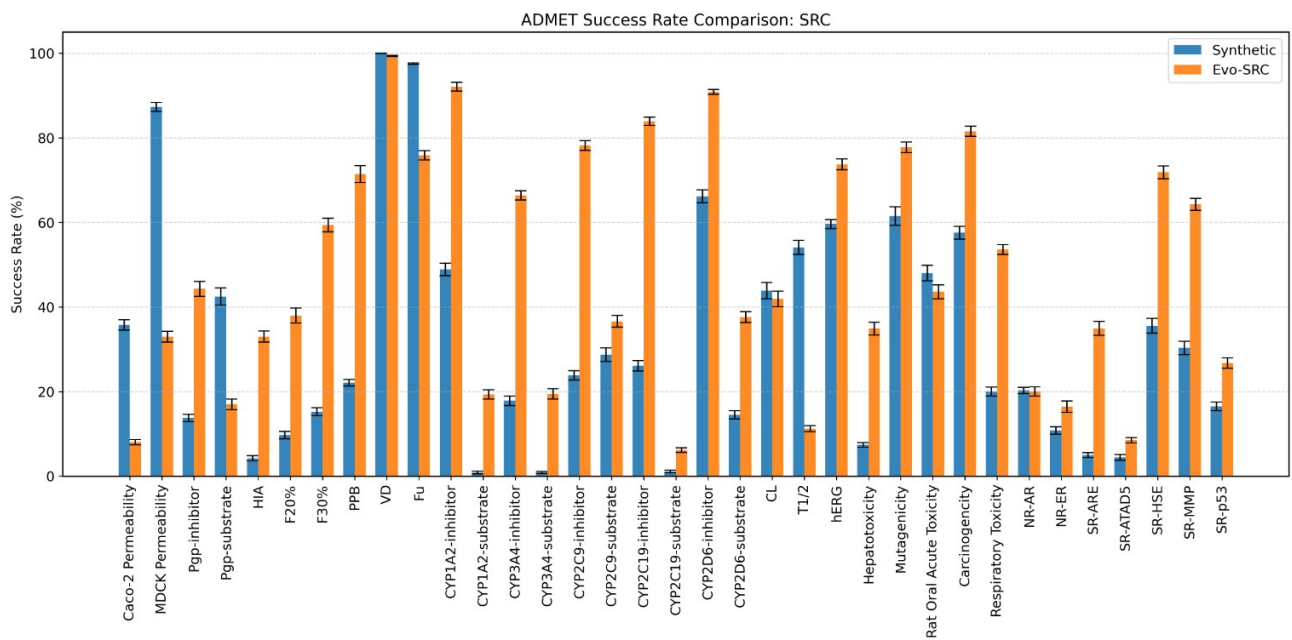

Supplementary Figure 49. Success rate statistics of 35 key ADMET properties (predicted using the enhanced ADMETlab 2.0) for the top 1k molecules ranked by QSAR scores under the SRC target. Molecules were obtained from random sampling (10 runs, 8000 molecules each) of the enumerated Evo space and the control Synthetic

space. Success rate is defined as the proportion of sampled molecules that satisfy the predefined property thresholds within a given space.

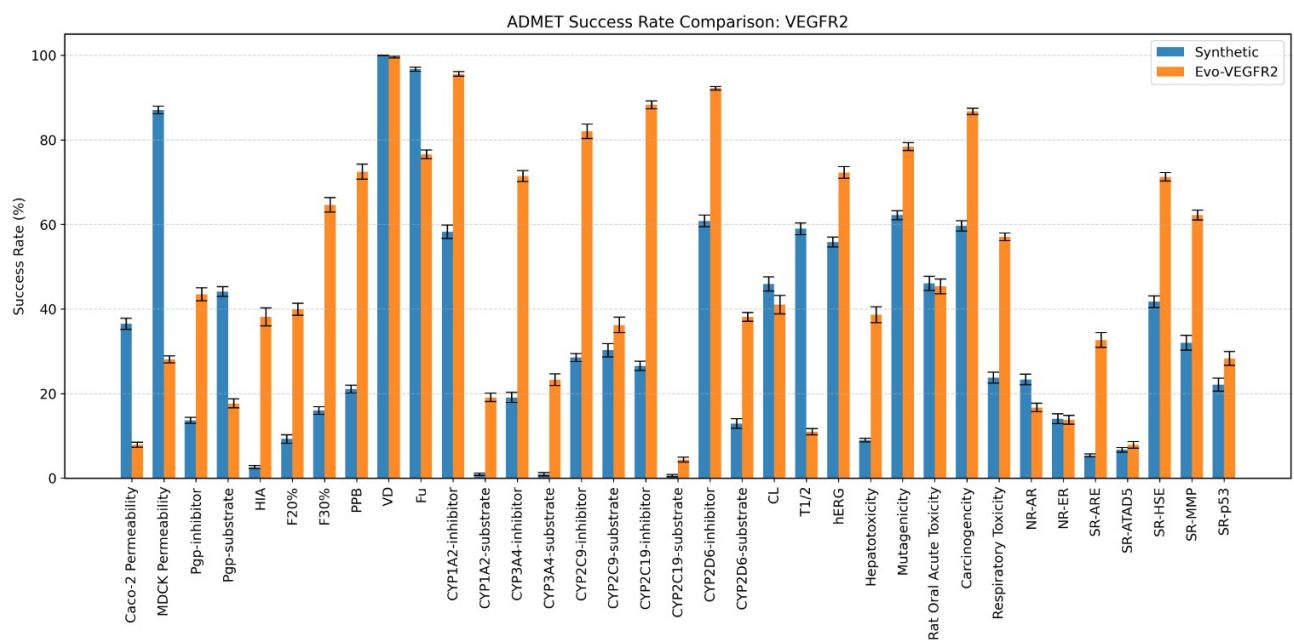

Supplementary Figure 50. Success rate statistics of 35 key ADMET properties (predicted using the enhanced ADMETlab 2.0) for the top 1k molecules ranked by QSAR scores under the VEGFR2 target. Molecules were obtained from random sampling (10 runs, 8000 molecules each) of the enumerated Evo space and the control Synthetic space. Success rate is defined as the proportion of sampled molecules that satisfy the predefined property thresholds within a given space.

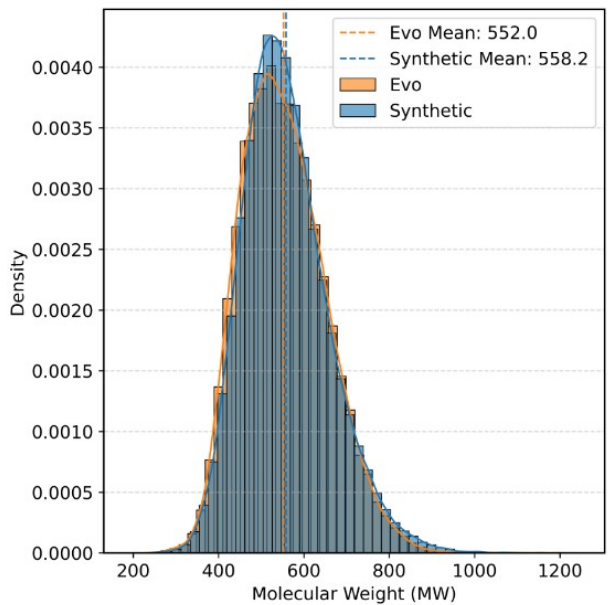
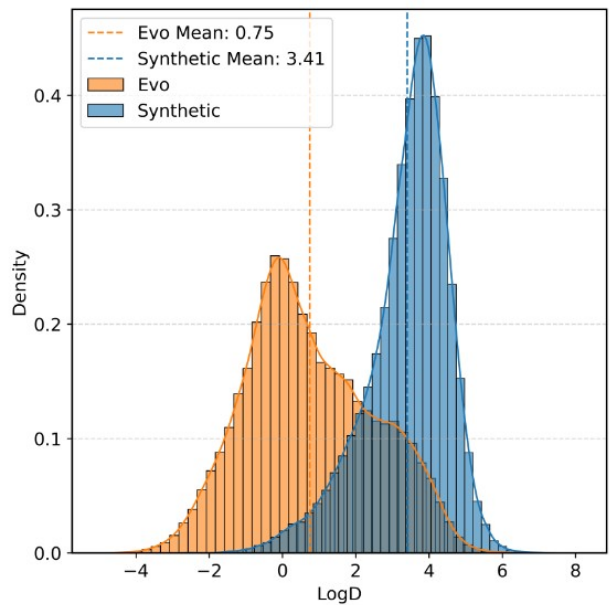
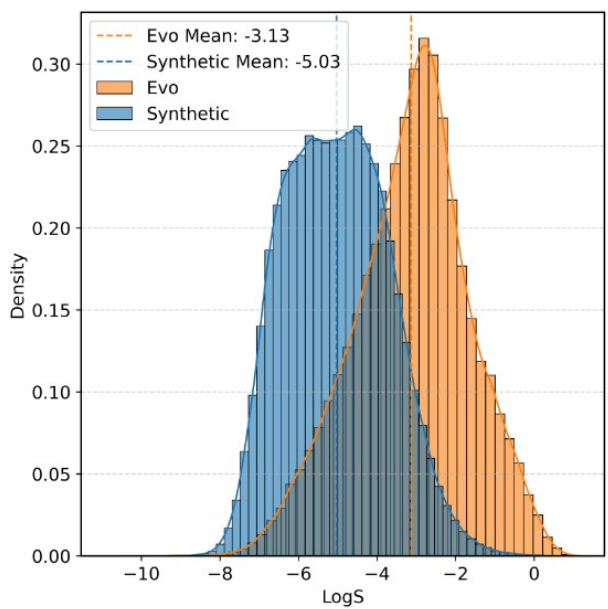

Supplementary Figure 51. Distribution of physicochemical properties for molecule sets obtained from random sampling of the enumerated Evo space and the control Synthetic space. Properties include molecular weight (MW), octanol–water distribution coefficient (LogD), and solubility (LogS).

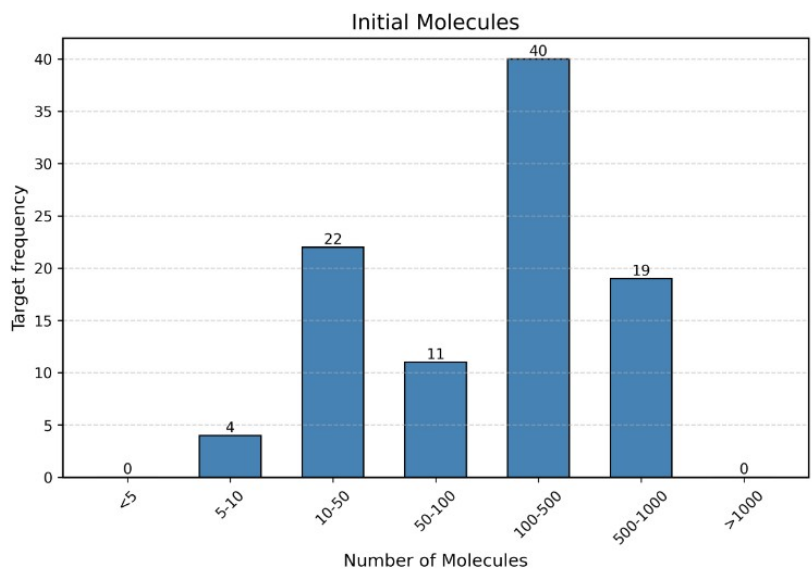

Supplementary Figure 52. Distribution of the initial number of molecules across the 96-target test set used in the Editing mode. The counts correspond to the final sets of molecules for optimization after data processing.

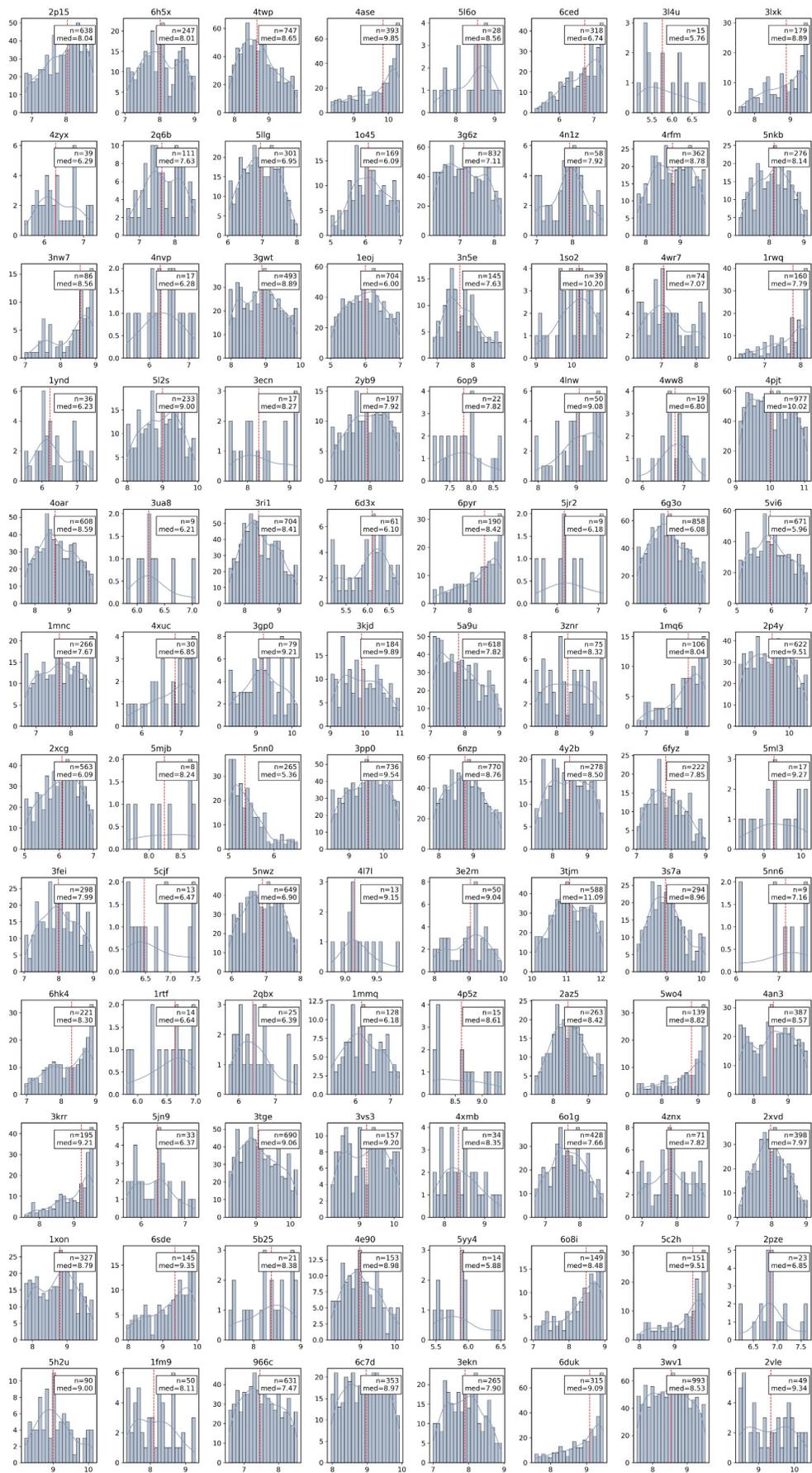

Supplementary Figure 53. Distribution of Uni-Dock scores across the 96-target test set in the Editing mode. During data processing, targets with an interquartile range (IQR) of Uni-Dock scores less than 2 kcal/mol were retained as the final valid targets.

#1

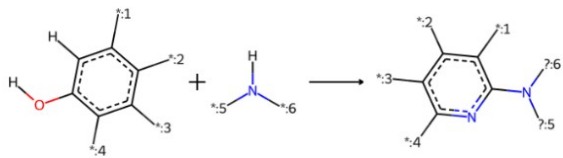

*J. Am. Chem. Soc.* 2022, 144, 39, 17797–17802

#2

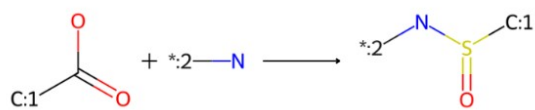

*Angew. Chem. Int. Ed.* 2019, *58*, 9054

#3

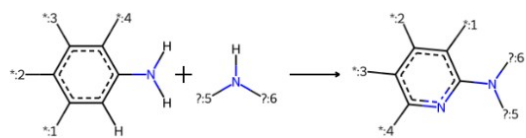

*J. Am. Chem. Soc.* 2022, 144, 39, 17797–17802

#4

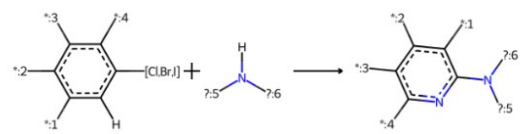

*J. Am. Chem. Soc.* 2022, 144, 39, 17797–17802

#5

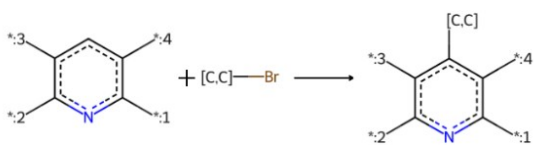

*Chem. Sci.,* 2024, 15, 12442-12450

#6

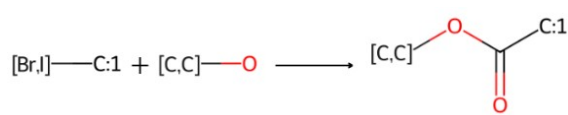

*Angew. Chem. Int. Ed.* 2024, 63, e202410928

#7

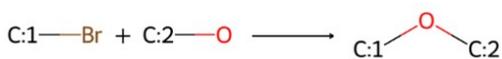

*Angew. Chem. Int. Ed.* 2021, 60, 20700

#8

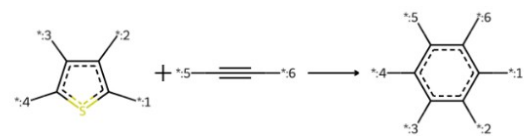

*Angew. Chem. Int. Ed.* 2019, 58, 12206

#9

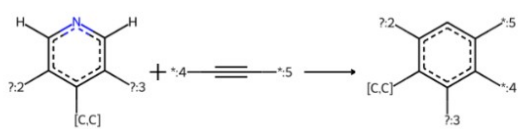

*Nat. Chem.* 2024, 16, 741–748

#10

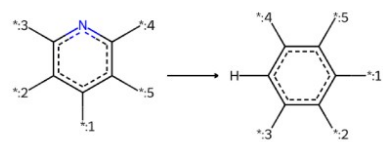

*Org. Lett.* 2024, 26, 20, 4280–4285

Supplementary Figure 54. Top 10 most frequent molecular editing reactions in the 96-target optimization tasks. The analysis considered molecules with score improvements greater than 0.5 kcal/mol for each target, and the occurrence frequency of different editing templates was calculated.

# Supplementary Tables

Supplementary Table 1. Table of average physicochemical properties of different natural product fragment libraries. We examined the natural product fragment library collected by Waldmann and colleagues, as well as those provided by Enamine and Life Chemicals. The evaluated properties were those in which natural products differ significantly from general synthetic molecules, including the fraction of sp³ carbons (Fsp³), number of oxygen atoms (O_Count), number of nitrogen atoms (N_Count), number of halogen atoms (Hal_Count), total number of rings (Num_Rings), number of aromatic rings (Num_AromaticRings), and number of non-aromatic rings (Num_NonAromaticRings).

|  | Enamine | Life Chemicals | Waldmann et al. |
|---|---|---|---|
| Fsp3 | 0.50 | 0.35 | 0.58 |
| O_Count | 1.86 | 2.22 | 2.33 |
| N_Count | 1.88 | 2.35 | 1.11 |
| Hal_Count | 0.29 | 0.23 | 0.08 |
| Num_Rings | 2.27 | 2.39 | 2.42 |
| Num_AromaticRings | 1.03 | 1.67 | 0.57 |
| Num_NonAromaticRings | 1.24 | 0.72 | 1.85 |

Supplementary Table 2. Table of average natural product–related physicochemical properties of sampled molecules from the Pseudo-NP space and the control synthetic chemical space across different targets. The evaluated properties were those in which natural products differ significantly from general synthetic molecules, including the fraction of sp³ carbons (Fsp³), number of oxygen atoms (O_Count), number of nitrogen atoms (N_Count), number of halogen atoms (Hal_Count), total number of rings (Num_Rings), number of aromatic rings (Num_AromaticRings), and number of non-aromatic rings (Num_NonAromaticRings).

|  | EGFR | | FGFR1 | | SRC | | VEGFR2 | |
|---|---|---|---|---|---|---|---|---|
|  | Pseudo-NP | Synthetic | Pseudo-NP | Synthetic | Pseudo-NP | Synthetic | Pseudo-NP | Synthetic |
| Fsp3 | 0.67 | 0.58 | 0.67 | 0.58 | 0.57 | 0.59 | 0.67 | 0.60 |
| O_Count | 5.26 | 4.63 | 5.90 | 3.93 | 7.03 | 4.43 | 6.37 | 4.63 |
| N_Count | 2.68 | 3.41 | 2.22 | 3.88 | 2.28 | 4.18 | 2.46 | 3.64 |
| Hal_Count | 0.17 | 1.30 | 0.36 | 1.10 | 0.13 | 0.74 | 0.50 | 0.84 |
| Num_Rings | 5.65 | 2.82 | 5.03 | 2.60 | 5.90 | 2.72 | 5.69 | 2.67 |
| Num_AromaticRings | 0.88 | 1.32 | 0.94 | 1.29 | 1.08 | 1.32 | 0.96 | 1.24 |
| Num_NonAromaticRings | 4.76 | 1.50 | 4.09 | 1.31 | 4.82 | 1.40 | 4.73 | 1.43 |

Supplementary Table 3. Comparison of the average number of Murcko scaffolds and average number of Circles before and after optimization by SpaceGFN across different targets. Both Murcko scaffold count and Circles count serve as metrics for assessing the diversity of molecular sets.

|  | Initial | Optimized | Improvement |
|---|---|---|---|
| Average Scaffold Number | 134.90 | 238.08 | 76.49% |
| Average Circles Number | 36.21 | 71.62 | 97.81% |